\documentclass[fleqn,usenatbib,useAMS]{mnras}
\usepackage{graphicx}	
\usepackage{amsmath}	
\usepackage{amssymb}	
\usepackage{multicol}        
\usepackage{bm}		
\usepackage{pdflscape}	
\usepackage{placeins}
\usepackage{float}
\usepackage{afterpage}

\DeclareRobustCommand{\VAN}[3]{#2}
\let\VANthebibliography\thebibliography
\def\thebibliography{\DeclareRobustCommand{\VAN}[3]{##3}\VANthebibliography}

\usepackage{graphicx}
\usepackage{float}

\usepackage{tabularx}
\usepackage{color}
\usepackage{url}

\usepackage{enumitem}
\usepackage{booktabs}
\usepackage{multirow}
\usepackage{subcaption}

\defcitealias{GamezMarin2025_PaperV}{Paper V}
\defcitealias{Gamez-Marin2024}{Paper IV}
\defcitealias{SantosSantos2024_anisotropic}{Paper III}

\newcommand{\zhigh}{z_{\rm high}}

\usepackage[T1]{fontenc}
\usepackage{ae,aecompl}

\usepackage{newtxtext,newtxmath}

\title[Kinematic satellite planes in TNG50]{A statistical look on kinematic planes of satellite galaxies II: The physics behind their early formation in TNG50 MW/M31-like galaxies}

\author[G\'amez-Mar\'in et al.]
{Mat\'ias G\'amez-Mar\'in,$^{1}$\thanks{E-mail: matias.gamez@estudiante.uam.es}
Rosa Dom\'inguez-Tenreiro,$^{1,2}$
Isabel Santos-Santos,$^{3}$
\newauthor
Diego Sotillo-Ramos,$^{1}$
Alexander Knebe$^{1,2,4}$
\\
$^{1}$Departamento de F\'isica Te\'orica, Universidad Aut\'onoma de Madrid, E-28049 Cantoblanco, Madrid, Spain \\
$^{2}$Centro de Investigaci\'on Avanzada en F\'isica Fundamental, Universidad Aut\'onoma de Madrid, E-28049 Cantoblanco, Madrid, Spain \\
$^{3}$Institute for Computational Cosmology, Department of Physics, Durham University, South Road, Durham, DH1 3LE, UK \\
$^{4}$International Centre for Radio Astronomy Research, University of Western Australia, 35 Stirling Highway, Crawley, Western Australia 6009, Australia}

\begin{document}
\label{firstpage}
\pagerange{\pageref{firstpage}--\pageref{lastpage}}
\maketitle

\begin{abstract}

We investigate the physical origin  of  kinematically persistent planes (KPPs) of satellite galaxies in a sample of 190 Milky Way (MW)/M31-like host-satellite systems drawn from the TNG50 simulation. 
Building on the identification of 46 early KPPs in a previous work, we analyse their formation in the context of the high-redshift evolution of the local Cosmic Web by tracking the deformation of the so-called Lagrangian Volumes (LVs) surrounding each system. Using a reduced tensor-of-inertia analysis, we characterise the time evolution of the principal directions of collapse and relate them to the clustering of satellite orbital poles. 
We find that in approximately 67\% of KPPs satellite orbital poles align with the LV direction of strongest collapse, $\vec{e}_3$, while a smaller fraction ($\sim20\%$) align with the intermediate axis, $\vec{e}_2$; alignments with the major axis are rare. These alignments are statistically distinct from random expectations and reflect the confinement of satellites to planar configurations normal to the corresponding LV principal directions.
We perform a kinematic analysis of satellite motion within KPPs, finding that vertical and radial motions relative to these KPPs decay early, leading to rotation-dominated, ``disky'' configurations. 
The characteristic timescales for satellites to settle onto a common orbital plane, for satellite
orbital pole clustering, and for LV shape evolution are found to be quasi-coeval, peaking at a Universe age T$_{\rm uni}\sim4$~Gyr, during the fast mass assembly phase of the host halo. 
These results support a scenario in which early KPPs are fossil remnants of high-redshift, anisotropic mass collapse driven by the local Cosmic Web  formation process in $\Lambda$CDM.

\end{abstract}



\section{Introduction}
\label{ch5:sec:intro}

The $\Lambda$CDM cosmological paradigm has been remarkably successful in explaining the formation and evolution of structures in the Universe, from the large-scale structure pattern of the Cosmic Web (CW) to galaxy clusters and individual galaxies. Nevertheless, discrepancies emerge when confronting model predictions with observations at smaller scales, especially in the regime of dwarf and satellite galaxies. Among these, one of the most long-standing challenges is the so-called 'satellite plane problem', considered as one of the main issues of the $\Lambda$CDM paradigm at short scales. The discovery that the satellites of the Milky Way (MW) are distributed along a thin plane  
\citep{Lynden76,Kunkel76}, and that a significant fraction of them share coherent orbital motions within that plane   \citep[e.g.,][]{Fritz18,SantosSantos2020I,Li21,Taibi24}  
have challenged  results from state-of-the-art cosmological simulations run within the $\Lambda$CDM model/paradigm  for decades \citep[see][ for a review]{Sales2022}. 

Similar planar configurations have been reported around M31 \citep{Koch06,McConnachie06,Metz07}, and other nearby galaxies beyond the Local Group \citep[e.g.,][]{Chiboucas13,Tully15,Muller17,Muller18,Heesters21,MartinezDelgado21,Paudel21,Karachentsev24}, giving rise to the question of whether such anisotropic and kinematically-coherent satellite distributions are common or exceptional features of galaxy formation.

A number of studies have examined the occurrence and persistence of satellite planes within cosmological simulations. \textit{Positional} planes -- i.e.,  sets of satellites whose positions can be fitted by a flattened ellipsoid along a time interval -- have been sometimes found, but their frequency used to be very low -- even less than 1\% depending on the methodology used by the authors.
On the other hand, 
positional planes exhibit continuously changing satellite membership, as some of their satellite members enter and leave the structures over relatively short timescales, -- of the order of the typical crossing time for these structures -- \citep[][]{Libeskind05,Libeskind09,Lovell11,Bahl14,Cautun15,Buck16,Ahmed17,SantosSantos2020I}.
While some positional planes -- or, more often, subsets of their satellites -- exhibit kinematic coherence \citep[e.g.,][]{Shao19,Samuel2021}, in some cases even over long timescales \citep[the so-called 'kinematically coherent planes of satellites'; see][]{Santos-Santos_2023,Gamez-Marin2024}, most positional planar configurations are short-lived. This transient nature has been reported in numerous studies \citep{Bahl14,Gillet15,Buck16,Maji17b,ZhaoXinghai2023,Xu23}.
However, in a recent study, \cite{Sawala25} suggested that these two apparently different predictions are a matter of perspective of the same prediction.

The series of works presented by \citet{SantosSantos2020I,SantosSantos2020II,Santos-Santos_2023,Gamez-Marin2024} (hereafter Papers I–IV) systematically addressed the plane of satellite problem using cosmological zoom-in simulations of 
two MW-like systems. These studies found that, while positional planes are indeed transient as a whole, a fixed subset (40\% in one case, and up to 80\% considering the two detected planes in the other system) of their member satellites exhibit long-term orbital coherence (for roughly half the Universe current age). \citet{Santos-Santos_2023} and \citet{Gamez-Marin2024} identified the so-called Kinematically-Persistent Planes (KPPs) of satellites: sets of satellites whose orbital poles, relative to the main galaxy, remain clustered around a fixed-in-time direction across long cosmic time intervals, see  its definition in the Glossary (Tab. \ref{tab:Glossary}) for more details.
In these systems, the concentration ellipsoid fitted to the satellites’ positions \citep[see][]{Cramer} remains thin and oblate over time, except for some short intervals.
The correspondence between positional and kinematic planes found in \citetalias{SantosSantos2024_anisotropic} suggests that the latter constitute a kind of `skeleton' of the observed positional planes.

While these works established the existence of KPPs and described their internal dynamical properties, the analysis was limited to two zoom-in simulations. \citet[][Paper V]{GamezMarin2025_PaperV} extended this framework to a statistically significant sample of MW/M31-like galaxies in the TNG50 of the IllustrisTNG simulation, quantifying the frequency of KPPs across different host-satellite systems. This study revealed that KPPs are present in approximately $24\%$ of host-satellite (HS) systems, increasing to about $40\%$ for systems with at least nine satellites. These results demonstrate that KPPs are not rare occurrences, and that those identified in 
\citetalias{GamezMarin2025_PaperV} at least, form early (around 4 Gyr in Universe age). 
Indeed, we note a tendency for 
recent studies based on large-volume  cosmological simulations to find higher frequencies of satellite planes than previously reported  \citep[e.g.,][]{HuTang24,Seo2024,Uzeirbegovic2024}.

\citetalias{GamezMarin2025_PaperV} made a comparison between KPPs and their extensions as coorbiting satellite sets or CS$(t)$ sets. 
A CS$(t)$ set for an HS  system at a Universe age T$_{\rm uni}=t$ is the set of its co-orbiting satellites  at this time, see Tab. \ref{tab:Glossary} for a more detailed definition.
Regarding its membership, the CS$(t)$ set consists of the satellites belonging to the kinematically persistent plane in this HS system, plus those non-KPP satellites that at time $t$ happen to temporarily co-orbit with the KPP.
Based on an analysis of the positional properties of KPPs and CS$(t)$ sets, \citetalias{GamezMarin2025_PaperV} confirmed that the former play the role of a kind of skeleton, ensuring the long term durability of the CS$(t)$ sets as positional planes.

Different theoretical frameworks have been proposed to explain the origin of planar and kinematically coherent satellite configurations. Some studies attribute their formation to 
processes occurring late relative to the epoch of early KPP formation in \citet{Gamez-Marin2024}, at T$_{\rm uni} \sim$ 4 Gyr. For example, 
group capture of satellites onto the central galaxy \citep{Lynden95,Li08,DOnghia:2008}, or satellites forming from tidal debris from past gas-rich interactions or mergers \citep[e.g.,][]{Hammer:2013,Kroupa15,Banik22}
, while others suggest that the dynamical influence of triaxial dark matter halos or rotating substructures could induce temporary alignments \citep[see, e.g.,][]{Shao19,Wang20}.  
Additionally, it has been proposed that the infall of a companion massive satellite -- similar to the Large Magellanic Cloud (LMC) -- could bring a group of correlated satellites, 
in such a way that it may generate a transiently aligned system \citep[see also discussions in \citetalias{SantosSantos2024_anisotropic};][]{Samuel2021,Garavito-Camargo21,Zhao2025}. 
Finally, the consequences of mergers  have also been subjected to investigation, either in large-volume simulations \citep[see, e.g.,][in the Illustris TNG simulation]{Kanehisa2023}, or in zoom-in experiments \citep{RodriguezCardoso2026}, linked to a GSE-like merger at early times.
 Although diverse, these scenarios struggle to reproduce both the strong spatial flattening and the  
kinematic coherence observed in the MW, motivating the search for an  
origin that can naturally account for these features, either at early cosmological ages -- T$_{\rm uni} \lesssim 4$ Gyr -- or later on.

Understanding the origin of these 
kinematic structures requires considering the environment in which galaxies form. In the cosmological context, the assembly of halos and galaxies occurs within the anisotropic flow field of the CW, emerging from the gravitational amplification of primordial density perturbations. The large-scale organization of these structures can be described within the framework of the Zel'dovich approximation \citep{Zeldovich:1970}, hereafter ZA. Due to multistreaming,\footnote{i.e., the phenomenum by which particles in the ZA cross paths, such that multiple particle streams, with different velocities, coexist at the same spatial point.} in ZA caustics disappear soon after their formation, contrary to results from numerical simulations.  
The Adhesion Model \citep[e.g.][hereafter AM]{Gurbatov:1989,Kofman:1990,Buchert:1992,Gurbatov:2012} refines the ZA picture by adding a viscosity term to the equations of motion, viscosity that vanishes everywhere except at caustics. In this way, the   multistreaming issue is solved and long-lasting caustics form.

The ZA combined with the AM describes the progressive formation of singularities --first walls, then filaments, and finally clumps-- that initially emerge on small scales and gradually grow, merge, and percolate as evolution proceeds. 
Mass elements flow from less dense regions towards caustics, thickening them at the expenses of matter in the former  \citep[i.e., the so-called migrant mass flows,][]{Kugel2024}, whose density decreases. Specifically, mass flows first towards walls causing three-dimensional voids, then mass flows along walls  towards filaments --either secondary filaments formed within walls, or those located at wall intersections-- feeding filaments and emptying walls. 
Subsequently, mass elements in filaments flow towards nodes or clumps, where matter accumulates. This hierarchical process continues until the Universe becomes dominated by the cosmological constant, whose repulsive effect eventually halts further large-scale collapse. Altogether, this sequence of events shapes the anisotropic cosmic environment in which galaxies form and evolve 
-- see \cite{Robles2026} for details -- , governing the development of structures even at the scales of the Local Volume \citep[see, e.g.,][for recent results]{Valade2024}.

Several studies have shown that the acquisition of angular momentum and the spin alignment of galaxies and their satellites are influenced by the surrounding CW \citep[e.g.,][]{Libeskind14,Tempel15,Welker18,Libeskind18,Welker20}. 
For example, \citet{Libeskind14}
and \cite{Dupuy22} found that subhalos exhibit preferential infall directions onto the host halo aligned with the principal axes of the local tidal and shear tensors. Indeed \citet{Libeskind15} reported correlations between the normal vectors to the satellite planes observed in the Local Group and the local shear tensor derived from peculiar velocity fields. Along these lines, \citet{Xu23} identified a halo in the TNG50-1 simulation showing properties akin to the MW’s plane of satellites, embedded in a sheet-like environment reminiscent of the MW’s own surroundings, and attributed the formation of its plane to the anisotropies of the local large-scale environment.

Building on the findings reported in \citetalias{SantosSantos2024_anisotropic}, \citetalias{Gamez-Marin2024} extended the analysis by connecting the formation of KPPs to the early evolution of the local CW. By characterizing the host environment in terms of the reduced Tensor of Inertia (TOI) using the so-called Lagrangian Volumes \citep[LVs,][]{Robles:2015,Robles2026}, they demonstrated that the same physical processes driving the early anisotropic collapse within the CW significantly affect satellite dynamics, shaping their orbital trajectories well before satellites become gravitationally bound to their central galaxy
and even prior to the first infall of all KPP satellite members into the host halo -- thereby implying an \textit{early} KPP formation scenario.
 In this scenario, orbital pole clustering --and thus the kinematic coherence observed in KPPs-- is established at high redshift, during an epoch when satellites and hosts are still in their early, violent formation phase. 
\citetalias{Gamez-Marin2024} therefore proposes a cosmological origin for satellite plane kinematics, where the observed in-plane coherence arises naturally from the anisotropic mass flows and early structure formation processes inherent to the $\Lambda$CDM framework, requiring no additional assumptions. It is worth remarking that the formation of kinematic planes (either short or long-lived ones) at late Universe ages can also be caused by other different mechanisms, such as the aforementioned group capture at late times and the presence of an LMC-like galaxy with its own satellite sub-system.
 
This study continues the investigation initiated in \citetalias{GamezMarin2025_PaperV},  
moving the focus to the physical processes driving KPP formation within the context of the surrounding CW. To this end, we analyze the large-volume, high-resolution TNG50 simulation, which enables a statistically significant characterization of HS systems embedded in diverse environments. In particular, we focus on the role of the local CW in shaping KPPs. 
By connecting the emergence of KPPs to the geometry and dynamics of the CW, this work aims to bridge the gap between the internal dynamics of satellite systems and the large-scale cosmological processes that govern their formation and evolution.

The paper is organized as follows: 
Sec.~\ref{ch4:sec:Simus} introduces the simulations used, and the selection criteria for host galaxies and their satellite sets.  
Sec.~\ref{sec:KPP-identification}  is devoted to reviewing the method used for KPP selection and the resulting fraction of early KPPs in the TNG50 simulation.
Sec.~\ref{ch5:sec:Vel_JstackPlane} presents the analysis of satellite motions with respect to the mathematical plane defined by the maximum co-orbitation axis, $\vec{J}_{\rm stack}$. The evolution of the mass density around the galaxy-to-be objects is discussed in Sec.~\ref{ch5:sec:LVs_intro}, where we introduce the concept of Lagrangian Volume (LV) and describe their main properties. Sec.~\ref{ch5:sec:AlignPDir} examines the alignment of satellite orbital poles relative to the LV's principal directions of compression, while Sec.~\ref{ch5:sec:TrajectoriesLV} focuses on how satellite positions and  velocities relate to the  dynamics of the LV.
An analysis of the distribution of the characteristic timescales arising from this work is presented in Sec.~\ref{ch5:sec:Timescales}, along with their mutual relationships.
Sec.~\ref{sec:Dicussion}, is devoted to compare our results with those obtained in \citetalias{Gamez-Marin2024} and others' made recently available, as well as a test of robustness.
Finally, Sec.~\ref{sec:SummConclu}, summarizes our main findings and conclusions.

This paper is complemented with one Supplementary File available online, where the information provided in this manuscript about properties of individual HS systems is extended to other 33 such systems out of the 46 where KPPs have been identified (hereafter KPP-HS systems) listed in Tab.~\ref{ch5:tab:DataTable2}. 
Figures in the Supplementary File are referred to as Figs.~A.1, A.2, and A.3.

In addition, a glossary is provided in Tab. \ref{tab:Glossary}, where definitions are provided for the main terms and concepts used throughout the paper.


\section{Simulation and host-satellite systems}
\label{ch4:sec:Simus}

In this work 
we make use of one of the IllustrisTNG simulations for galaxy formation \citep{Marinacci18,Naiman18,Nelson18,Pillepich18a,Springel18}. Details on this simulation have been given in \citetalias{GamezMarin2025_PaperV}. For the sake of completeness, a brief summary is given here.

We adopt the highest resolution flagship of IllustrisTNG, the TNG50-1 (hereafter TNG50) simulation \citep{Pillepich19,Nelson19b}. TNG50 is one of the few simulations able to resolve and follow the evolution of low-mass -- i.e., dwarf satellites --  galaxies in a large cosmological context. This is a necessary condition to statistically study the origin and frequency of KPPs around central galaxies with a well-resolved and numerous satellite population. 
Details on the parameters of the cosmological model and simulation can be found in Tab.~\ref{ch5:tab:CMSimu}, including the masses of the dark matter and baryonic particles.

\begin{table}
\centering
\footnotesize 
\caption{Some parameters of the  cosmological model and simulation blocks for the TNG50 simulation.
}
\begin{tabular}{|  l | l | l | l |}
\hline
  \multicolumn{2}{|c |}{Cosmological model} & 
 \multicolumn{2}{c |}{Simulation} \\
\hline
    Parameter & Value & Parameter & Value \\
\hline
   $\Omega_{\rm m}$ & 0.3089 & $z_{\rm high}$ & 20.05  \\
   $\Omega_{\rm b}$ & 0.0486 & $L_{\rm box}$ [cMpc] & 51.7 \\
   $\Omega_{\Lambda}$ & 0.6425 & $m_{\rm bar}$ [M$_{\odot}$] & $8.5\times10^4$ \\
   $H_0$ [km/s/Mpc]        &        67.74 & $m_{\rm DM}$  [M$_{\odot}$] &  $4.6\times10^5$ \\

\hline
\end{tabular}
\label{ch5:tab:CMSimu}
\end{table}

The TNG50 simulation employs the \textsc{AREPO} code \citep{Springel10AREPO} to follow coupled magneto-hydrodynamics and self-gravity, implementing prescriptions for star formation, stellar feedback, metal enrichment and black hole physics. See \citet{Weinberger20} and references therein for more details.
Groups in the TNG database are identified using the Friends-of-Friends algorithm on particles, while subhalos are structures within groups identified using the \textsc{Subfind} code \citep{SpringelWhite01}.

Our analysis refers to the MW/M31-like galaxies presented in \citet{Pillepich24_MWM31} as host galaxies, and study their satellite populations. 
Different studies have been presented about this particular sample of galaxies, analyzing  e.g.:  its satellite populations \citep{Engler21}; mergers and disk survival \citep{SotilloRamos:2022}; stellar bulges and halos \citep{Gargiulo:2022}.

Host (also termed central) galaxies in HS systems were selected after imposing several conditions that had to be simultaneously met:
\begin{enumerate}[label = {\roman*)}, wide, left=0pt, labelsep=0em,labelindent=0pt]
    \item Galaxies' stellar masses within 30 kpc from their centre must be in the range $M_{\ast} = 10^{10.5}-10^{11.2}$ M$_{\odot}$.
    \item Galaxies must have a disk-like stellar morphology 
   \item 
   Galaxies must have no neighbouring galaxy with a stellar mass of $M_{\ast}\geq10^{10.5}$ M$_{\odot}$ within a 500 kpc distance at $z=0$. Additionally, 
   the host halo mass is limited  to values below $M_{\rm 200c}<10^{13}$M$_{\odot}$. 
   This helps to exclude systems that have undergone major mergers at low redshift, as well as MW/M31-like galaxies in compact groups.
\end{enumerate}

These conditions return a total of 198 MW/M31-like galaxies, 
from which we exclude a subgroup of 8 central galaxies that are FoF satellites of a more massive galaxy,
keeping a sample of 190 MW/M31-like galaxies, hereafter the P24 sample. 
We recall that we aim at characterizing the local density field evolution surrounding these systems and relate it to the formation of KPPs. Such characterization would be strongly affected by the presence of a more massive companion.

For each host galaxy, we consider as satellites candidates  those galaxies gravitationally-bound to the host halo at $z=0$\footnote{Imposing the requirement that satellites persist to $z=0$ is consistent with the kinematic \textit{persistence} property that we seek to characterize. Furthermore, the only presently available observational data on planes of satellite galaxies pertain to systems at $z=0$.}, 
according to the \textsc{Subfind} halo finder.
We apply a mass cut of a minimum of 10 gravitationally bound  stellar particles to each satellite, which translates in a satellite stellar mass lower-limit of $M_{\rm \ast,cut}=8.5 \times 10^5$ M$_{\odot}$\footnote{The fraction of satellites whose $z=0$ halo has less than 50 dark matter particles is only 0,22\%. 
}.

To confirm that a given galaxy is a satellite, we follow back in time its trajectory from $z=0$ to high redshift. 
We demand that satellites have infall into the host halo at least 500 Myr prior to T$_{\rm uni}(z=0)$. Backsplash galaxies are included in the sample if they remain bound at $z=0$ as well.

The P24 satellites are of cosmological origin, as guaranteed by the use of the SubhaloFlag in the TNG Subfind catalog. This selection excludes spurious baryonic fragments (flag = 0) in favor of genuine satellites (flag = 1) that are consistently traced along the merger trees to ensure reliable evolutionary histories.

According to the selection criteria described above, the average number of satellites per host galaxy in the P24 sample  is $N_{\rm sat, aver}$ = 11$^{+9}_{-4}$, a rather high dispersion extending up to more than 40 satellites. 
Our results are consistent with \citet{Engler21}, despite these authors applying more restrictive selection criteria to their satellite sample.


\section{KPP identification and their frequency of occurence in TNG50}
\label{sec:KPP-identification}

KPPs in  HS systems have been identified through the so-called `Scanning of Stacked Orbital Poles Method', or $\vec{J}_{\rm stack}$ method, see \citet{Santos-Santos_2023}.
For a given host-satellite system,  we first determine the $\vec{J}_{\rm stack}$  direction as an axial vector\footnote{No distinction is made between those satellites rotating in the same and those rotating  in the opposite sense.}, fixed in time, around which a maximum of the satellite member orbital poles cluster over time,  within an aperture angle of $\alpha_{\rm crit} = 36.^{\circ}87$, see details in \citetalias{SantosSantos2024_anisotropic}. The $\vec{J}_{\rm stack}$-plane in a given HS system is defined as the plane normal to $\vec{J}_{\rm stack}$, going through the central galaxy centre of mass (c.o.m.), see Tab. \ref{tab:Glossary}.
KPP satellite members are then identified through an analysis of their pole conservation and clustering. They are asked to stay within an angle aperture of  $\alpha_{\rm crit}$ from $\vec{J}_{\rm stack}$ for at least 4 Gyr between a Universe age
 T$_{\rm uni}\simeq 6$ Gyr (a typical value for the halo mass stabilisation timescale in KKP-HS systems, $T_{\rm no-fast}$\footnote{$T_{\rm no-fast}$ timescale is defined as the Universe age  when the violent, fast-phase of halo mass assembly ends. See \citetalias{GamezMarin2025_PaperV} for more details.}, see values in Tab. \ref{ch5:tab:DataTable2}) up to $z=0$.

In this work we focus only on ``significant'' KPPs,  i.e., those with  more than $N_{\rm KPP}^{\rm min}$ = 5 satellites members, and where the fraction of satellites in KPPs 
($f_{\rm KPP} = \frac{N_{\rm KPP}}{N_{\rm sat}}$) represents at least a $25\%$ of the total satellite population.
No KPP structures fulfilling the previous requirements have been identified in HS systems with $N_{\rm sat}<9$, 
making a total of 67 HS systems in the P24 sample. Such low-$N_{\rm sat}$ systems are considered as \textit{non-eligible} for our study, and excluded for further analyses\footnote{We have checked that these selection criteria for significant KPPs do not imply any stellar mass bias between KPP and non-KPP satellites.}.

HS systems with $N_{\rm sat}\geq9$ (123 of them) will be referred to as the \textit{eligible} sample.  46 of them host KPPs according to the selection criteria just mentioned, resulting in a frequency of occurrence of 24\%  or 37\% relative to all (190) or just to the elegible (123) systems in the P24 sample, respectively.

All KPP-HS systems are listed in Table \ref{ch5:tab:DataTable2}, where we give their halo identity numbers at $z=0$ and several parameters to be defined throughout this work. 
Besides, in Table 1 of \citetalias{GamezMarin2025_PaperV} we can find the values of important parameters such as: 
the total number of satellites identified,  $N_{\rm sat}$, and the fraction of them in KPPs, $f_{\rm KPP} = \frac{N_{\rm KPP}}{N_{\rm sat}}$, regardless of the sense of rotation.

KPP-HS systems are ordered in Tab.~\ref{ch5:tab:DataTable2} 
roughly according to their satellite richness: the first block includes those KPP-HS systems with $f_{\rm KPP}>30\%$, the second block those with $25\% \leq f_{\rm KPP}\leq 30\%$, and the third block includes those systems with a low $N_{\rm sat}<20$ (note that the latter are biased against low $f_{\rm KPP}$ values).

We recall here that \citetalias{GamezMarin2025_PaperV} showed that most early KPP-HS systems show well-defined two-phase mass assembly histories, reflecting faster evolution than nonKPP-HS systems, which in turn sustain their dynamical activity for longer cosmic periods. Moreover, within KPP-HS systems, KPP satellites are statistically-distinguishable from non-KPP satellites in that they show larger pericentric distances and higher specific orbital angular momenta, characteristics which make them more resilient to survive to $z=0$.


\section{The $\vec{J}_{\rm stack}$-plane formation from the velocity and position spaces perspective}
\label{ch5:sec:Vel_JstackPlane}

\subsection{Velocity space}
\label{ch5:sec:VelSpace}
 
In Papers IV and V the kinematic-morphological parameter $\kappa_{\rm rot}$ is used to analyze the disky or velocity-dispersion dominated character of KPP sets of satellites in two
KPP-HS systems, their time evolution and the differences relative to non-KPP satellites.

The $\kappa_{\rm rot}(t)$ parameter is defined at cosmic time T$_{\rm uni}=t$ as the median value, over all satellites in a given system, of the fraction of  kinetic energy in ordered rotational motion about the host galaxy.
Satellites are treated as massive point particles, and velocities  are measured in a cylindrical coordinate system centred on the host galaxy c.o.m., with the $Z$-axis aligned with the $\vec{J}_{\rm stack}$ vector. In this frame, rotational motion corresponds to the azimuthal ($\phi$) component of the velocity.

In an analogous way, we further decompose the satellite kinetic energy into vertical and radial components. We define the $\kappa_{\rm z}(t)$ as the median fraction of kinetic energy parallel to $\vec{J}_{\rm stack}$, and $\kappa_{\rm rad}(t)$ as the median fraction associated with radial motions in the same cylindrical reference frame, i.e. motions toward or away from the $\vec{J}_{\rm stack}$ axis.
Formally, for the $i$-th satellite member -- $i$ = 1,2, ..., $N_{\rm KPP}$ --, the $\kappa_{z, i}(t)$ is given by,
\begin{equation}
\kappa_{z,i}(t) = (v_{z, i}(t) / v_i(t) )^2
\label{krot}
\end{equation}

where $v_{z, i}(t)$ is the $i$-th satellite vertical velocity relative to the system centre of velocity, and $v_i$ is the magnitude of its  velocity at T$_{\rm uni} = t$.
An equivalent definition -- through median values -- is applied to $\kappa_{\rm rad}(t)$,  using the radial velocity component. 
Finally,  $\kappa_{\mathrm{rot},i}(t)$ refers to the $i$-th satellite tangential motion, $v_{\phi, i}$, as mentioned.
Together, these quantities provide insights into the trajectories of KPP satellites before they settle onto the $\vec{J}_{\rm stack}$-plane and ultimately form the KPP.

\begin{figure*}
\centering
\includegraphics[width=0.95\linewidth]{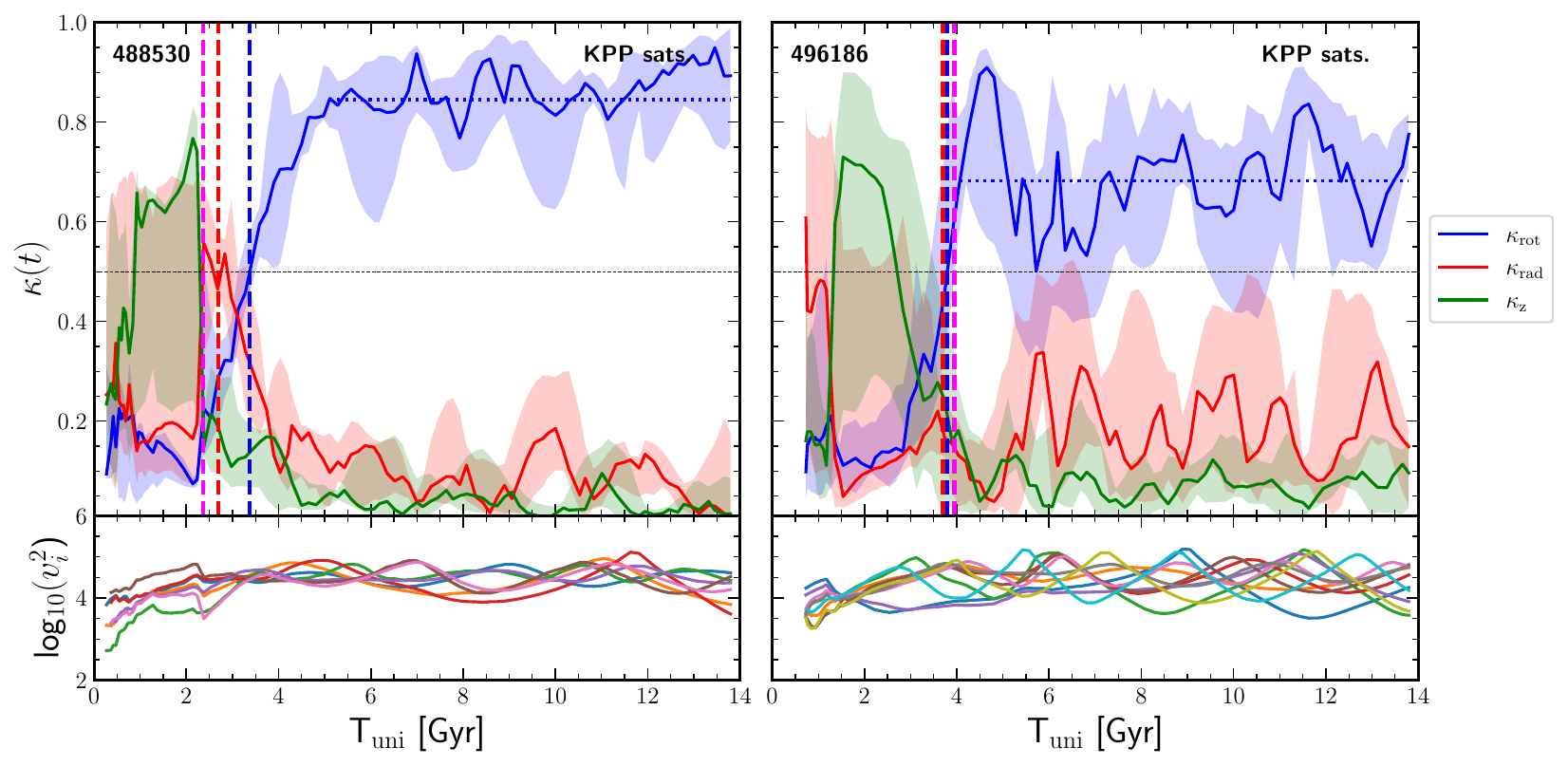}\\
\caption{\textit{Upper Panels}: Median values and percentiles for the satellite kinetic energy fractions in a cylindrical coordinate system for KPP satellites in two examples of KPP-HS systems. Blue, red and green curves stand for $\kappa_{\rm rot}$, $\kappa_{\rm rad}$ and $\kappa_z$, respectively (see Sec.~\ref{ch5:sec:VelSpace} for their definitions).
The coordinate system is centred at the centre of mass and velocity of the host galaxy, with its $Z$ axis aligned with the corresponding $\vec{J}_{\rm stack}$ vector of each KPP-HS system.
Blue, magenta and red dashed vertical lines mark the $T_{\rm Krot}^{0.5}$, $T_{\rm cluster}^{\rm Jstack}$ and $T_{\rm dJs}$ timescales for each system. A blue dotted horizontal segment marks the $\kappa_{\rm rot}^{\rm eq}$ value for each system, with the starting point at $T_{\rm Krot}^{\rm eq}$.
\textit{Lower Panels}: The  velocity square for each of the early KPP satellites in the two example HS systems.}
\label{ch5:fig_VI:vel_ratios_KPPHS}
\end{figure*}

In Fig.~\ref{ch5:fig_VI:vel_ratios_KPPHS} and upper rows of Fig.~A.1, we show the evolution of the $\kappa_{\rm rot}(t)$, $\kappa_{\rm z}(t)$ and $\kappa_{\rm rad}(t)$ curves, represented as blue, green and red lines, respectively, for KPP satellites in 33 KPP-HS systems.
Shaded bands represent their corresponding interquartile range. The dotted, horizontal line marks the $\kappa_{\rm rot}(t) = 0.5$ value, taken from \cite{Sales:2012} as the separation value between a ``disky'' rotation-dominated satellite system from a dispersion-dominated one.
Its intersection with the $\kappa_{\rm rot}(t)$ curve provides the $T_{\rm Krot}^{0.5}$ timescale for the HS system, listed in Tab.~\ref{ch5:tab:DataTable2}, while a blue dotted horizontal segment marks their corresponding $\kappa_{\rm rot}^{\rm eq}$ value (see definition below). 
The lower rows of Fig.~A.1 shows the respective curves for the non-KPP satellite population of each KPP-HS system.
In the lower panels of Fig.~\ref{ch5:fig_VI:vel_ratios_KPPHS} we plot the $[v_i(t)]^2$ curves (velocity square) for each of the KPP satellites of the two example HS systems. 
A common pattern can be appreciated in the evolution of a satellite  velocity:
i), First, a phase driven by the local Universe expansion.
ii),  A bound regime, once the satellite has been captured by the host, with oscillating velocity values (maxima at pericenter, minima at apocenter). These maxima and minima have almost constant values, as expected in a Keplerian system.
 And, 
iii), in between, a satellite infall period towards the plane and the host.

The three $\kappa_{\rm rot}(t)$, $\kappa_{\rm z}(t)$  and $\kappa_{\rm rad}(t)$ curves show similar patterns across time for KPP sets, among the different HS systems: 
i) $\kappa_{\rm rot}(t)$ is an increasing curve at early times, with KPP satellite sets transitioning from velocity-dispersion-dominated configurations at high redshift to disky configurations from $T_{\rm Krot}^{0.5}$  onwards,
ii) $\kappa_{\rm z}(t)$ curves  tend to have relatively high values at early Universe ages, indicating that satellites follow trajectories whose velocity has a rather high $v_{z}$ component. 
The $\kappa_{\rm z}(t)$  value then decreases, either rapidly (over less than one Gyr), or more gradually, until it reaches a low and typically stable value, apart from occasional fluctuations.

Two main well-marked possibilities emerge as the $\kappa_{\rm z}(t)$ curve reaches low values. First, the reduction in the $\kappa_{\rm z}(t)$ curve is coeval with a rapid increase in $\kappa_{\rm rot}(t)$, reaching values up to $\kappa_{\rm rot}=0.5$. In this case, the vertical blue-dashed line drawn through $T_{\rm Krot}^{0.5}$ almost coincides with the moment when the $\kappa_{\rm z}(t)$ curve reaches a low, stable value. Some of the KPP-HS systems that show this behavior are IDs (at $z=0$) \#436932, 475619 and 496186, see panels in Fig.~A.1.
This suggests that KPP  satellites adopt disk-like configurations 
while the vertical motion relative to the $\vec{J}_{\rm stack}$-plane decreases.
In other systems (e.g. ID\# 428177, 432106 and 470345), the decrease in the $\kappa_{\rm z}$ values of KPP satellites is accompanied by an increase in their $\kappa_{\rm rad}(t)$ curve, indicating that satellites transition to trajectories with a higher radial velocity component relative to the 
axis $\vec{J}_{\rm stack}$. This results in a delay between the stabilisation of $\kappa_{\rm z}(t)$ to low values, and the establishment of disky motion, at $T_{\rm Krot}^{0.5}$.

A third interesting situation clearly distinct from the aforementioned ones is observed in systems like ID\# 422754, 430864, 456326, 506720 and 547293, where the $\kappa_{\rm rad}(t)$ curve is high at early Universe ages, while the $\kappa_{\rm z}(t)$ curve remains consistently low from high redshift, with only minor fluctuations. 
In these cases, KPP satellites are at low distances from the $\vec{J}_{\rm stack}$-plane since high redshift, initially moving with a dominant radial velocity component, and adopting a disky configuration as they approach the host halo.

Lastly, intermediate situations exist where neither $\kappa_{\rm z}(t)$ nor $\kappa_{\rm rad}(t)$ curves are dominant at early times. This behavior suggests that in these cases  satellites move in trajectories whose velocities have an important radial component while having a non-negligible velocity component perpendicular to the $\vec{J}_{\rm stack}$-plane before settling the disky configuration.

Once KPP satellites acquire their disky configuration, from $T_{\rm Krot}^{0.5}$ onwards the $\kappa_{\rm rot}$ curve continues to increase while the  $\kappa_{\rm z}$ and $\kappa_{\rm rad}$ components decrease,  until it stabilizes at $\kappa_{\rm rot}^{\rm eq}$ (except for fluctuations), marking the systems' final configuration 
at a Universe age $T_{\rm Krot}^{\rm eq}$\footnote{It is worth noting that axisymmetric systems where energy can be dissipated while angular momentum is conserved tend to have planar orbits and to circularize them. Indeed, in such systems, circular orbits have minimum energy at a given angular momentum \citep[see Chapter 3 in][]{BinneyTremaine08}. An effective viscosity -- the same responsible for dark matter sticking and long-lasting caustic formation in dark-matter-only simulations, see \citet{BuchertDominguez:1998} for a tentative modelling -- and interactions with substructure might  explain such dissipation process  in our case.}.
It is worth noting from Fig.~ \ref{ch5:fig_VI:vel_ratios_KPPHS} (compare the corresponding upper and lower panels) that this final stationary configuration at  $T_{\rm Krot}^{\rm eq}$
is reached when the system enters a periodic Keplerian regime.

In contrast, non-KPP satellites in KPP-HS systems exhibit no clear pattern in the evolution of the $\kappa_{\rm z}(t)$ , $\kappa_{\rm rad}(t)$ and  $\kappa_{\rm rot}(t)$ curves (see the bottom row panels of Fig.~A.1 for each KPP-HS system). Their kinetic energy appears evenly distributed among all components.
However, in roughly 50\% of the cases, the radial component $\kappa_{\rm rad}$ shows a slight decrease from high redshift, potentially reflecting the collapse of the attraction basin where the HS system forms, similar to the behavior of KPP satellites.


\subsection{Evolution of (proto-)satellite distances to the $\vec{J}_{\rm stack}$-plane}
\label{ch5:sec:DisPlaneJstack}

\begin{figure*}
\centering
\includegraphics[width=0.64\linewidth]{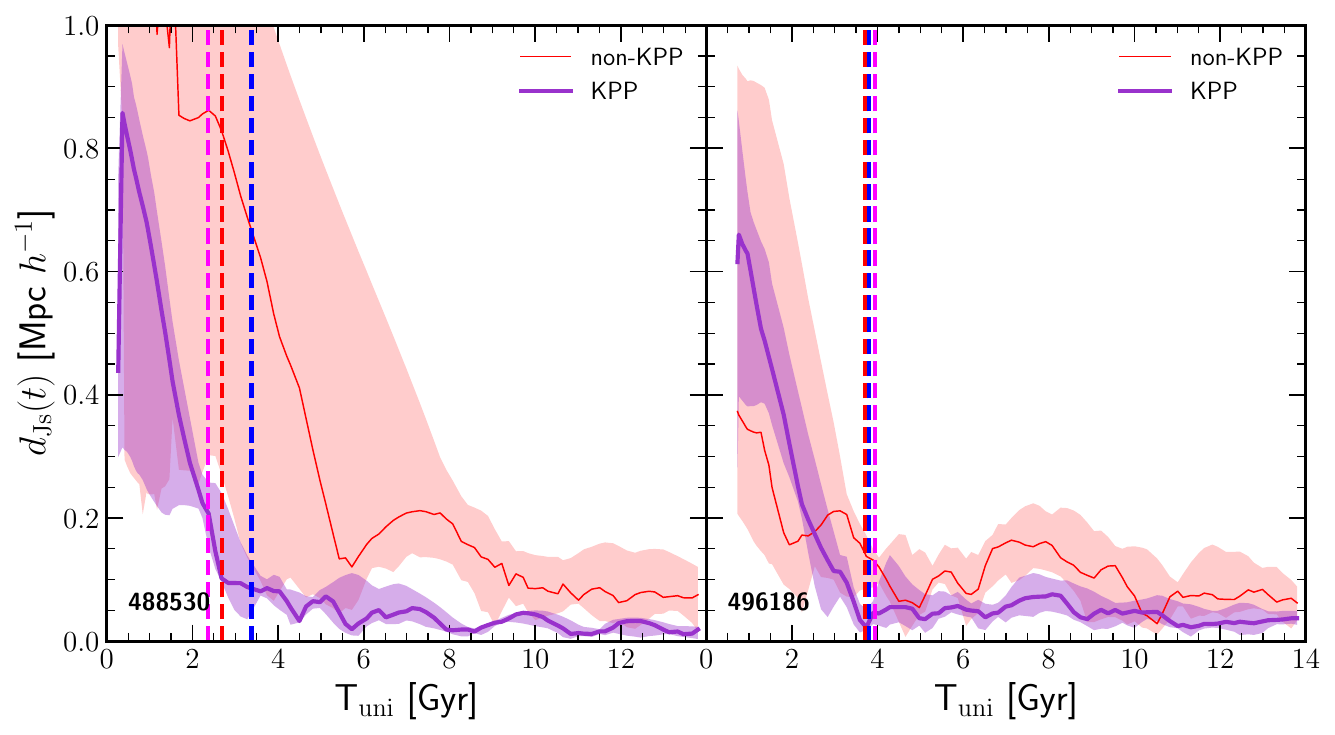}
\includegraphics[width=0.35\linewidth]{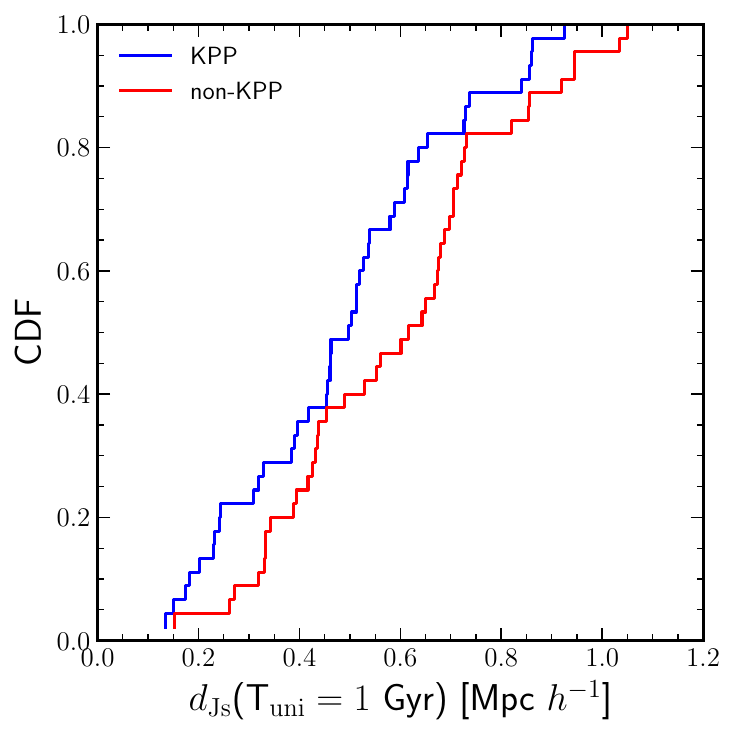}
\caption{Left and middle panels: comoving distance curves to the $\vec{J}_{\rm stack}$-plane ($d_{\rm Js}(t)$) for satellites in two KPP-HS systems. Purple curves and shaded regions stand for the median and interquartile ranges of the individual distance values, $d_{i, \mathrm{Js}}$, of KPP satellites at each timestep. Red curves and shades represent the same values  for the non-KPP satellite sets. Blue, magenta and red dashed vertical lines mark the $T_{\rm Krot}^{0.5}$, $T_{\rm cluster}^{\rm Jstack}$ and $T_{\rm dJs}$ timescales, respectively.  Right panel: Cumulative distribution functions for the median values of the KPP and non-KPP satellite sets distances at Universe age  T$_{\rm uni} = 1$ Gyr.}
\label{ch5:fig_VI:dist_Jstack_KPPHS}
\end{figure*}

The aforementioned decrement of $\kappa_z$ suggests that KPP satellites have their vertical motion component highly diminished, that is, they get confined to the $\vec{J}_{\rm stack}$-plane. This is a very relevant result concerning their orbital pole clustering. 
To dig deeper into this, for each KPP-HS system we calculate the comoving\footnote{Although not explicitly written across this work, it is worth noting the time dependence of the Hubble constant $h\equiv h(t)=\dot{a}(t)/a(t)$, with $a(t)$ being the expansion factor at a time T$_{\rm uni}=t$.} distance to the so-called $\vec{J}_{\rm stack}$-plane for each of the $j= 1, 2, ..., N_{\rm KPP}$ satellites in the KPP set, $d_{j, \mathrm{Js}}(t)$. 
We have determined, at different timesteps $t$, the median values, $d_{\rm Js}(t)$, of the individual $d_{j, \mathrm{Js}}(t)$ for the 
satellites belonging to a given KPP, as well as for satellites outside KPPs.
In Fig.~\ref{ch5:fig_VI:dist_Jstack_KPPHS} and second row panels in Fig.~A.1  we plot the evolution of $d_{\rm Js}(t)$ for KPP satellites 
(purple lines) as well as for the non-KPP ones (red lines). Shaded bands indicate the interquartile ranges.

Except for some KPP-HS systems, most $d_{\rm Js}(t)$ curves display a similar pattern.
At high redshift, satellites show a fast decrease in their distance to the $\vec{J}_{\rm stack}$-plane, 
followed by eventual stabilisation at lower redshifts, aside from minor fluctuations. 
The transition is fast in most cases, with  most systems displaying a distinct 'knee' feature marking the turning point in the trajectory evolution.  
This transition  defines a timescale, $T_{\rm dJs}$, representing the Universe age when the $d_{\rm Js}(t)$  curve shows its knee or when it reaches its equilibrium minimum (in case the knee is not that clearly marked or the transition is smoother) marking the collapse of the $\vec{J}_{\rm stack}$-plane\footnote{See sec.~8.1 in \citetalias{Gamez-Marin2024} for details on the concept and definition of `plane collapse' based on collapse events suffered by halos as described by the Spherical Collapse Model \citep{Padmanabhan93}.}. 
Values of $T_{\rm dJs}$ for the 46 HS systems in P24 sample are listed in Tab.~\ref{ch5:tab:DataTable2}, and its distribution is displayed as a cumulative distribution function (CDF) in Fig.~\ref{ch5:fig_VI:CDF-timescales}.

Interestingly, some KPP-HS systems show unusually shorter $d_{\rm Js}(t)$ distances at very high redshift, i.e., KPP (proto-)satellites are close to their $\vec{J}_{\rm stack}$-plane since high redshift. This is often due to low energy in the  $\kappa_{\rm z}$ component and high value of the $\kappa_{\rm rad}$ component.

Lastly, we examine the $d_{\rm Js}(t)$ curves for non-KPP satellites. 
Several important differences stand out compared to KPP satellites. Although non-KPP satellites also exhibit a two-phase structure of their curves, they often come from further away distances than KPP satellites, reaching the $\vec{J}_{\rm stack}$-plane at later times. 
 Indeed,  in the right panel of Figure \ref{ch5:fig_VI:dist_Jstack_KPPHS} the CDF for the $d_{\rm Js}$ distances at T$_{\rm uni}$ = 1 Gyr are plot both for KPP and non-KPP satellites. The KS test indicates that they are different at a $\sim$ 95\% of confidence.
Also, the dispersion (shaded region) is generally higher than the one that KPP satellite members show.


\subsection{A Comparison of early 
KPP establishment using different analysis tools}
\label{ch5:sec:ComparingTools_Jsplane}

We compare the $T_{\rm dJs}$  timescale (Universe age when KPP satellite positions settle into the $\vec{J}_{\rm stack}$-plane), the $T_{\rm Krot}^{0.5}$ timescale (Universe age when a 50\% of the median satellite kinetic energy is provided by satellite rotation, thereby when the $\kappa_z$ energy component becomes subdominant,) 
and the $T_{\rm cluster}^{\rm Jstack}$ timescale  (Universe age when KPP satellites establish themselves in the $\vec{J}_{\rm stack}$-plane
in terms of orbital pole clustering). 
These three timescales are marked as red-, blue- and magenta-dashed vertical lines in Figs.~\ref{ch5:fig_VI:vel_ratios_KPPHS} and \ref{ch5:fig_VI:dist_Jstack_KPPHS}, as well as in the different plots of Fig.~A.1, respectively, and their distributions are analyzed in Sec.~\ref{ch5:sec:Timescales}.    

In the CDFs (binning-free) plots in  Fig.~\ref{ch5:fig_VI:CDF-timescales}, we see that the median values  for the $T_{\rm cluster}^{\rm Jstack}$, $T_{\rm dJs}$ and $T_{\rm Krot}^{\rm 0.5}$ timescale distributions are similar ($T_{\rm cluster}^{\rm Jstack}=3.9^{+0.7}_{-0.5}$ Gyr, $T_{\rm dJs}=4.2^{+1.0}_{-0.6}$ Gyr, and $T_{\rm Krot}^{0.5}=5.1^{+0.9}_{-0.9}$ Gyr, where the $\pm$ uncertainties are the corresponding quartiles).
Interestingly, these distributions peak at  T$_{\rm uni}\sim 4.0$ Gyr, indicating that the $\vec{J}_{\rm stack}$-plane formation is roughly coeval when described through orbital pole clustering, satellite plane incorporation, and fraction of kinetic energy in rotational motions, respectively.
This is an important result that supports the robustness of our analyses.

In summary, we have identified that $\vec{J}_{\rm stack}$ defines a privileged direction, as well as a corresponding privileged plane, the $\vec{J}_{\rm stack}$-plane, which passes through the host centre and is normal to $\vec{J}_{\rm stack}$. Along this direction, the related kinetic energy fraction, $\kappa_z$, for most KPP satellite sets decreases towards low values from high redshift. The $d_{\rm Js}$ curves show generally two phases: an infall phase, and a second phase when satellites have been incorporated into the plane. Exceptions occur for KPP satellite sets whose $\kappa_{\rm z}$ values are already low at high redshifts. 
Once a set of satellites stabilizes at low $\kappa_{\rm z}$ values and settles into the $\vec{J}_{\rm stack}$-plane, a consequence is that their orbital poles cluster. 
Therefore, the incorporation of satellites onto the $\vec{J}_{\rm stack}$-plane, either with a reorientation of satellite orbital poles or not, is what is ultimately responsible for the clustering of orbital poles, i.e., the KPP formation. After satellite incorporation, angular momentum conservation and maximum plane circularization ensures the persistence of the configuration.

The question that arises is what are the physical processes behind the $\vec{J}_{\rm stack}$-plane formation.




\section{LV Analysis}
\label{ch5:sec:LVs_intro}

\subsection{Method}

We characterise the evolution of the local density environment of each HS system by analyzing the deformations of its corresponding Lagrangian Volume \citep[LV, ][]{Robles:2015,Robles2026}. 
This LV is built at  redshift $\zhigh=20.05$  -- the earliest snapshot available for TNG50 -- by identifying the particles that are within a spherical volume of radius $R_{\rm LV}=K\times R_{200c}(z=0)$, with $K=15$\footnote{In Sec.~\ref{ch5:sec:K_LV_change} we discuss the convenience of this choice.}, 
around the  position (at $\zhigh$) 
of the c.o.m of the dark matter particles that are gravitationally-bound to the analysed host halo at $z = 0$.
Particles enclosed within these initially spherical regions are then followed forward in time to $z=0$.  
The construction of these LVs aims at isolating the environment around HS systems and characterising the anisotropic collapse of the (proto-)structures that determines the final distribution and kinematics of the mass distribution.

The deformation and shaping of LVs are analyzed at each timestep $t$ using the reduced\footnote{See the justification in sec.~5 in \citetalias{Gamez-Marin2024}.} TOI, $I_{i,j}^{\rm r}$ \citep{Cramer}, defined as:
\begin{equation}
   I_{ij}^{\rm r} =\sum_{n}m_n\frac{(\delta_{ij}r_{n}^2 - r_{i,n}r_{j,n})}{r_{n}^2}, \hspace{0.5cm} n=1, ..., N
   \label{reducedI}
\end{equation}
where $r_{n}$ is the distance of the $n$-th LV particle to the LV center of mass and $N$ is the total number of DM particles enclosed within the LV.
This approach allows us to extract the principal directions of deformation, $\vec{e}_{i}(t)$ $(i = 1,2,3)$, and characterise the inertia ellipsoid through its principal axes \textit{a(t), b(t)} and \textit{c(t)}, derived from the eigenvalues ($\lambda_i(t)$ with $\lambda_1(t)\leq\lambda_2(t)\leq\lambda_3(t)$) of the $I_{i,j}^{\rm r}$, so that $a\geq b\geq c$:
\begin{eqnarray}
	a = \sqrt{\frac{\lambda_2 - \lambda_1 + \lambda_3}{2M_{\mathrm{DM}}}}, \qquad
	b = \sqrt{\frac{\lambda_3 - \lambda_2 + \lambda_1}{2M_{\mathrm{DM}}}}, \\ \nonumber
	c = \sqrt{\frac{\lambda_1 - \lambda_3 + \lambda_2}{2M_{\mathrm{DM}}}},
	\end{eqnarray}
	where $M_{\rm DM}$ is the DM mass of a given LV. 
	We denote the directions of the principal axes of inertia by $\vec{e}_i$, $i=1,2,3$, where $\vec{e}_1$ correspond to the major axis $a$,
	$\vec{e}_2$ to the intermediate axis $b$, and $\vec{e}_3$ to the minor axis $c$.

\subsection{Projections}
 
\begin{figure}
\centering
\includegraphics[width=0.85\linewidth]{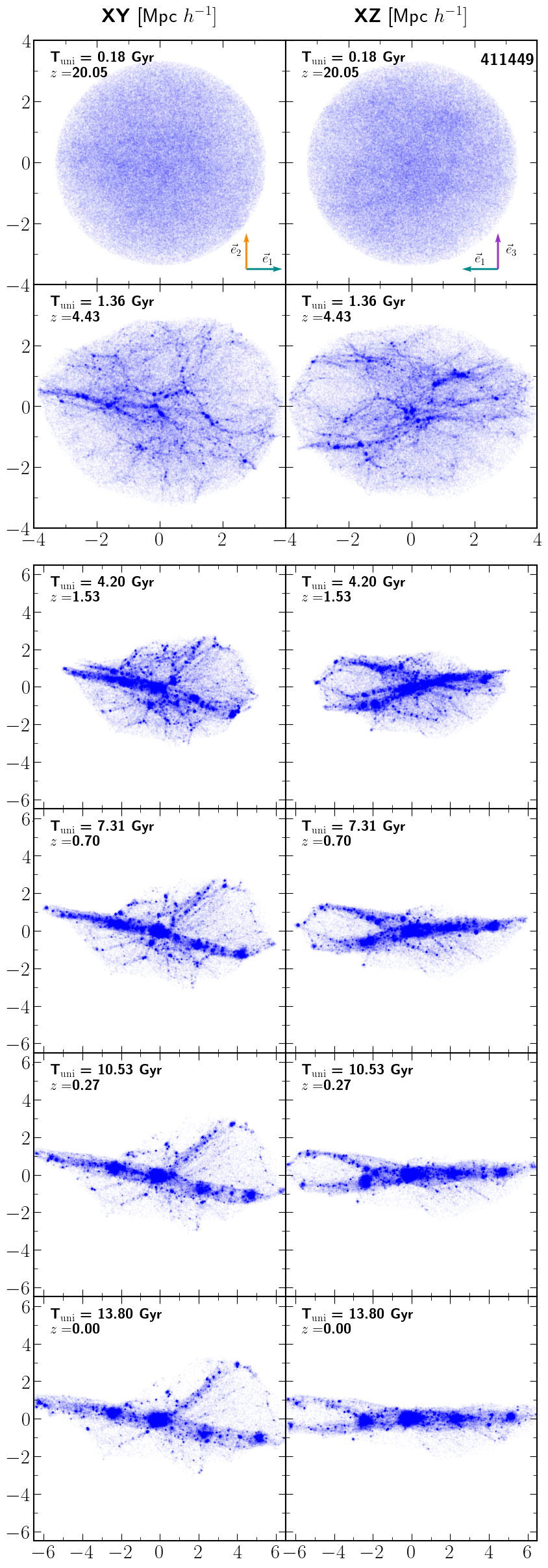}
\caption{Projections of the LV   around the  ID \# 411449 KPP-HS, from $z_{\rm high}$ (T$_{\rm uni} = 0.18$ Gyr) onward. Two LV projections along fixed axes are shown for each snapshot, one along the $\vec{e}_3(z=0)$ principal direction onto the $XY$ plane (first column), and the second one along the $\vec{e}_2(z=0)$ principal direction onto the $XZ$ plane (second column). The snapshot redshift $z$ and Universe age T$_{\rm uni}$ are given at the top left corner in each panel. 
}
\label{ch5:fig_VI:LV_projection}
\end{figure}

 Fig.~\ref{ch5:fig_VI:LV_projection}  illustrates the projections of the dark matter particles of a particular LV with ID \# 411449.
 The projections are made along the LV principal directions $\vec{e}_3$ (left column) and $\vec{e}_2$ (right column) at $z= 0$. These fixed projections, presented sequentially from $z_{\rm high}$ to $z=0$, illustrate the dynamical evolution of the LV over cosmic time.
 Panels (a) in Fig.~A.2 extend this information by showing the evolution of 33 LVs built around their corresponding central galaxies (one per page).

As expected, the high-redshift projections indicate  spherical configurations. 
Then, most  LVs undergo a pronounced global contraction along the $\vec{e}_3(t)$ direction over time, hereafter the direction of maximum collapse. Sequentially (or simultaneously in most cases)  they are globally stretched along the $\vec{e}_1(t)$ direction. These trends are more clearly shown in the right  
 column of the projections. 

The behavior along the $\vec{e}_2(t)$ direction, however, is more variable. Depending on the LV, it can exhibit stretching, compression, or relative stability.
Of particular interest are the  global  mass flows that take place along this direction at later times (after those along the direction of maximum compression), as they can alter the morphology of the LV. 

As explained in Sec.~\ref{ch5:sec:intro}, the cause of this large-scale evolution, as describe by the Zel’dovich approximation \citep[][]{Zeldovich:1970} and Adhesion Model \citep[see e.g.][ and references therein]{Gurbatov:1989,Kofman92,Gurbatov:2012}, is the  non-linear evolution of the density perturbations of the density field. This leads to the formation of singularities (first walls, then filaments, finally clumps), initially manifesting at small scales. As evolution proceeds, they are smoothed out, after having  grown, percolated and  merged. At the same time, the same process occurs  repeatedly at increasingly larger scales, driving the hierarchical evolution of cosmic structures. 
At any scale, mass elements in walls flow towards filaments, either secondary\footnote{Secondary filaments are coplanar filaments forming in 2-dimensional structures (walls).} filaments or filaments forming at wall intersections. These mass flows feed the filaments as walls get weaker or dissapear; then mass elements in filaments flow towards clumps, where mass piles up.

Interestingly, the edge-on projections along $\vec{e}_2(t)$ of the LVs indicate that many of these substructures exhibit a rather coplanar arrangement. This coplanarity was already reported in \citet{Cautun:2014}.


\subsection{Evolution of the Principal Directions}
\label{ch5:sec:PDirEvol}

\begin{figure}
\centering
\includegraphics[width=\linewidth]{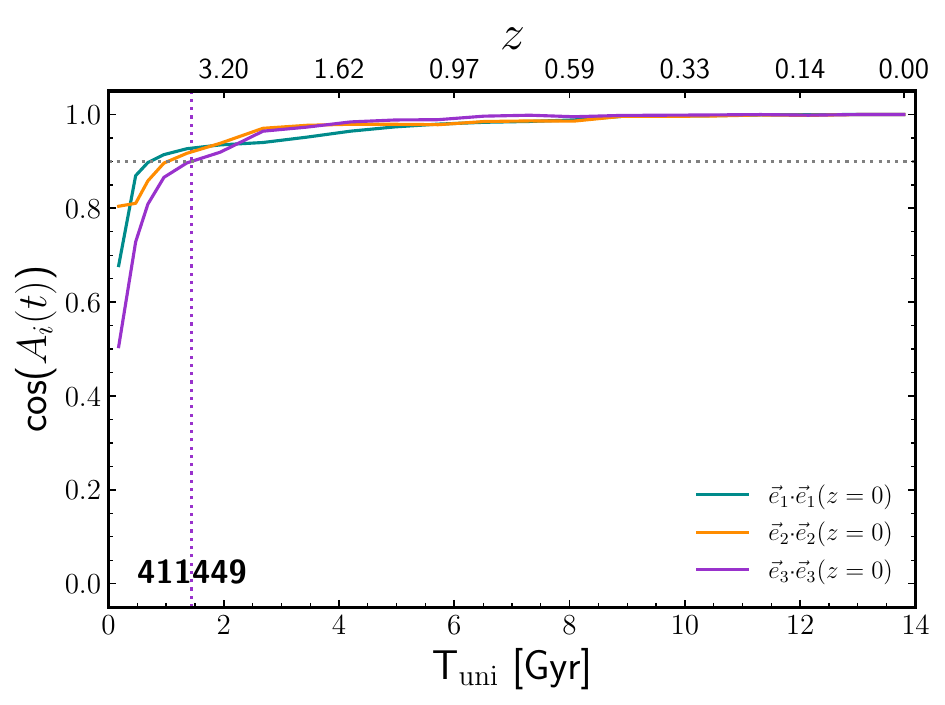}
\includegraphics[width=\linewidth]{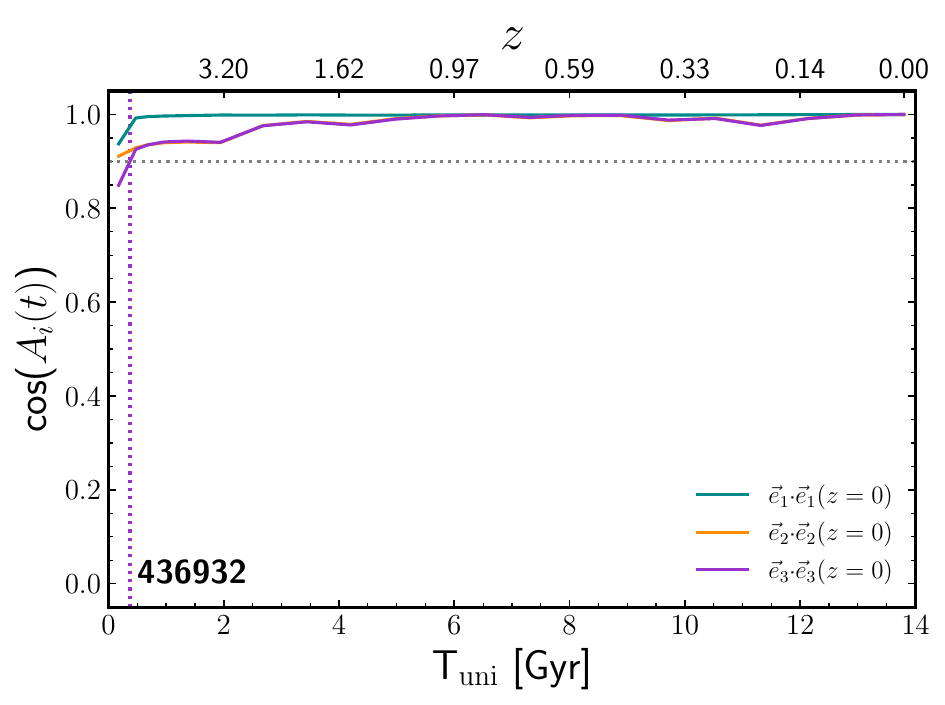}
\caption{Evolution of the cosine of the angle $A_i$ formed by the eigenvectors $e_{i}(t)$ and $e_{i}(z=0)$ with $i = 1,2,3$.
Upper horizontal axes give the redshift scale, while the lower ones stand for the Universe age T$_{\rm uni}$. The vertical dotted line stands for the freezing-out timescale $T_{\rm dir}^{e3}$, moment when the $\vec{e}_{3}(t)$ principal direction becomes fixed within a 10\% threshold (cos($A_3(t)$)$\geq 0.9$, marked as a horizontal dotted line). Colour codes are provided in the legends.}
\label{ch5:fig_VI:axes_freezing_out}
\end{figure}

The orientation of the principal directions may evolve over time due to large-scale forces and torques.
To quantify these changes, in Fig.~\ref{ch5:fig_VI:axes_freezing_out} and panels (b) in Fig.~A.2 
we  present, for the 33 example KPP-HS systems, the cosine of the angle $A_i$ formed by each principal directions at a Universe age T$_{\rm uni}=t$, $\vec{e}_i(t)$, and their corresponding directions at $z=0$.
Different colours stand for the three principal directions. 
We consider that these directions become fixed at the Universe age when they stabilize within a 10\% threshold (i.e. cos$(A_i)\geq0.9$, see \cite{Robles:2015}), implying that subsequent mass rearrangements at LV scales occur along well-defined, stable axes -- within a threshold--. This allows us to define the 'freezing-out' timescale  $T_{\rm dir}^{ei}$ for each principal direction of the LV. 
We will only focus on the freezing-out of the $\vec{e}_3$ direction, $T_{\rm dir}^{e3}$, as this timescale quantifies the moment when the overall strongest mass flow halts. This timescale is represented for each LV as a vertical dotted line in these figures, and their values are shown in Tab.~\ref{ch5:tab:DataTable2}, and
Figs.  \ref{ch5:fig_VI:CDF-timescales} and \ref{ch5:fig_VI:Timescales_visual_distrib}.
We observe that, while some LVs fix their directions at early Universe ages (even before T$_{\rm uni}\leq 1$ Gyr), other LVs show changes in their principal directions for long periods of time.

We have confirmed that reorientations or jumps in the principal directions occur in systems where several massive halos appear at low redshift in the projections of the LV evolution (see Fig.~A.2, e.g. ID\# 505586). This suggests that these irregularities or jumps are caused by the corresponding LV encompassing more than one attraction basins (or part of them) where the different massive halos are assembled. 
In particular, the collapse of the attraction basin within the global LV causes perturbations to the evolution of the environment of the host galaxy under study. In Sec.~\ref{ch5:sec:AlignPDir} we will see that, despite sudden changes of the principal LV directions, the physical principal direction the $\vec{J}_{\rm stack}$ axial vector is aligned to remains the same.


\subsection{Shape evolution}
\label{ch5:sec:ShapeEvol}

The shape evolution of LVs can be characterised through the time evolution of their semi-axes \textit{a(t)}, \textit{b(t)} and \textit{c(t)}.  
These parameters can be combined to compute various derived quantities, including the axis ratios \textit{b(t)/a(t)} and \textit{c(t)/a(t)}, and the triaxiality parameter $T$ \citep{Franx:1991}, which is defined at a Universe age T$_{\rm uni}=t$ as:
\begin{equation}
    T(t) = \frac{1-b(t)^2/a(t)^2}{1-c(t)^2/a(t)^2}
\label{TshapeDef}    
\end{equation}
Note that a system with $c/a>0.9$ has an almost spheroidal shape, whereas objects with $c/a$ below this threshold and $T<0.3$ are oblate spheroids.
In extremely oblate systems $b/a\rightarrow1$ and in flattened elliptical structures $c/a\rightarrow0$ (i.e. $c \ll a$, with any $b/a$ value, this is, any triaxiality). 
Thus, three distinguished $T$ regions can be defined: $0.7 \le T \le 1$, $0.3 \le T \le 0.7$ and $T \le 0.3 $, corresponding to  prolate, triaxial and oblate shapes.
Finally, extremely prolate systems are those with  \textit{b/a} < 0.4, and filamentary-shaped configurations have $c,b\ll a$, close to the $T$=1 line ~\citep{GonzalezGarcia:2009}.

\begin{figure}
\centering
\includegraphics[width=\linewidth]{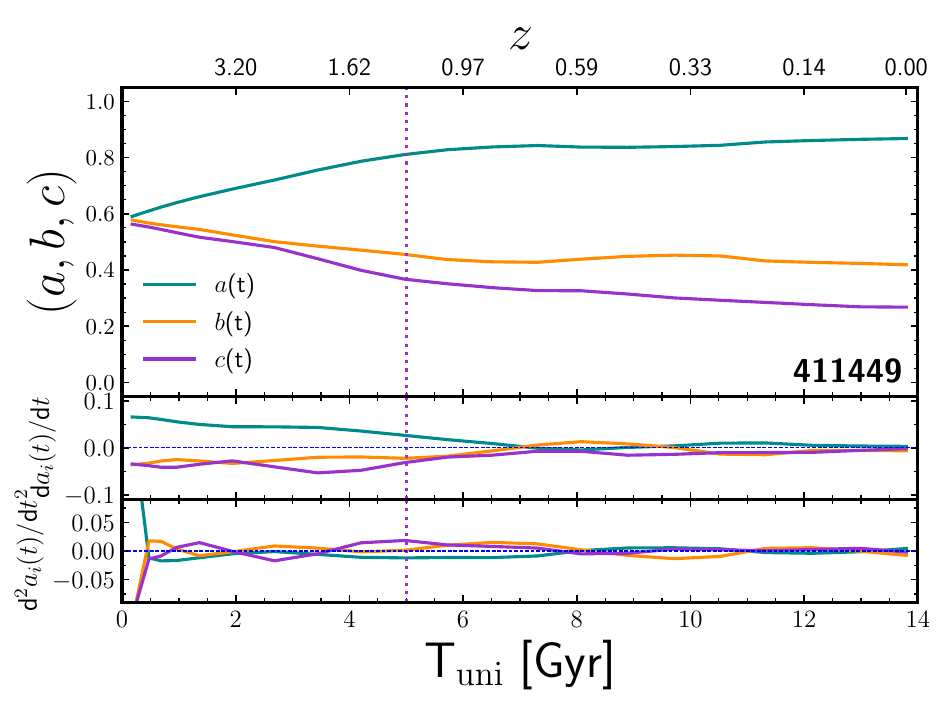}
\includegraphics[width=\linewidth]{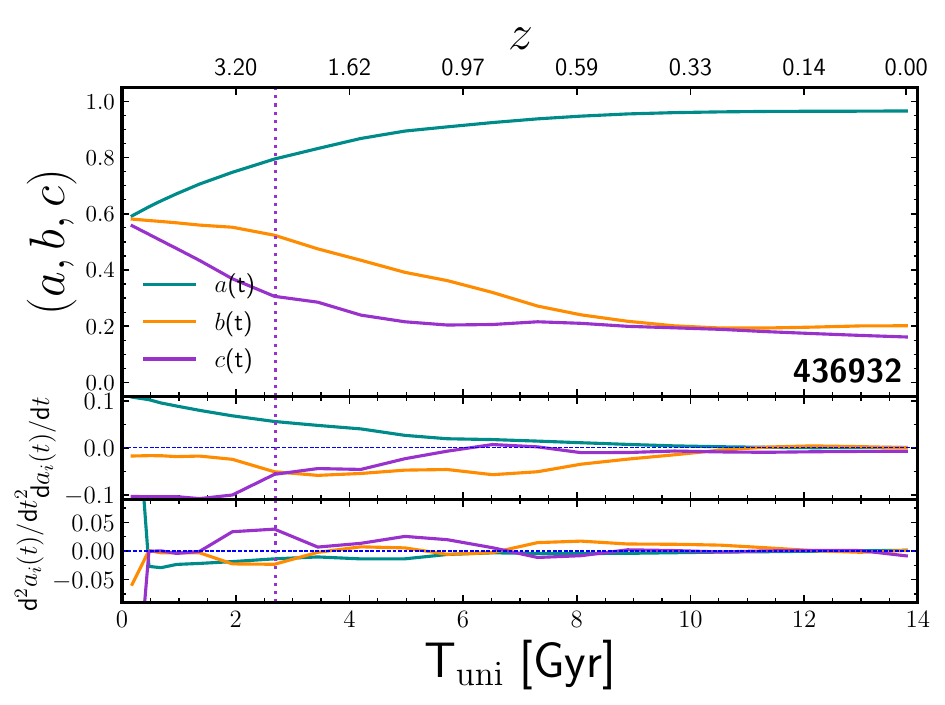}
\caption{Evolution of the LV principal axes $a(t), b(t)$ and $c(t)$, $a(t) > b(t) > c(t)$, colour-coded as depicted in the legend. Middle and bottom panels display the first and second time derivatives of these parameters, respectively. The vertical dotted line stands for the  $T_{\rm shape}^{e3}$ timescales, moment when a discontinuity at early times appears in $c(t)$.}
\label{ch5:fig_VI:LV_axes_evol}
\end{figure}

Fig.~\ref{ch5:fig_VI:LV_axes_evol} and panels (c) in Fig.~A.2 show the \textit{a(t)}, \textit{b(t)}, and \textit{c(t)} functions and their corresponding first and second time  derivatives for the LVs associated with 33 KPP-HS systems. A general trend can be observed:
during the early Gyrs of cosmic evolution, the slopes (i.e. the time derivatives) of $a(t)$, $b(t)$ and $c(t)$ remain relatively constant for many analyzed LVs. While the major axis $a(t)$ exhibits rapid growth during this early times, reflecting strong global stretching along $\vec{e}_1(t)$, the minor axis $c(t)$ decreases significantly, corresponding to strong global compression along the $\vec{e}_3(t)$ direction. On the other hand, the intermediate axis $b(t)$ of the ellipsoid, either grows, deforming the initially-spherical LV into an oblate or wall-like structure, or remains roughly stable, maintaining a triaxial shape.
The rapid decrease of $c(t)$ at early times then slows down, at a given time T$_{\rm uni}$.
This change is either manifested as a clear transition in the slope of the $c(t)$ curve, visible as a discontinuity in its time derivative, or as a gradual transition, in which case the $c(t)$ slope shows a smoother change extended in time, leading to a continuous derivative curve. 
This behavior allows us to define the timescale 
$T_{\rm shape}^{e_3}$, the Universe age when either a sharp discontinuity appears in $c(t)$  at early cosmic times, or the time when the $c(t)$ derivative first begins to change noticeably (as expressed by important changes in the second derivative).  
The $T_{\rm shape}^{e3}$ values for the 46 analyzed LVs are given in Tab.~\ref{ch5:tab:DataTable2}, and statistically analyzed in Sec.~\ref{ch5:sec:Timescales}.

\begin{figure}
\centering
\includegraphics[width=0.9\linewidth]{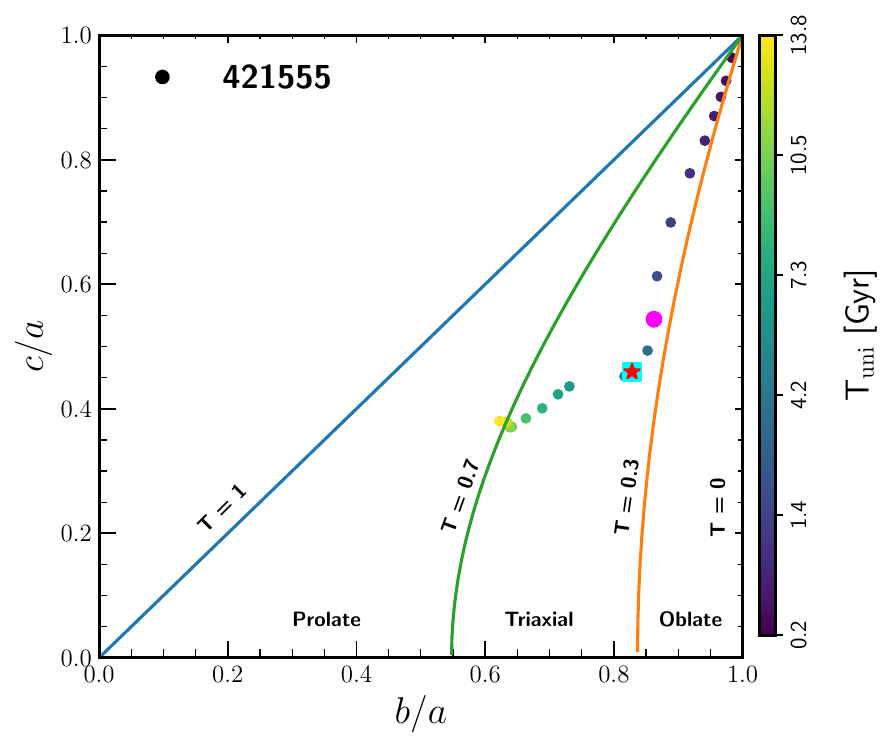}
\includegraphics[width=0.9\linewidth]{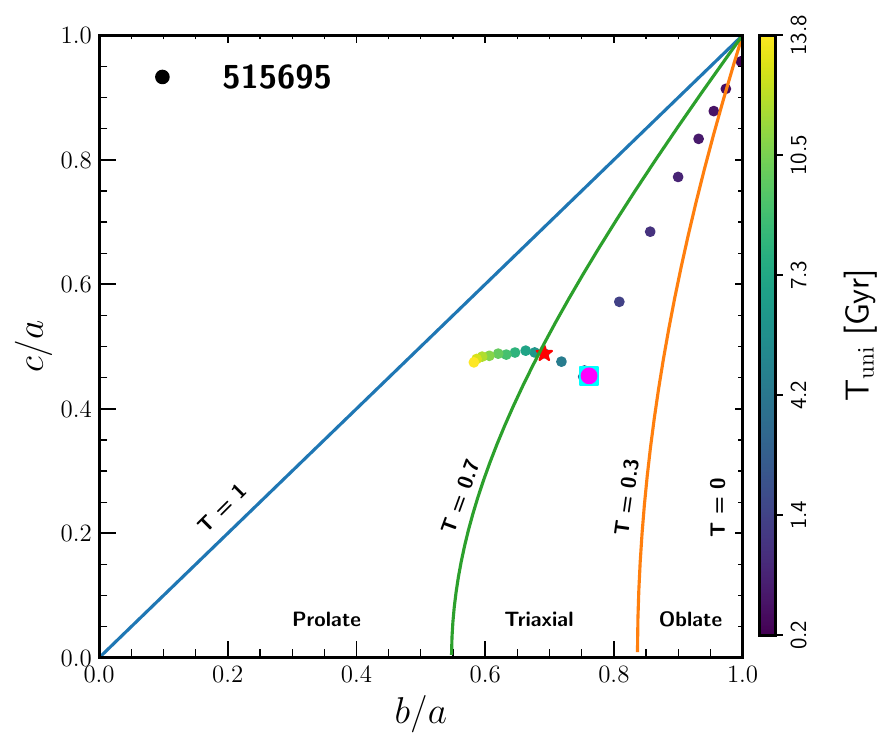}
\includegraphics[width=0.9\linewidth]
{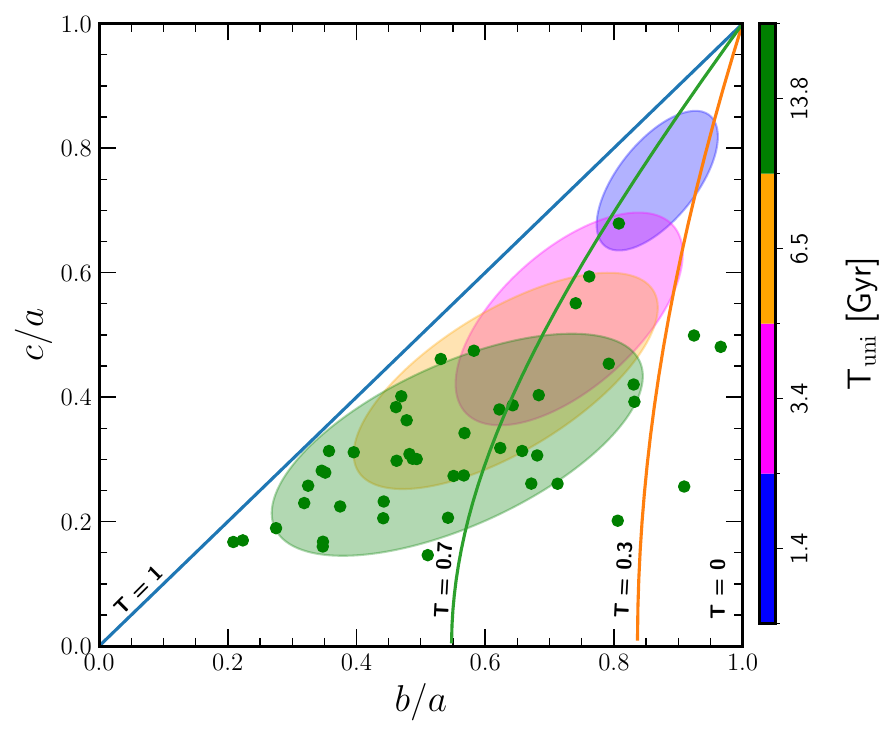}
\caption{Upper and middle panels: Evolution of the axis ratios  $b(t)/a(t)$ and $c(t)/a(t)$ for two KPP-HS systems. Outputs are colour-coded according to the corresponding Universe age for each point, as illustrated in  the colourbar at the right of the panel. The $c/a$ versus $b/a$ plane is splitted into three regions, according to the values the $T$ shape parameter takes on them, see the blue, green and orange  iso-$T$ curves and corresponding labels. Legends on the bottom of the plot indicate the LV shapes at each of the three regions.
The outputs closest to the $T_{\rm cluster}^{\rm Jstack}$,  $T_{\rm dJs}$ and $T_{\rm shape}^{e3}$ timescales have been highlighted in magenta, red, and cyan, respectively.
Bottom panel: Green points represent the $z=0$ values for the 46 LVs where KPPs are formed. Ellipses are the 1.5$\sigma$ 
concentration ellipses, calculated through a Huber-weighted orthogonal regression method, of results at T$_{\rm uni}$ = 1.4, 3.4, 6.5 Gyr  and at $z=0$ (blue, purple, orange and green colours, respectively).}
\label{ch5:fig_VI:ba_ca_rat}
\end{figure}

The overall shape evolution of LVs can be effectively captured by analyzing the time evolution of their axes ratios \textit{b(t)/a(t)} and \textit{c(t)/a(t)}, as well as the triaxiality parameter $T(t)$. This approach provides insights into the morphological transitions experienced by LVs as they evolve under the influence of large-scale dynamics.
In the upper and middle panels of Fig.~\ref{ch5:fig_VI:ba_ca_rat} we plot 
 the \textit{c(t)/a(t)} versus \textit{b/(t)/a(t)} ratios across time for two LV examples. Points in the plots are colour-coded to indicate the evolution of these ratios over time, as indicated by the colourbar at the right-hand side of each plot. To categorize the LV morphologies, we employ the $T$-shape criterion illustrated in \citet{GonzalezGarcia:2009}. Continuous lines mark the boundaries of the iso-$T$ regions -- blue ($T=1$) and green ($T=0.7$) lines mark the limit of prolate 
 regions, green and orange ($T=0.3$) lines delimit triaxial regions, and orange with the vertical axis denote the boundary of oblate (flattened) regions, as far as $c/a<0.9$. 
Similar plots are displayed in panel
 (d) of Fig.~A.2, for each of the 33 KPP-HS systems selected to represent the P24 sample.

At high redshifts, all LVs start as spherical structures by construction, with $c/a$ = $b/a$ = 1, with points locating at the upper-right corner of the plot.
As they evolve, the LVs diverge into different regions of the $c(t)/a(t)$ vs $b(t)/a(t)$ plane. 
The paths of the ($c(t)/a(t)$, $b(t)/a(t)$) points in the plot show, in most cases, a two-phase pattern. 
The first early-time phase is regular,  while the second one is more complex.  In general,  the separation between the two phases  is clearly marked by a `knee' feature.  To gain insights into the LV shape evolution, the outputs corresponding to timesteps closest to the $T_{\rm cluster}^{\rm Jstack}$,  $T_{\rm dJs}$ and $T_{\rm shape}^{e3}$ timescales have been highlighted in magenta, red, and cyan, respectively.
It turns out that most of them fall within the first, regular phase,  close to the `knee', and in general at least one of  these timescales are coeval with the `knee' feature, being in most cases $T_{\rm shape}^{e3}$ (see panels (d) in Fig.~A.2).
In the first phase the evolution of an LV in the $c(t)/a(t)$ vs $b(t)/a(t)$ plane proceeds from top to bottom (as the $c(t)/a(t)$ ratio decreases), and from right  to left (as the $b(t)/a(t)$ ratio decreases as well), resulting in LVs becoming increasingly triaxial (most of them) or even prolate in this first phase.
However, in some cases (e.g., KPP-HS ID\# 473329 and 550149), LVs evolve into wall-like structures (more oblate).

In the second phase, the ($c(t)/a(t)$, $b(t)/a(t)$) point patterns are more irregular and complex-behaved, as mentioned. 
Most  LV shapes become increasingly prolate at low redshift, while some of them exceptionally transition between different shapes. For instance, some LVs with an initially decreasing $b(t)/a(t)$ ratio during the fast phase become more triaxial after a prolate phase, or more wall-like after a triaxial phase (e.g., ID\# 402555, 514829, 543114, 547293).  
Such particular LV evolution is typically detected when a massive halo, distinct from the host one,  emerges at low redshift within the LV. This particular feature suggests that the LV encompasses more than one significant gravitational attraction basin. As previously mentioned, these peculiar evolutions in the LV morphology appear to have no remarkable impact on satellite alignment in KPPs. 

The two-phase structure found in the LV patterns of shape evolution  is reminiscent of the two phases  found in the distance curves, $d_{\rm Js}(t)$, as well as in the $a(t)$ and $c(t)$ principal axes evolution.  `Knee' features generally indicate abrupt changes. In these cases, the change comes from the $c(t)/a(t)$ ratio becoming  slow-changing, while the $b(t)/a(t)$ ratio becomes faster-changing after the knee,  see Fig.~\ref{ch5:fig_VI:ba_ca_rat}.
These changes in the $c(t)/a(t)$ and $b(t)/a(t)$ time derivatives are  associated to the end of the rapidly decreasing (stretching) phase of the $c(t)$ ($a(t)$) axes leading to a more flattened structure nearly normal to the $\vec{e}_3$ principal direction.  This is one key ingredient for the establishment of satellite pole clustering, 
as we will see in the next sections for the P24 sample, see also \citetalias{Gamez-Marin2024}.

The bottom panel of Fig. \ref{ch5:fig_VI:ba_ca_rat} reports on the collective shape evolution of the 46 KPP-HS LVs.
The sequence of the four 1.5$\sigma$ concentration ellipses goes from T$_{\rm uni} = 1.4$ to 3.4 (the median value of $T_{\rm shape}^{e3}$, purple), to 6.5 Gyr (the value of $T_{\rm Krot}^{\rm eq}$, the Universe age when KPPs reach their maximum rotation support),
and then to  $z=0$ (blue, purple, orange and green).  This sequence reflects what has just been explained for the two LV evolution examples in this same Figure: shape evolution is strong until $T_{\rm shape}^{e3}$ = 3.4 Gyr, but only mild from that point until $z=0$. In particular, the global shape evolution is fast between 1.4 and 3.4 Gyr, showing i.e., the  global $c/a$ decrement towards, in most cases,   flattened structures.
On the other hand, points at $z=0$ indicate an important shape variability in the P24 sample, with a dominance of prolate systems (63\%), including highly filamentary ones (i.e., with $c \sim b \ll a $, close to the $T=1$ line, around 9 LVs), and a scarcity of oblate $z=0$ LVs (6.5\%). Both the distribution of point locations at $z=0$ and the evolution meant by the ellipses, are consistent with results shown in fig.~8 of \cite{Robles:2015} for LVs; see also \cite{Robles2026}.




\section{Alignments with the LV principal directions}
\label{ch5:sec:AlignPDir}

In Sec.~\ref{ch5:sec:Vel_JstackPlane} we have shown that orbital pole clustering comes about as the KPP satellites-to-be are incorporated onto the $\vec{J}_{\rm stack}$-plane, both in velocity and position spaces. $\vec{J}_{\rm stack}$-plane.
In this section, we explore the relationship between the clustering of satellite orbital poles into KPPs and the principal directions of deformation around these systems across time.
Similar to the $\vec{J}_{\rm stack}$-plane, we define the $\vec{e}_i$-planes as the planes normal to $\vec{e}_i(t)$ ($i=1,2,3$), passing through the c.o.m. of the LV at a Universe age T$_{\rm uni} = t$.


\subsection{Alignments of the principal directions with satellite orbital poles}
\label{ch5:sec:JorbEiAlign}
 
\begin{figure*}
\centering
\includegraphics[width=0.9\linewidth]{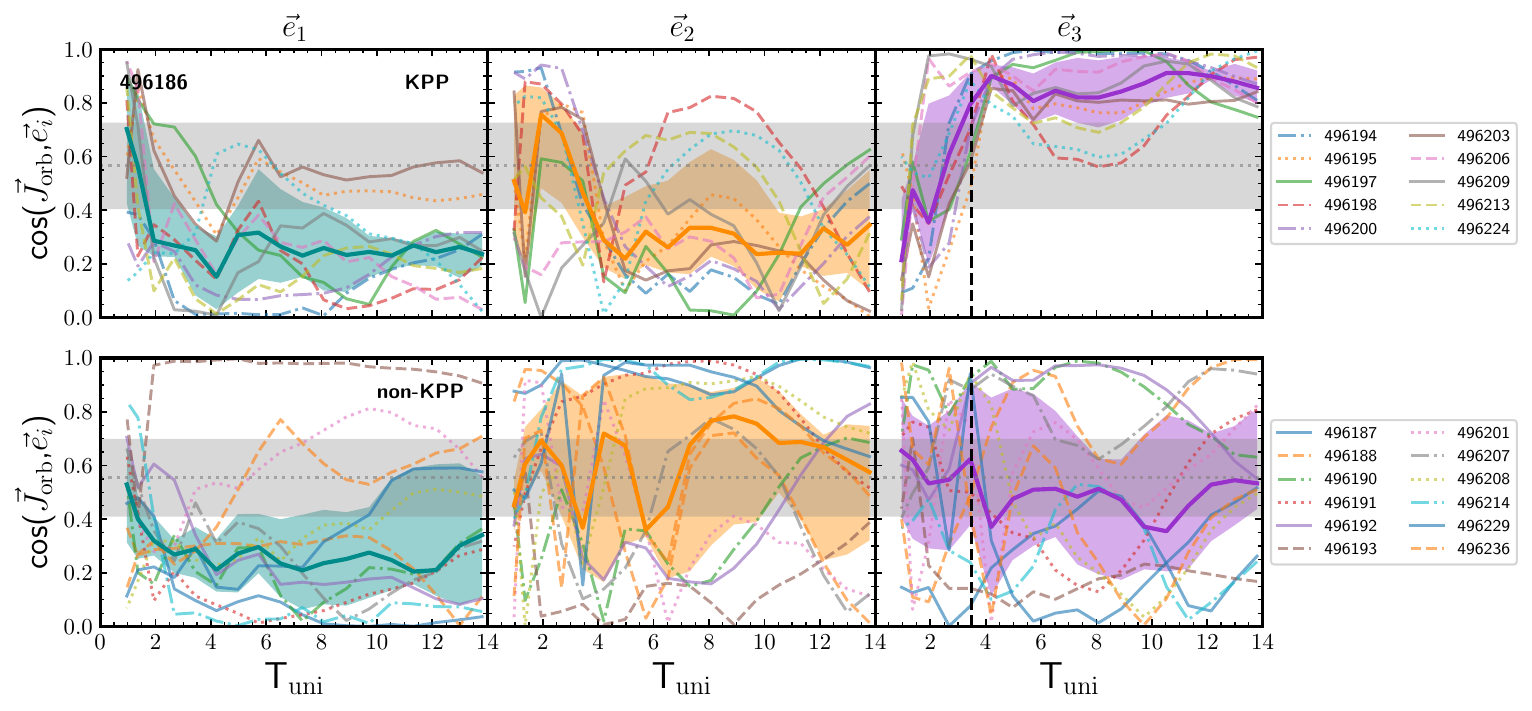}
\caption{Time evolution of the angles between the $\vec{J}_{\rm orb}$ of satellites with respect to the principal directions of the LVs, $\vec{e}_i(t)$, $i=1,2,3$, for one KPP-HS system. 
The cos($\vec{J}_{\rm orb}, \vec{e}_i(t)$) are shown for satellites belonging to KPPs, and outside these structures as well, see legends.
Thin lines correspond to individual satellites, with subhalo IDs, colours and line-types as encoded on the right of the panels. 
Thick lines and shaded bands indicate the median and interquartile ranges at T$_{\rm uni} = t$, respectively.
Grey dotted lines and shaded areas stand for the median alignment angle, Ali$_{\rm ran}(t)$, and their interquartile ranges for a randomly oriented orbital pole distribution with the same number of satellites as the compared subset of satellites (KPP or non-KPP), in this particular case, 10 and 12 for the KPP and non-KPP satellite sets, respectively. Black dashed vertical lines mark the alignment $T_{\rm align}^{ei}$ timescale of KPP satellite orbital poles with the best aligned $\vec{e}_i$ principal direction.}
\label{ch5:fig_VI:Jorb_ei_alignments}
\end{figure*}

To this end, for the 46 KPP-HS systems presented in Tab.~\ref{ch5:tab:DataTable2}, we measure the alignment signal cos($\vec{J}_{\rm orb},\vec{e}_i(t)$) -- where $\vec{J}_{\mathrm{orb},k}$ is the orbital angular momentum of the $k$-th satellite\footnote{We recall that the alignment signals are measured such that $0^{\circ}\leq\alpha(\vec{J}_{\rm orb},\vec{e}_i(t))\leq90^{\circ}$.}-- of individual satellite orbital poles with the three principal directions of their corresponding LV at T$_{\rm uni}=t$ 
($\vec{e}_i(t)$), dividing our satellite samples into KPP and non-KPP satellites. 
This is presented in Fig.~\ref{ch5:fig_VI:Jorb_ei_alignments} and panels (e) of Fig.~A.2 for a total of 33 KPP-HS systems. Upper (lower) panels display the alignment signal for KPP (non-KPP) satellites. Thin coloured lines with different line styles stand for different individual satellites.
Each satellite maintains consistent colour and line coding throughout figures corresponding to a given HS system. 
Thick coloured lines depict the respective median alignment signal for the  overall sample of KPP or non-KPP satellites, while shaded areas are the respective interquartile ranges.
Medians and shaded areas are colour-coded according to the corresponding $\vec{e}_i(t)$ axis: alignment with $\vec{e}_3(t)$ (purple), with $\vec{e}_2(t)$ (orange), and with $\vec{e}_1(t)$ (blue).
Grey dotted lines and shaded areas stand for the median alignment angle, Ali$_{\rm ran}$, and their interquartile ranges for a time-independent randomly oriented orbital pole distribution with the same number of satellites as  the subset under consideration (KPP or non-KPP). 
In this particular case, these satellite numbers are 10 and 12 for the KPP and non-KPP satellite sets, respectively.

A clear alignment signal is evident for KPP satellites, whereas no such alignments are observed for non-KPP satellites. Indeed, while in the latter case the alignment curves are compatible with Ali$_{\rm ran}(t)$, given its interquartile ranges, in the former case the  alignment signal is clearly distinguisable from random results once alignment comes about at $T_{\rm align}^{ei}$.
We define this timescale as the time when the median alignment satisfies cos$(\vec{J}_{\rm orb},\vec{e}_i)\geq$ cos($\alpha_{\rm crit}$) = 0.8, following the alignment criterion described by \citet{Fritz2018}, i.e., we have chosen to use the same alignment criteria applied throughout the paper, see e.g.  Section  \ref{sec:KPP-identification}, and widely used in previous papers by different authors. The  angle $\alpha_{\rm crit} =  36.^{\circ}87$
encloses a spherical‑cap solid angle amounting to <10\% of a hemisphere.

The fact that there is alignment between the orbital pole direction of a satellite and the $\vec{e}_i$ principal direction is independent of there being alignment between the poles of individual satellite among themselves, which we then define as KPP satellite members. Hence, the alignment with LV principal directions is a non-trivial result that emerges independently from the KPP selection criteria.

The corresponding values of the $T_{\rm align}^{ei}$ timescale are provided in Tab.~\ref{ch5:tab:DataTable2}, and their distribution is shown in Fig.~\ref{ch5:fig_VI:CDF-timescales} and \ref{ch5:fig_VI:Timescales_visual_distrib}, with a median and quartile range of $3.3^{+1.8}_{-1.7}$ Gyrs. The $\vec{e}_i$ principal direction that best aligns with KPP satellite orbital poles in each KPP-HS system is presented in the 'Axis $\vec{e}_i$' entry in Tab.~\ref{ch5:tab:DataTable2}.

Notably, only 4 KPP-HS systems (8.7\% of the total sample) lack alignment of their KPP satellite poles with any principal direction of their LVs, and, as a result, no $T_{\rm align}^{ei}$ is associated with these systems (see Tab.~\ref{ch5:tab:DataTable2}). 
For the remaining 42 KPP-HS systems, the principal direction driving the clustering of KPP satellite orbital poles is $\vec{e}_3$ in 31 KPP-HS systems, representing the 67.4\% of the total KPP-HS sample.
A common result inferred from these panels is that alignments with the $\vec{e}_3(t)$ direction develop progressively during the fast phase of mass assembly of the host halo. 
Except for two KPP-HS systems (ID\# 416713 and 547293), such alignments are generally weak or absent at high redshift. In fact, for 20 out of the 31 KPP-HS aligned with $\vec{e}_3$, the clustering of KPP satellite orbital poles around $\vec{e}_3(t)$ emerges after their turn-around time (i.e. their first apocentre time, $T_{\rm apo1}$).
This finding suggests that in most of these systems orbital pole clustering around $\vec{e}_3(t)$ arises when (proto-)satellites are small substructures that follow the mass flows collapsing onto the (mathematical) plane defined by $\vec{e}_3(t)$, shaping the local environment. 
These results are consistent with findings for KPP satellites in the Aq-C$^{\alpha}$ simulation system and KPP2 satellites in the PDEVA-5004 simulation\footnote{In the PDEVA-5004 simulation, two KPPs were identified, one with satellite orbital poles aligned with $\vec{e_1}$ (which we refer to as KPP1) and another with orbital poles aligned with $\vec{e}_3$ (KPP2). For details see \citetalias{SantosSantos2024_anisotropic} and \citetalias{Gamez-Marin2024}.}. 

Additionally, we find that 11 KPP-HS systems exhibit strong alignment of their KPP satellite orbital poles with their environment's $\vec{e}_2(t)$ and $\vec{e}_1(t)$ principal directions. Specifically, 9 systems are aligned with $\vec{e}_2(t)$, and 2 with $\vec{e}_1(t)$, corresponding to 19.5\% and 4.3\% of the total KPP-HS sample, respectively. Interestingly, except for two of them (KPP-HS ID\# 461785 and 503437), satellites forming these KPPs display a strong alignment with their respective $\vec{e}_i(t)$ direction since high redshift, showing significantly lower values of $T_{\rm align}^{ei}$ compared to most KPP satellites aligned with $\vec{e}_3(t)$. 
This early alignment suggests that the trajectories of these satellites were already constrained close to the mathematical plane defined by their corresponding $\vec{e}_i(t)$, moving along nearby mass flows (generally aligned with the dominant component $\vec{e}_3(t)$).
This finding highlights the influence of early Universe dynamics in shaping the alignment of satellites within their local environments. We will explore this issue in forthcoming sections, focusing on the relationship between satellite trajectories and the principal directions of deformation of the local CW during the early stages of the Universe. 

To finish this subsection, we come back to the collective behaviour of alignments. To quantify the visual impressions of a clear distinction between Ali$_{\rm ran}(t)$ and the alignment signal for KPPS,
for each set of KPP and non-KPP satellites in a given HS-system, we have calculated the average distance between the curves Ali$_{\rm ran}(t)$ and cos($\vec{J}_{\rm orb}, \vec{e}_i(t)$)$_{\rm median}$ after alignment is established at
$T_{\rm align}^{ei}$. With these average distances, a CDF is calculated both for KPP and non-KPP satellite sets.  The KS test indicates that they are statistically distinguishable at a  confidence level higher than a 99.9\%. This is a very important result, that quantitatively puts light on the role the principal directions of the local TOI play  as drivers of satellite pole clustering at high redshift.


\subsection{A visual illustration of KPP formation}

\begin{figure*}
\centering
\subfloat{\includegraphics[width=0.4\linewidth]{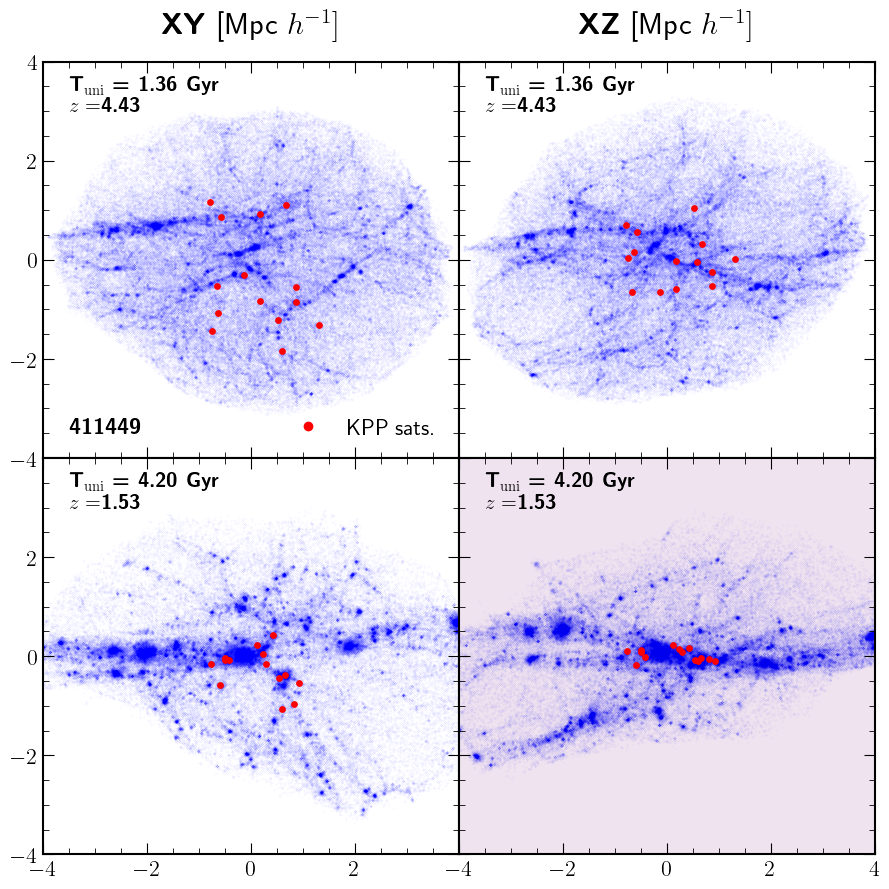}
{\vspace{0.2cm}\hspace{0.1cm}(a)}}
\subfloat{\includegraphics[width=0.4\linewidth]{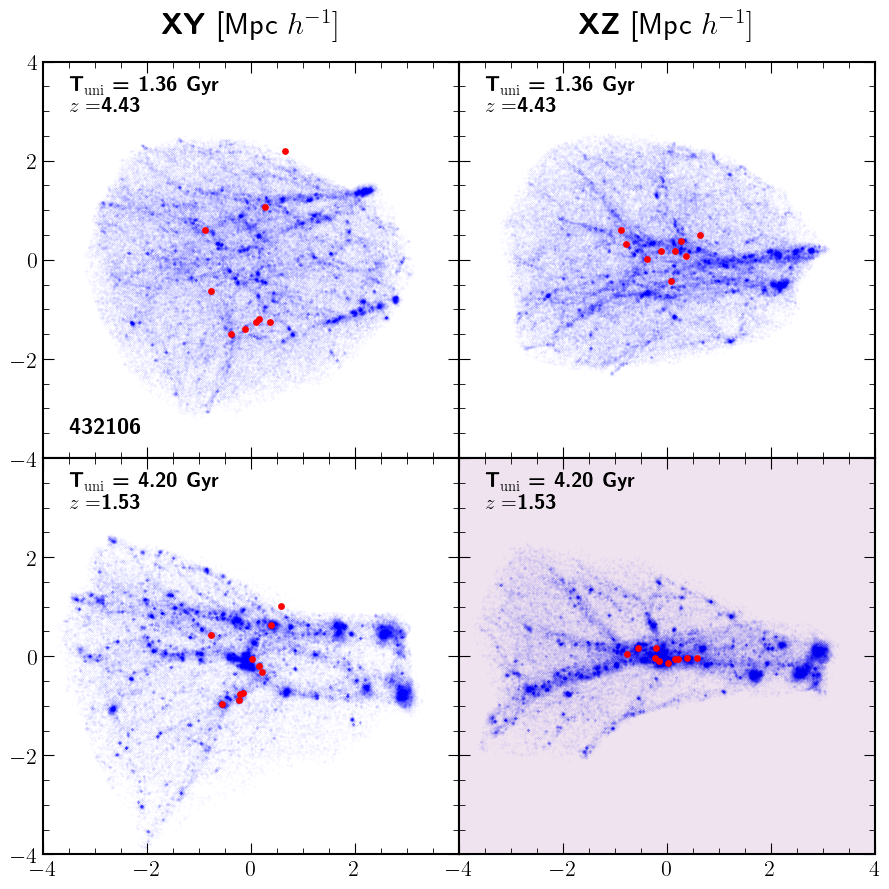}
\vspace{0.2cm}{\hspace{0.1cm}(b)}}
\\
\subfloat{\includegraphics[width=0.4\linewidth]{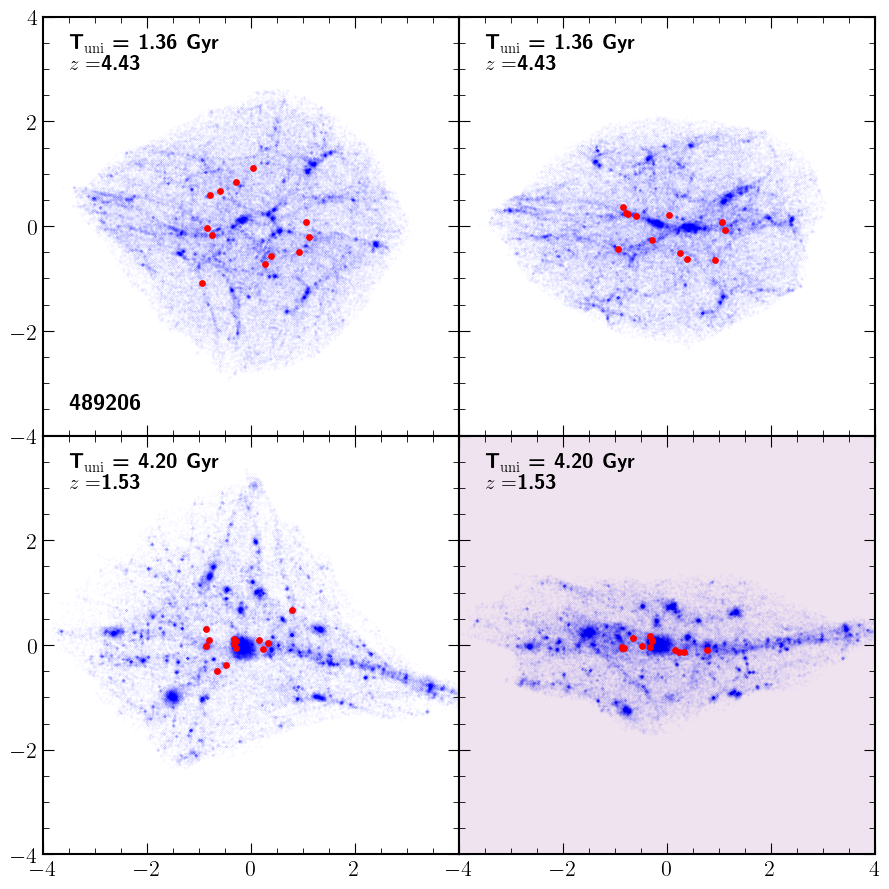}
{\vspace{0.2cm}\hspace{0.1cm}(c)}}
\subfloat{\includegraphics[width=0.4\linewidth]{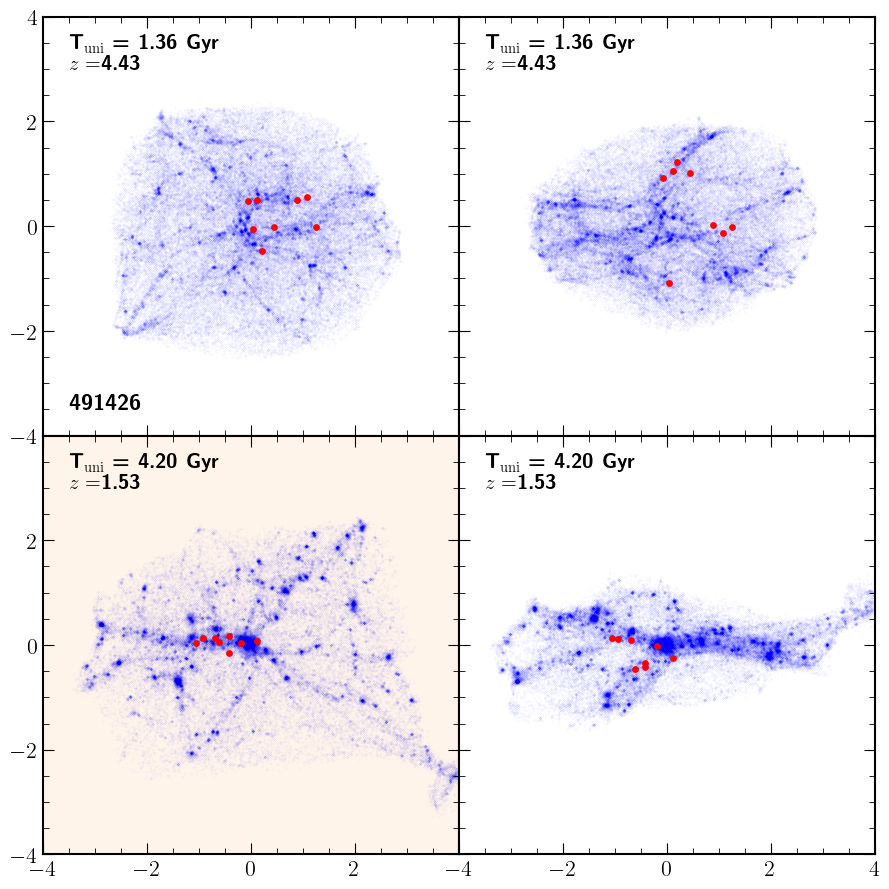}
{\vspace{0.2cm}\hspace{0.1cm}(d)}}

\caption{Projections along the $\vec{e}_3(t)$ (left panels) and $\vec{e}_2(t)$ (right panels) of the density field around the host galaxy formation site for 4 KPP-HS systems (with ID\#s indicated in the bottom left corner in their left panels) for two different values of the Universe age given in the top left corner in each panel.  
KPP satellites are marked as red circles in each panel, at the location of their most gravitationally-bound particle.
KPP-HS ID\# 491426 forms its KPP in the $\vec{e}_2$-plane direction, while the rest form it in the $\vec{e}_3$-plane, see shaded panels.
}
\label{ch5:fig_VI:Projections+satellites}
\end{figure*}

An illustration of the interplay between the CW and KPP formation is shown in Fig.~\ref{ch5:fig_VI:Projections+satellites}, where we plot the projections along the $\vec{e}_3(t)$ (i.e., onto the $XY$ plane, left panels) and $\vec{e}_2(t)$ (onto the $XZ$ plane, right panels) principal directions around four KPP-HS systems at two early Universe ages. The locations of their KPP (proto-)satellite centres of mass are marked as red points.
We see, in the projections along $\vec{e}_2(t)$, the global early collapse of the environment along the $\vec{e}_3(t)$ direction, with substructures (among which satellite-to-be objects are included) following the collapse of the environment along the $\vec{e}_3(t)$ axis. Indeed, we clearly see, for the first three KPP-HS systems plotted in Fig.~\ref{ch5:fig_VI:Projections+satellites}, that by T$_{\rm uni}\sim4$ Gyr satellites have been incorporated to the $\vec{e}_3$-plane and KPPs are already in place (panels highlighted with light purple shade).
At this time, most KPP satellites have not been incorporated into the halo yet, as the  positional extent of the KPP system exceeds the halo $R_{200c}$ scale. The bottom left projection in panel (d), tinted with light orange shade, shows a KPP already formed in  the $\vec{e}_2$-plane. 

These panels illustrate the global effects of the forces and torques produced by the whole mass distribution -- organized into a forming CW -- driving the incorporation of satellites to their corresponding $\vec{e}_i$-plane, constraining their trajectories within it and eventually leading to KPP formation.
(Proto-)satellites come from small, dense mass elements as well. These small mass elements progressively collapse forming satellites, and are incorporated into more massive structures that shape the local CW. 

\subsection{Alignments of the $\vec{J}_{\rm stack}$ axis with the LV principal directions}
\label{ch5:sec:JstackEi_Align}

In addition to the orientation of individual satellite orbital poles with respect to the principal directions of the LV, we also examine the orientation of the axis of maximum co-orbitation $\vec{J}_{\rm stack}$ with respect to these directions.

\begin{figure}
\centering
\includegraphics[width=\linewidth]{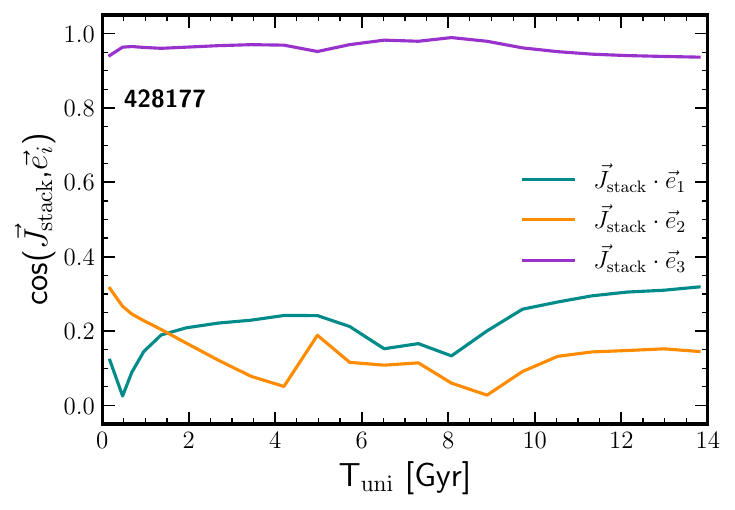}
\caption{Alignment signals between the co-orbitation axis $\vec{J}_{\rm stack}$ and the LV principal directions at each time, $\vec{e}_i(t)$, for a KPP-HS system. Colour codes refer to the alignment signal with the $\vec{e}_3(t)$ (purple), $\vec{e}_2(t)$ (orange) and the $\vec{e}_1(t)$ (blue) principal directions.}
\label{ch5:fig_VI:Jstack_ei_LVs}
\end{figure}

In Fig.~\ref{ch5:fig_VI:Jstack_ei_LVs} (and panels (f) in Fig.~A.2) we display, for the 33 example KPP-HS systems shown in this study, the values of cos($\vec{J}_{\rm stack},\vec{e}_i(t)$), with $i=1, 2, 3$, represented by blue, orange and purple curves, respectively. 
As expected, the majority of KPP-HS systems have their $\vec{J}_{\rm stack}$ vectors aligned at high redshifts with the same principal axis that individual KPP satellite poles become aligned with at early Universe ages.  
This relevant result explains the physical basis 
for the existence of $\vec{J}_{\rm stack}$: at high redshift, $\vec{J}_{\rm stack}$ points close to $\vec{e}_i(t)$, which means that the apparent role of $\vec{J}_{\rm stack}$ in driving the clustering of satellite orbital poles is actually inherited from the local CW formation and evolution at early times.

It is worth noting that a few KPP-HS systems exist  where the KPP satellite orbital poles align with a specific LV principal direction, but their $\vec{J}_{\rm stack}$ does not. This discrepancy arises due to the angular aperture allowed in our definition of co-orbitation ($\alpha_{\rm co-orbit} \leq 36.87^{\circ}$), which allows KPP satellites to align with a particular $\vec{e}_i(t)$ direction while the $\vec{J}_{\rm stack}$ vector remains only partially aligned. Examples of such systems are KPP-HS systems ID\# 475619 and 483594.




\section{Satellite trajectories relative to the LV principal directions}
\label{ch5:sec:TrajectoriesLV}

So far, we have analyzed KPP satellites' trajectories from their velocity and positional space perspectives.
As explained in Sec.~\ref{ch5:sec:DisPlaneJstack}, KPP satellites tend to rapidly collapse onto the plane defined by their axis of maximum co-orbitation $\vec{J}_{\rm stack}$,  drastically lowering their velocity component along it, and, consequently, their $\kappa_z$ component.
In addition to this, in Sec.~\ref{ch5:sec:AlignPDir} we analyzed how most KPP satellites have their orbital poles clustered around the principal directions of compression of the LVs, hence forming the KPP in the plane defined by them. 
This suggests that the LV's dynamics might be driving orbital pole clustering and shaping the trajectories of satellites.
Therefore, in this section we analyze how satellite trajectories, and in particular how their instantaneous velocities, $\vec{v}_i(t)$, relate to the LV dynamics.

 
\subsection{Satellite distances to the $\vec{e}_i$-plane}
\label{ch5:sec:distplane_ei}

\begin{figure*}
\centering
\includegraphics[width=0.64\linewidth]{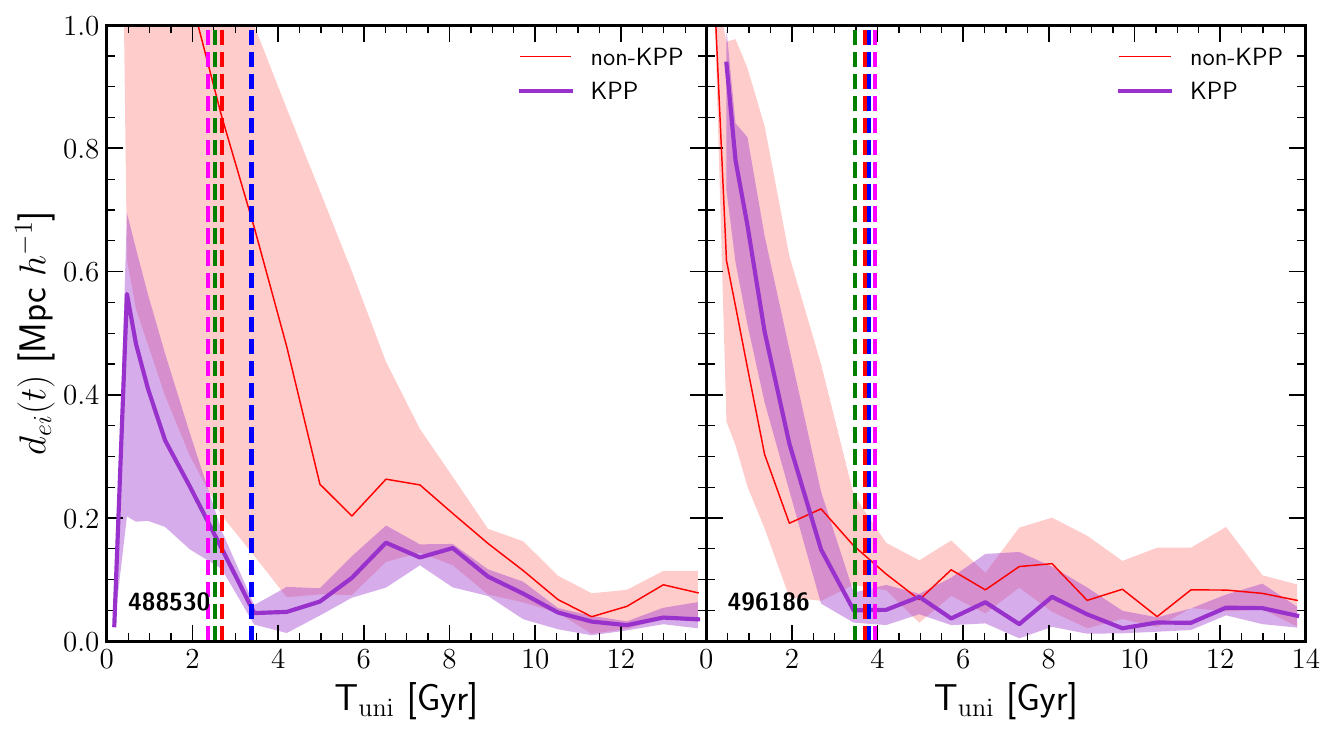}
\includegraphics[width=0.35\linewidth]{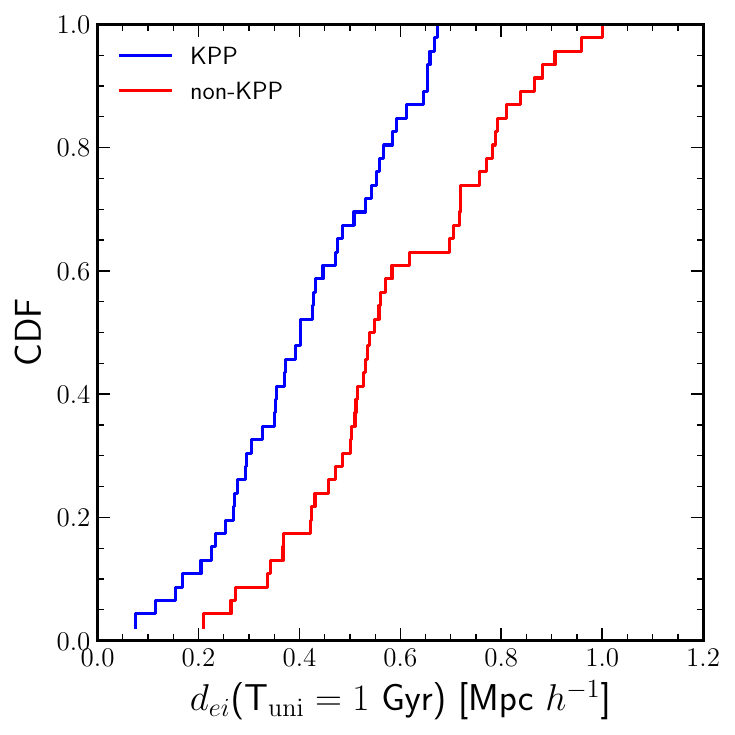}
\caption{Left and middle panels: comoving distance of satellites to the $\vec{e}_3$-plane as a function of time. Purple curves and shaded regions stand for the median and interquartile ranges of the individual distance values of KPP satellites at each timestep. Red curves and shades represent the same values  for the non-KPP satellite sets. As in Fig.~\ref{ch5:fig_VI:dist_Jstack_KPPHS}, blue, magenta and red dashed vertical lines mark the $T_{\rm Krot}^{0.5}$, $T_{\rm cluster}^{\rm Jstack}$ and $T_{\rm dJs}$ timescales, respectively, while the green dashed lines represent the $T_{\rm align}^{ei}$ timescale for each KPP-HS system. Right panel: CDFs for the median values of the KPP and non-KPP satellite sets $d_{ei}(t)$ distances at Universe age T$_{\rm uni}=1$ Gyr}.
\label{ch5:fig_VI:dist_eiz_plane_KPPHS}
\end{figure*}

In an analogous way as in Sec.~\ref{ch5:sec:DisPlaneJstack}, here we calculate the distance to the so-called $\vec{e}_i$-plane for each of the $j= 1, 2, ..., N_{\rm KPP}$ satellites in the KPP set, $d_{j, i}(t)$. 
For each KPP-HS system, we consider and work with the $\vec{e}_i(t)$ principal direction best aligned with the KPP satellite orbital poles (see  the 'Axis $\vec{e}_i$' entry in Tab.~\ref{ch5:tab:DataTable2}  and third row panels in Fig.~A.1).  

In the third row panels in Fig.~A.1, we plot the evolution of the median distance $d_{ei}(t)$, $i=1,2,3$, for KPP satellites in the 33 KPP-HS systems taken as examples. The curves are colour-coded by the LV principal direction around which the KPP aligns: purple ($\vec{e}_3(t)$), orange ($\vec{e}_2(t)$), blue ($\vec{e}_1(t)$). Black curves stand for KPPs that do not fully align with any particular direction -- in these cases, $d_{ei}(t)$ is evaluated relative to $\vec{e}_3(t)$, the direction of maximum collapse. 
For comparison, we also plot the distance $d_{ei}(t)$ of the non-KPP satellite set (red lines) to the same $\vec{e}_i$-plane as their KPP satellites counterparts. Shaded bands indicate the interquartile ranges.
Fig.~\ref{ch5:fig_VI:dist_eiz_plane_KPPHS} shows two example KPP-HS systems that align with $\vec{e}_3$.

These plots indicate that the $d_{e3}(t)$ curves (purple)  display a similar pattern as the $d_{\rm Js}(t)$ curve 
(shown in Fig.~\ref{ch5:fig_VI:dist_Jstack_KPPHS} for the same KPP-HS systems, and in the second row panels of Fig.~A.1), with KPP satellites rapidly collapsing coevally with the $\vec{e}_3$-plane. This allows us to define the timescale $T_{\rm dei}$, which is the Universe age when the $d_{ei}(t)$ curve (in this case the $d_{e3}(t)$ curve) reaches its final, globally stable value.
As in the case of the $\vec{J}_{\rm stack}$-plane, most KPP-HS systems show that, as KPP satellites approach the $\vec{e}_3$-plane at a time $T_{\rm dei}$, they progressively reorient their orbital poles and align them with the $\vec{e}_3(t)$ direction. As reported in Sec.~\ref{ch5:sec:VelSpace}, this establishment within the $\vec{e}_3$-plane is accompanied by a decrease in their $\kappa_z$ values in favor of $\kappa_{\rm rot}$ (with their trajectories progressively becoming more disky). 

Remarkably, some KPP-HS systems whose satellite orbital poles align with the $\vec{e}_3(t)$ principal direction show unusually shorter $d_{e3}(t)$ distances at very high redshift. At the same time, they often show low energy in the  $\kappa_{\rm z}$ component, indicating that (proto-)satellites are already close to the $\vec{e}_3$-plane since high redshift, with their orbital poles almost aligned with the $\vec{e}_3(t)$ direction since early times.

As previously mentioned, orange (blue) distance curves mean that the KPP satellite poles align with the $\vec{e}_2(t)$ ($\vec{e}_1(t)$) principal direction 
(i.e., $\vec{J}_{\rm stack}$ aligns with $\vec{e}_2(t)$ ($\vec{e}_1(t)$) at high redshift). 
In these cases, they present larger values of $T_{\rm dei}$ than in the case of alignments with the $\vec{e}_3$-plane, see CDFs in Fig.~\ref{ch5:fig_VI:CDF-timescales} and Fig.~\ref{ch5:fig_VI:Timescales_visual_distrib}.
These KPP satellite sets are generally at smaller distances from their corresponding $\vec{e}_i$-plane, and/or have a slower collapse onto it.

As for non-KPP satellites, their curves show an analogous behavior to their $d_{\rm Js}$ curves. However, non-KPP satellites come, on average, from larger distances than KPP ones.
The CDF for their distances to the $\vec{e}_i$-plane -- where  $\vec{e}_i$ is the principal direction the corresponding KPP satellites are aligned to -- at T$_{\rm uni}$ =1 Gyr, is drawn in the right panel of the Figure (red line), alogside with the CDF for the KPP members $d_{ei}(t)$ (blue curve). The CDFs difference stands out, with non-KPP satellites coming form further away from their $\vec{e}_i$-planes than KPP ones at a KS test confidence limit of 99.6\%.

Finally, KPP satellite sets not aligned  with any of the LV principal directions (black curves in Fig.~A.1) show a similar evolution of their $d_{e3}(t)$ distance as KPP satellites aligned with $\vec{e}_3(t)$. This suggests that KPP satellites generally follow the directions of maximum compression of their local environment at early times. We note that these systems are characterised by having late dynamical activity presumably due to the presence of a massive halo nearby, impeding the proper isolation and characterization of the surrounding environment.

It is worth remarking that when using $\vec{J}_{\rm stack}$ to study (proto-)satellite motions towards the $\vec{J}_{\rm stack}$-plane (see Figs~\ref{ch5:fig_VI:dist_Jstack_KPPHS} and A.1) we follow and characterise \textit{(proto-)satellites flows} towards this plane. 
In turn, when we study LV deformations through the $\vec{e}_i(t)$ principal directions, we follow and characterise \textit{DM mass flows} within LVs (i.e., the so-called mass-migration flows). 
Alignments found between $\vec{J}_{\rm stack}$ and principal directions $\vec{e}_i(t)$ just reflect the fact that (proto-)satellites are close to  follow flows of migrant mass in their  way towards establishing themselves as KPPs.  In other words, (proto-)satellites are but part of migrant mass flows that shape the local CW elements at high redshift: flattened structures in the first place, and then, in most cases, filaments and, finally,  clumps (HS systems).
Therefore, both satellite trajectory and DM structure formation are a consequence of the sticking nature of matter, as phenomenologically catched by the AM.

We finish this subsection by noting that differences between the $d_{ei}(t)$ and the $d_{\rm Js}(t)$ curves in Fig.~A.1 are a consequence of misalignments between $\vec{J}_{\rm stack}$ and the $\vec{e}_i(t)$ principal directions (as depicted in panels (g) in Fig.~A.2) and differences between the LV's and host galaxy's centres of mass. These differences are more prominent in systems whose LVs overlap with other LVs, and those that encompass the formation of a massive halo different from that of the analyzed systems.


\subsection{Alignment of satellite velocities with principal directions: high redshift flows towards $\vec{e}_i$-planes}
\label{ch5:sec:vel_ei_LVs}

\begin{figure*}
\centering
\includegraphics[width=0.95\linewidth]{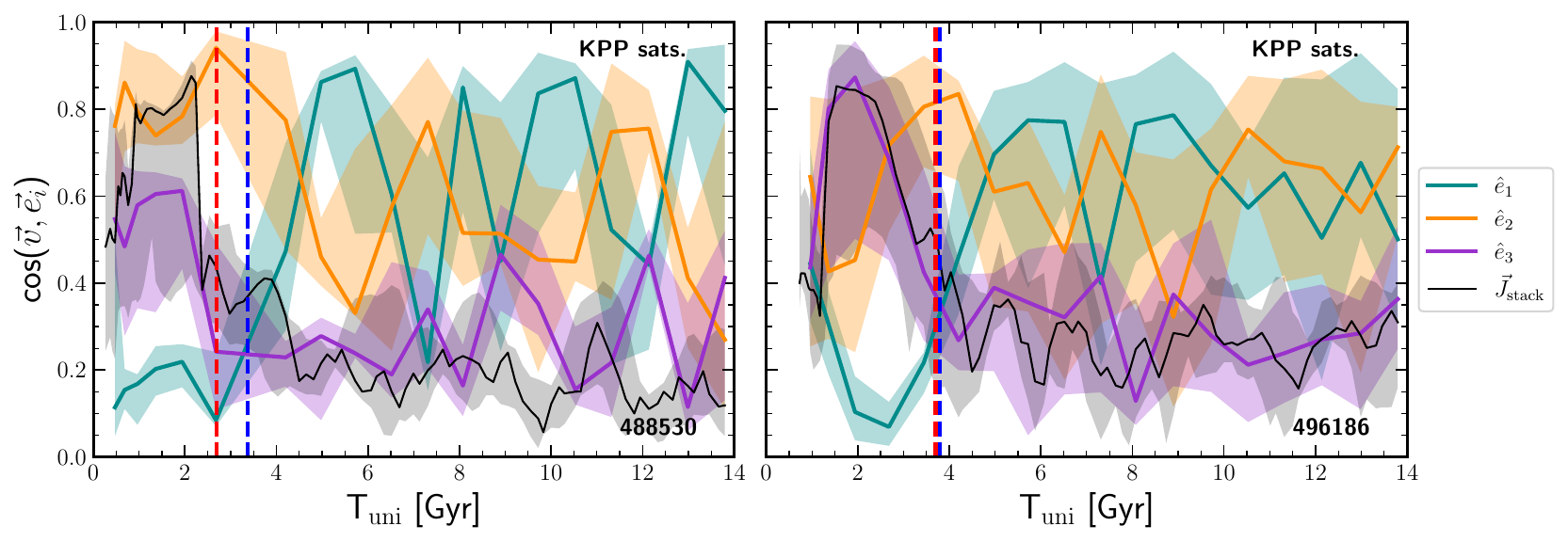}
\caption{Median values $A_i$ of the normalized projections of KPP satellite velocity vectors onto the three principal directions $\vec{e}_i(t)$. See legend for the representative colour codes. Shaded areas stand for the interquartile ranges. Black continuous lines and shades stand for the cos$(\vec{v}_j,\vec{J}_{\rm stack})$ median values and interquartile ranges. Blue and red dashed lines mark the $T_{\rm Krot}^{0.5}$ and $T_{\rm dJs}$ timescales, respectively}.
\label{ch5:fig_VI:vel_LV_KPPHS}
\end{figure*}

In this section we  examine how satellite trajectories (i.e., instantaneous velocities $\vec{v}_j(t)$) relate to LV dynamics. To this end, for each satellite  $j=1, 2, …  N_{\rm sat}$ in each HS system, its alignment signals cos$(\vec{v}_{j}$,$\vec{e}_i)\equiv A_{j,i}$  with the 3 principal directions $\vec{e}_i(t)$, $i=1,2$ and 3, have been calculated. 
Early velocity alignments and their changes provide relevant information about high redshift (proto-)satellite trajectories (i.e., how the satellite moves relative to the principal directions) and whether or not these trajectories change (i.e., they are bent) as the KPP forms.

In Fig.~\ref{ch5:fig_VI:vel_LV_KPPHS} and Fig.~A.3  we depict the median and quartile values of the $A_{j,i}$ alignment signals for the satellite sets at each timestep, denoted $A_i$. Specifically, panels in Fig.~\ref{ch5:fig_VI:vel_LV_KPPHS}, and upper panels in Fig.~A.3, display $A_i$ for KPP satellites, while the lower panels in Fig.~A.3 show the alignment signals for non-KPP satellites. We also plot the alignment signal between the velocity vectors $\vec{v}_j(t)$ of satellites with their axis of maximum co-orbitation, $\vec{J}_{\rm stack}$.
This signal confirms that the $\vec{v}_j(t)$ vectors  lie in their corresponding  $\vec{J}_{\rm stack}$-planes, as expected.

In general, plots in Fig.~\ref{ch5:fig_VI:vel_LV_KPPHS} and A.3 reveal that most KPP satellite trajectories are roughly aligned with the $\vec{e}_3(t)$  principal direction (i.e., median of $A_3>0.7$ or so) at early times. This is an indication that (proto-)satellite velocity has a high component in the direction of strongest collapse, irrespective of whether they eventually form their KPP in the $\vec{e}_3$-plane or not.  In other cases, however, the median of $A_{j,i}$ for the KPP set of a given HS system takes its maximum for $\vec{e}_2$ (see left panel of Fig.~\ref{ch5:fig_VI:vel_LV_KPPHS}). A third possibility is that  both the alignment signals for $\vec{e}_2$ and $\vec{e}_3$ are very similar.

It is worth noting that a satellite velocity alignment with a given principal direction $\vec{e}_i(t)$ at T$_{\rm uni}=t$ excludes  that the satellite pole is simultaneously aligned with $\vec{e}_i(t)$ and conversely. Thereby, an early velocity alignment with $\vec{e}_3(t)$ demands a change in the velocity direction (i.e., a trajectory bending) for the satellite to be incorporated onto the $\vec{e}_3$-plane later on. As shown in Fig.~\ref{ch5:fig_VI:vel_LV_KPPHS} (right panel), such a bending event is signalled by a decrease in $A_{3}$ (purple lines) accompanied by an increase in either the $A_2$ or  $A_1$ medians (orange or blue lines).  
We also note that, once a satellite becomes bound to its host and begins to orbit, the alignments of its velocity vector fluctuate. A clear example of this case is in the right panel of Fig.~\ref{ch5:fig_VI:vel_LV_KPPHS} from T$_{\rm uni}~\sim4$ Gyr onwards. 

With this in mind, we now analyze the different possibilities implied by the early values of the medians $A_{1}$, $A_{2}$ and $A_{3}$ in these plots. 

\begin{enumerate}[label = {\roman*)}, wide, left=0pt, labelsep=0em,labelindent=0pt]
    \item When the $A_3$ value is initially high and long-lasting, then satellites in the KPP-HS system under study remain in a close-to-normal motion relative to the $\vec{e}_3$-plane, with their poles aligned with the $\vec{e}_2$ (or $\vec{e}_1$) principal directions. Examples are KPP-HS systems with ID\# 430864, 483594, 491426, 515695 in Fig.~A.3 (all of them having their orbital poles aligned to $\vec{e}_2$ from their $T_{\rm cluster}^{\rm Jstack}$ timescale onwards).
    
    \item In the case $A_3$ is initially high but short-lasting, this signals that a bending event has occurred involving  a given satellite set.  From this point onwards, involved satellites have their poles clustered and aligned with the $\vec{e}_3$ principal direction. Some examples are KPP-HS systems with ID\# 432106, 436932, 473329, 501725,  514829, 519311,  among others.

    \item A third possibility is that the $A_3$ median is the maximum of all $A_{i}$ alignments but is not quite high ($A_3\lesssim 0.7$), as is the case e.g. for the KPP-HS systems with ID\# 447914, 470345, 505586, 550149. In these cases, bending of their trajectories  and  their incorporation onto the $\vec{e}_3$-plane are more frequent than in the previous cases. Thus, their poles become clustered  as they  align with the $\vec{e}_3$ direction.

    \item In some cases the $A_{3}$ and $A_{i}$, $i= 1$ and/or 2, are initially similar and relatively high (e.g. for KPP-HS systems ID\#  402555, 428177, 470345,  489206).  In these cases, satellite trajectories are also bent to eventually have orbital poles aligned with the $\vec{e}_3$ principal direction.

    \item Let us turn to analyze the cases of HS systems  where  the median  $A_{2}$  is the first to dominate, i.e., it is the highest of the three medians of  $A_{j,i}$, ($i=1, 2$ and 3) at early cosmic times.
    This defines a set of satellites whose velocity components are preferentially parallel to the $\vec{e}_2$ principal axis, that is, they lie in the $\vec{e}_3$-plane. Some of them have their poles aligned with $\vec{e}_3$ from very early times ($T_{\rm align}^{ei}< 2$ Gyr). KPP-HS systems with ID\# 422754, 456326, 488530, 535774 (not shown in the Supplementary File), 547293 are in this situation (see left panel of Fig.~\ref{ch5:fig_VI:vel_LV_KPPHS} and Fig.~A.3). In Tab.~\ref{ch5:tab:DataTable2} we can see that these systems have their poles aligned with the $\vec{e}_3$ principal direction (see `Axis $\vec{e}_i$' column). 
    There is no case in which the median $A_1$ dominates first for a set of KPP satellites.    
\end{enumerate}


A common pattern for some KPP satellite sets in Fig.~A.3 (upper panels) 
-- clearly illustrated in Fig. \ref{ch5:fig_VI:vel_LV_KPPHS}, right panel --  is that   alignments with the $\vec{e}_3(t)$ (purple), $\vec{e}_2(t)$ (orange) and $\vec{e}_1(t)$ (blue) clearly dominate  one after the other.
In other words, in these cases satellite velocity components are preferentially  aligned first with the $\vec{e}_3(t)$ direction,   then along  the $\vec{e}_2(t)$ direction enclosed in the $\vec{e}_3$-plane (after a  bending event),  and finally preferentially along the  $\vec{e}_1(t)$ principal direction  within the prolate structure, presumably after a second bending event.
Satellites in these KPP-HS systems clearly mark the expected successive mass flows shaping their environment according to the TOI analysis.
Some examples of these patterns can be found in KPP-HS systems with ID\# 432106, 436932, 496186, 514829, 550149. All of them lead to KPP configurations whose  orbital poles are eventually aligned with the $\vec{e}_3(t)$ principal direction.

Another pattern  at high redshifts consists of  satellite  velocities  first aligned to $\vec{e}_2(t)$ within the $\vec{e}_3$-plane  (orange curves) and then to $\vec{e}_1(t)$ (blue curves). Some examples are KPP-HS systems whose ID\#  are 488530, 547293. According to Tab.~\ref{ch5:tab:DataTable2} these are aligned to $\vec{e}_3$, and with $T_{\rm align}^{ei} = 2.52$ Gyr and 0.78 respectively, rather early values.

In summary, the analysis  presented in this section indicates that (proto-)satellite velocities most frequently exhibit an initial alignment with $\vec{e}_3$ (the direction of maximum compression). This preferential alignment arises from the most common temporal ordering in which mass inflows of migrating material become dominant.
Consequently, an alignment of the orbital poles with $\vec{e}_3$ at very early times -- where “very early” refers to epochs preceding the onset of their clustering at $T_{\rm cluster}^{\rm Jstack}$ -- is physically disfavoured. 


\section{Timescale Statistics}
\label{ch5:sec:Timescales}


\begin{table*}
\caption{Table containing an overview of the properties presented in this paper regarding HS systems  where KPPs have been detected (KPP-HS systems). The information is categorized in terms of host information, KPP information, LV information, LV alignments, and properties of satellite trajectories. These timescales and properties give information about the process of KPP formation, and the possible relationship with the shaping  of their environment at early Universe ages. Timescales are given in Gyr and $M_{\rm subh}$ in units of M$_{\odot}$ (logarithmic scale). Numbers within brackets below the column headers indicate the subsection where the corresponding header definition can be found. See Glossary in Table \ref{tab:Glossary} as well. 
}
\scriptsize
\vspace{0cm}
\begin{tabular}{|l | c  | c | c | c | c | c | c | c | c | c | c | c | c | c | c | c |}
\hline
\hline
\multicolumn{2}{|c|}{Host information} & \multicolumn{2}{ c|}{KPP information} & \multicolumn{4}{ c|}{LV information} & \multicolumn{2}{ c|}{LV alignments}   &  \multicolumn{5}{c|}{Trajectory properties} \\
\hline
\multicolumn{1}{|c|}{Host ID\#} & \multicolumn{1}{c|}{$T_{\rm no-fast}$} & \multicolumn{1}{c|}{$T_{\rm cluster}^{\rm Jstack}$} &  \multicolumn{1}{c|}{$T_{\rm dJs}$} & \multicolumn{1}{c|}{$M_{\rm DM}$} & \multicolumn{1}{c|}{$R_{\rm LV}$} & \multicolumn{1}{c|}{$T_{\rm dir}^{e3}$} & \multicolumn{1}{c|}{$T_{\rm shape}^{e3}$} & \multicolumn{1}{c|}{Axis $\vec{e}_i$} &  \multicolumn{1}{c|}{$T_{\rm align}^{ei}$} &  \multicolumn{1}{c|}{$T_{\rm apo1}$} & \multicolumn{1}{c|}{$\kappa_{\rm rot}^{\rm eq}$} & \multicolumn{1}{c|}{$T_{\rm Krot}^{\rm  0.5}$} & \multicolumn{1}{c|}{$T_{\rm Krot}^{\rm eq}$} & \multicolumn{1}{c|}{$T_{\rm dei}$} \\
    & [\ref{sec:KPP-identification}]& [\ref{ch5:sec:ComparingTools_Jsplane}]& [\ref{ch5:sec:DisPlaneJstack}]& [\ref{ch5:sec:LVs_intro}]& [\ref{ch5:sec:LVs_intro}]& [\ref{ch5:sec:PDirEvol}]& [\ref{ch5:sec:ShapeEvol}]& [\ref{ch5:sec:AlignPDir}]& [\ref{ch5:sec:JorbEiAlign}] & [\ref{ch5:sec:JorbEiAlign}] & [\ref{ch5:sec:VelSpace}]& [\ref{ch5:sec:VelSpace}]& [\ref{ch5:sec:VelSpace}]& [\ref{ch5:sec:distplane_ei}]\\ 

\hline
Med. \& Quar.  &  $6.2^{+1.0}_{-0.7}$ & $3.9^{+0.7}_{-0.5}$ &  $4.2^{+1.0}_{-0.6}$ &  -- & -- & $3.8^{+3.7}_{-3.4}$ & $3.4^{+1.4}_{-0.7}$ & -- & $3.3^{+1.8}_{-1.7}$ & $2.7^{+0.7}_{-0.6}$  & $0.68^{+0.1}_{-0.1}$ & $5.1^{+0.9}_{-0.9}$ & $6.5^{+1.2}_{-1.4}$ & $3.9^{+0.9}_{-0.5}$\\   
\hline
402555 & 4.8 & 3.27 & 4.1  & 13.10 & 2.96 & 0.43 & 5.0 & $\vec{e}_3$ & 5.72 & 2.14 & 0.65 & 4.78 & 4.94 & 4.1 \\
411449  & 6.8&  4.53 & 3.6 & 13.25 & 3.36 & 1.44 & 5.0 & $\vec{e}_3$ & 6.62 & 2.68  & 0.77 &  4.91  & 7.26 & 4.25 \\
430864  & 7.3 & 5.47& 4.9 & 12.98 & 2.74 & 0.47 & 5.0 & $\vec{e}_2$ & 0.63 & 3.44  & 0.75  & 5.57 & 7.76 & 4.9 \\
432106   & 5.5 & 3.93 & 3.2 & 13.10 & 2.95 & 0.18 & 5.0 & $\vec{e}_3$ & 3.34 & 3.37  & 0.65  & 5.89 & 7.52 & 3.0 \\
433289  & 5.4  & 4.42 & 5.2 & 13.18 & 3.14 & 10.4 & 3.5 & $\vec{e}_3$ & 6.21 & 1.94  & 0.73  & 5.53  & 6.37 & 4.6 \\
436932 & 5.5  &  3.41 & 5.8 & 13.13 & 3.00 & 0.37 & 2.7 & $\vec{e}_3$ & 3.28 & 1.94  & 0.82  & 3.09 & 4.96 & 2.25  \\
447914  & 6  & 3.49 & 5.4 & 13.19 & 3.22 & 4.69 & 2.7 & $\vec{e}_3$  & 2.49 & 3.44  & 0.56 & 8.35 & 8.58 & 5.5 \\
468590  & 7  & 5.70 & 4.3 & 13.06 & 2.87 & 0.18 & 1.9 & $\vec{e}_3$ & 9.34 &   3.28  & 0.58  & 7.99 & 8.28 & 3.5  \\
470345  & 6  & 4.20 & 3.9  & 13.05 & 2.88 & 3.30 & 3.4 & $\vec{e}_3$ & 4.64 & 2.83  & 0.57  & 5.38 & 5.62 & 4.5 \\
471248  & 6  & 3.49 & 4.0 & 13.06 & 2.91 & 0.47 & 4.2 & -- & -- & 0.72  &  0.87    & 4.28  & 6.05 & 4.0 \\
483594  & 6.5  & 4.10 & 3.8 & 13.04 & 2.83 & 0.41 & 1.9 & $\vec{e}_2$ & 0.60 & 2.34  & 0.61  & 4.28 & 6.08 & 5.1 \\
488530  & 5.3  & 2.36 & 2.7 & 12.89 & 2.50 & 10.5 & 2.6 & $\vec{e}_3$  &  1.69    & 2.52 & 0.84  & 3.38 & 5.10 & 3.4 \\
489206  & 5  & 3.81 & 3.7 & 12.96 & 2.67 & 0.18 & 2.0 & $\vec{e}_3$  & 3.28 & 2.14  & 0.67  & 4.33 & 6.09 & 2.7 \\
491426  & 4.5  & 4.62 & 4.0 & 12.91 & 2.58 & 0.42 & 4.2 & $\vec{e}_2$ &0.58 & 2.76  & 0.71  & 5.24 & 8.54 & 4.6 \\
496186 & 6  & 3.95 & 3.7 & 12.96 & 2.67 & 10.85 & 5.0 & $\vec{e}_3$ & 3.48 & 1.81    & 0.68  & 3.80 & 4.07 & 3.5  \\
501208 & 6 & 3.19 & 3.2 & 12.89 & 2.54 & 3.12 & 2.7 & $\vec{e}_3$ & 3.0 &   2.14   & 0.75  & 3.33 & 4.46 & 2.0  \\

\hline
\hline
416713 & 7.3  & 3.56 & 5.6 & 13.23 & 3.31 & 4.81 & 2.7 & $\vec{e}_3$ & 1.35 & 3.66 & 0.59  & 6.69 & 9.18 & 6.5 \\
419618 & 8.6 & 4.51 & 6.9 & 13.19 & 3.15 & 4.65 & 5.6 & $\vec{e}_2$ & 0.84 &   3.74 & 0.78  & 6.0 & 8.78 & 5.6 \\
421555 & 7.1  & 4.42 & 5.8 & 13.21 & 3.27 & 1.18 & 5.6 & $\vec{e}_3$ & 4.65 & 3.44 & 0.74  & 5.85 & 7.10 & 4.8  \\
422754 & 6.7  & 3.28 & 3.6 & 13.14 & 3.11 & 3.50 & 4.2  & $\vec{e}_3$ & 1.70 & 3.74 & 0.69  & 7.37 & 7.62 & 4.2 \\
428177 & 6.5  & 5.16 & 5.4 & 12.99 & 2.79 & 0.38 & 4.2 & $\vec{e}_3$ & 5.15 & 2.98 & 0.50  & 7.65 & 9.02  & 3.4 \\
456326 & 6.3  & 3.75 & 1.7 & 13.10 & 2.99 & 5.35 & 9.0 & $\vec{e}_3$ & 1.71 & 2.53 & 0.64  & 4.2  & 5.34  & 2.6 \\
461785 & 9.2 & 3.90 & 5.2 & 13.11 & 3.02 & 9.23 & 5.7 & $\vec{e}_1$ & 10.04 & 3.05 & 0.77  & 6.38 & 7.64  & 7.3 \\
\hline
\hline
473329 & 6  & 3.87 & 2.1 & 12.91 & 2.60 & 0.90 & 3.3 & $\vec{e}_3$ & 3.98 &   4.03  & 0.63  & 6.40  & 8.61 & 3.5 \\
475619 & 5.25  & 2.61 & 3.8 & 13.10 & 2.92 & 13.45 & 1.9 & $\vec{e}_1$ &0.76 & 1.81 & 0.76  & 3.96 & 4.64 & 5.0 \\
479290 & 7.8  & 3.63 & 5.0 & 12.98 & 2.72 & 0.18 & 2.7 & $\vec{e}_3$ & 6.52 & 4.65 & 0.64  & 6.37 & 8.50 & 4.3 \\
482155 & 8.4  & 1.34 & 4.4 & 13.01 & 2.77 & 0.18 & 2.7 & $\vec{e}_2$ & 1.03 & 2.19 & 0.69  & 2.03 & 2.51 & 3.8 \\
494011 & 7.4  & 5.54 & 2.6 & 12.85 & 2.49 & 5.23 & 4.2 & $\vec{e}_3$ & 5.40 & 2.68 & 0.79  & 5.70 & 7.56 & 4.2 \\
498522 & 7.2  & 4.02 & 4.3 & 12.94 & 2.64 & 4.67 & 3.8 & $\vec{e}_3$ & 3.96 & 2.53 & 0.65  & 5.09 & 5.18 & 4 \\
501725 & 5.8  & 4.82 & 3.9 & 12.85 & 2.44 & 8.31 & 3.5 & $\vec{e}_3$ & 4.78 & 2.46 & 0.70  & 4.36  & 4.72  & 3.4 \\
503437 & 7.3  & 6.12 & 7.4 & 12.88 & 2.47 & 11.1 & 4.6 & $\vec{e}_2$ & 8.07 & 2.68 & 0.77  & 6.01 & 7.87 & 7.3 \\
505586 & 9  & 3.32 & 3.4 & 12.85 & 2.40 & 12.5 & 2.3 & $\vec{e}_3$ & 4.76  &   1.69 & 0.80  & 3.14 & 3.64 & 2.7 \\
506720 & 7  & 5.25 & 8.1 & 12.86 & 2.49 & 4.09 & 5.7 & $\vec{e}_3$ & 3.13  &   5.11 & 0.52 & 10.48 & 11.45  & 10.5 \\
514829 & 5.7 & 4.01 & 4.2 & 12.81 & 2.39 & 4.14 & 2.6 & $\vec{e}_3$ & 3.10 &   3.74 & 0.73  & 4.0 & 6.67 & 2.6 \\
515695 & 5.6 & 3.69 & 6.5 & 12.83 & 2.39 & 0.54 & 3.4 & $\vec{e}_2$ & 0.89  & 2.14 & 0.62 & 5.06  & 7.06 & 5.7 \\
517271 & 5.5  & 4.79 & 4.3 & 12.82 & 2.41 & 4.23 & 3.4 & $\vec{e}_3$ & 5.69 & 3.44 & 0.66  & 5.09 & 5.49 & 4.0  \\
519311 & 7.5 & 5.74 & 4.8 & 12.82 & 2.39 & 0.18 & 3.4 & -- & -- & 2.24 & 0.84   &  5.89 & 6.18 & 2.0  \\
523889 & 5 & 3.99 & 4.0 & 12.67 & 2.18 & 0.18 & 2.6 & -- & -- & 3.44 & 0.51  & 6.88 & 6.90 & 2.6  \\
529365 & 4.5 & 3.96 & 3.1 & 12.77 & 2.31 & 8.05 & 3.1 & $\vec{e}_2$ & 0.47 &   2.98 & 0.71   & 4.24 & 4.38 & 3.5 \\
530330 & 4  & 3.72 & 2.9 & 12.80 & 2.39 & 0.18 & 3.4 & $\vec{e}_3$ & 6.46 &    2.83 & 0.62   & 3.85 & 5.26 & 3.0 \\
531910 & 7.3  & 5.81 & 5.2 & 12.72 & 2.20 & 12.0 & 4.3 & $\vec{e}_3$ & 3.75 & 3.59 & 0.63   & 5.87 & 7.77 & 3.8 \\
535774 & 6.7  & 1.27 & 4.1 & 12.71 & 2.15 & 7.84 & 2.6 & $\vec{e}_3$ & 1.88 & 2.14 & 0.82   & 4.35 & 4.63 & 2.6 \\
543114 & 4.5  & 5.22 & 3.5 & 12.72 & 2.20 & 3.0 & 3.4 & -- & --  & 1.46  & 0.63  & 3.79 & 5.90 & 3.5 \\
544408 & 7.2  & 5.61 & 6.4 & 12.67 & 2.15 & 0.58 & 4.2 & $\vec{e}_2$ & 1.59 & 1.94 & 0.76   & 4.12 & 5.25 & 3.4 \\
547293 & 6.1  & 3.24 & 1.9 & 12.65 & 2.07 & 6.43 & 4.9 & $\vec{e}_3$ & 0.88 & 1.94 & 0.71  & 4.03 & 5.06 & 1.4 \\
550149 & 5.3 & 3.04 & 4.2 & 12.64 & 2.08 & 0.18 & 3.4 & $\vec{e}_3$ & 3.13 &   3.36 & 0.65  & 6.52 & 6.95 & 3.5 \\

\hline
\hline
\end{tabular}
\label{ch5:tab:DataTable2}
\end{table*}


Throughout this paper we have studied several timescales related to the emergence, deformation and shaping of the local environment around HS systems, as well as other timescales concerning the trajectories and kinematics of KPP satellites in the context of their orbital pole clustering, some of them discussed in \citetalias{GamezMarin2025_PaperV} as well. These timescales are summarized in Tab.~\ref{ch5:tab:DataTable2}.
In this section, we compare these timescales with each other and examine their statistical distributions and possible interconnections.

\begin{figure*}
\centering
\includegraphics[width=\linewidth]{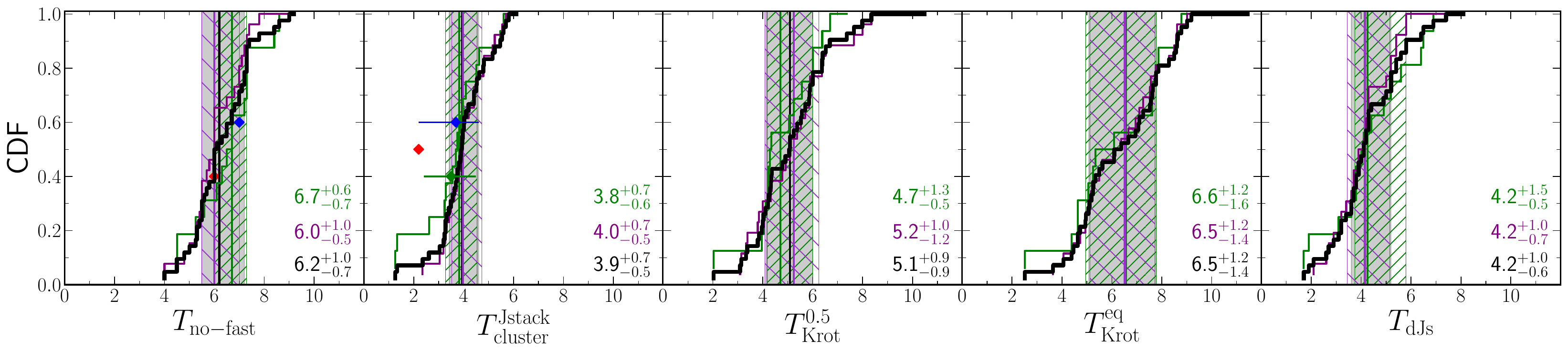}
\includegraphics[width=\linewidth]{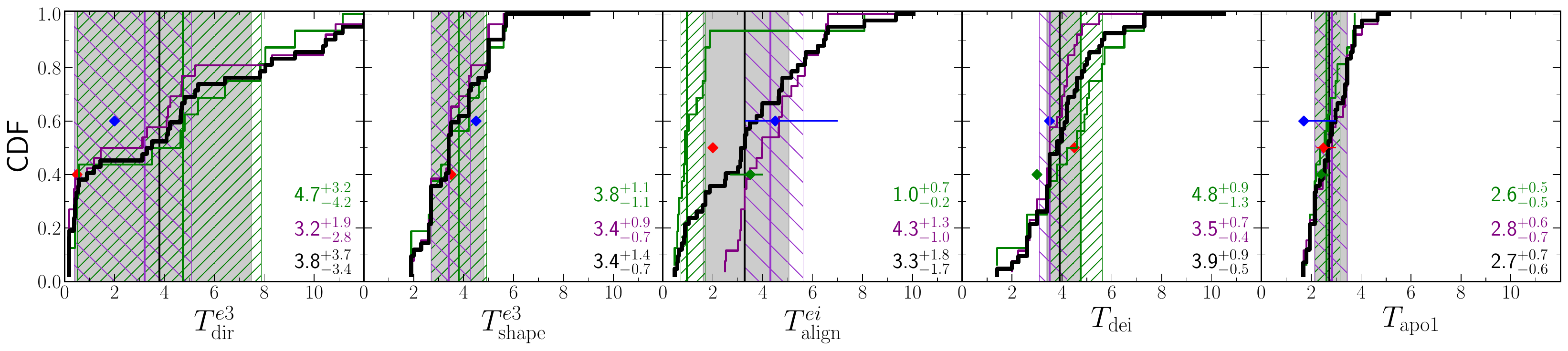}
\caption{Cumulative distribution functions of the different timescales analyzed in this work. 
Black lines show the total distributions, and green and purple thin lines show results separating KPP-HS systems into different populations as decribed in Sec.~\ref{ch5:sec:Timescales}.
The median and interquartile ranges for each of the analyzed distributions are given in the lower right corner of each plot, as well as visually depicted as thick lines and shaded vertical bands respectively, colour-coded according to the colour assigned to each KPP-HS population 
differentiated in Sec.~\ref{ch5:sec:Timescales}.
Diamonds with horizontal error bars in some panels stand for the corresponding values of the KPPs (when this is the case) in the Aq-C$^{\alpha}$ (blue) and the PDEVA-5004 simulations (red, or red and green when distinguishing between the KPP1 and the KPP2 planes), as presented in tab.~2 of \citetalias{Gamez-Marin2024}.}
\label{ch5:fig_VI:CDF-timescales}
\end{figure*}

\begin{figure}
\centering
\includegraphics[width=\linewidth]{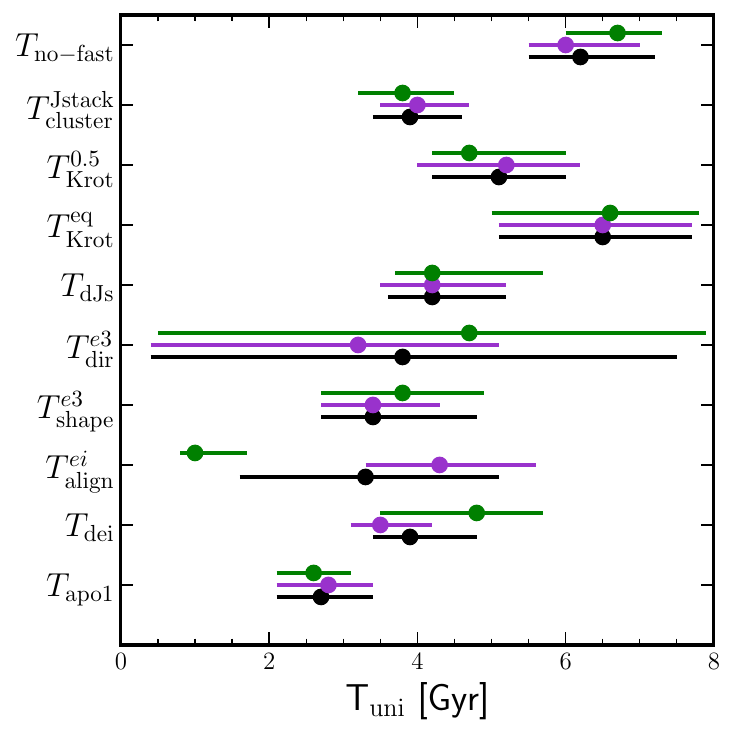}
\caption{Representation of the distributions of the different timescales analyzed in this paper. Median values and interquartile ranges are represented by filled points with horizontal error bars. For each timescale, three distributions are drawn, corresponding to the three KPP-HS populations considered in Sec.~\ref{ch5:sec:Timescales} as indicated by their colour. 
}
\label{ch5:fig_VI:Timescales_visual_distrib}
\end{figure}

To avoid possible binning effects, the statistical properties  are reported through  CDFs.  For each timescale, its CDF  has been calculated and it is represented in Fig.~\ref{ch5:fig_VI:CDF-timescales} along with its median values (vertical lines) and interquatile ranges (shaded areas), displayed on the lower right corner of each panel. This is of particular relevance, as it provides the dispersion of KPP formation timescales due to using large volume cosmological simulations, instead of zoom-in ones (as presented in \citetalias{Gamez-Marin2024}). 
As a visual guide and a summary of the information provided in Fig.~\ref{ch5:fig_VI:CDF-timescales}, in Fig.~\ref{ch5:fig_VI:Timescales_visual_distrib} the median values and interquartile ranges for the 10 timescales are marked by segments with filled circles.

The colour-coding is defined as follow: green  CDFs (and segments and filled circles in Fig. ~\ref{ch5:fig_VI:Timescales_visual_distrib}) stand for those HS systems whose corresponding $\vec{J}_{\rm stack}$ vector and KPP satellite orbital poles align with either the  $\vec{e}_1(t)$ or the $\vec{e}_2(t)$  principal directions, as well as those that align with $\vec{e}_3(t)$ at times $T_{\rm align}^{ei}<2$ Gyr; purple CDFs refer to those HS systems whose satellite orbital poles align with the $\vec{e}_3(t)$ principal direction at times $T_{\rm align}^{\rm e3}\geq2$ Gyr; black CDFs do not distinguish between these two possibilities. The distributions of those KPP-HS systems whose satellites do not align with any principal direction are not represented in the CDFs, leaving a total of 42 KPP-HS systems in these plots.

We first report on the general statistical properties provided by the overall CDFs (medians and dispersions as black curves and grey shades in Fig.~\ref{ch5:fig_VI:CDF-timescales}, respectively), and black circles and segments in Fig.~\ref{ch5:fig_VI:Timescales_visual_distrib}:

\begin{enumerate}[label = {\roman*)}, wide, left=0pt, labelsep=0em,labelindent=0pt]
    \item $T_{\rm cluster}^{\rm Jstack}$, $T_{\rm dJs}$, $T_{\rm dei}$ and $T_{\rm dir}^{e3}$ CDFs show very close median values around T$_{\rm uni}\sim4$ Gyr.  However, while the three first CDFs show a low dispersion, the dispersion in $T_{\rm dir}^{e3}$ is remarkably large.
    
    \item The black CDFs of $T_{\rm no-fast}$ and $T_{\rm Krot}^{\rm eq}$ show median values close to 6 Gyr. They are close to $T_{\rm Krot}^{0.5}$ when percentiles are considered.

    \item The median values of $T_{\rm shape}^{e3}$ and $T_{\rm align}^{ei}$ show slightly lower values than the timescales mentioned in i), with values around 3.3 and 3.4 Gyr, respectively, although their large dispersions make them consistent with each other.

    \item 
    The satellites' first apocentre timescale, $T_{\rm apo1}$, occurs at earlier times, having a median value of 2.7 Gyr, with a low dispersion ($\leq 0.8$ Gyr).
\end{enumerate}

Summing up, we find three main blocks of timescales for the overall population of KPP-HS systems (black CDFs): i), those timescales related to plane collapse, KPP formation and LV evolution, having median values around $\sim4$ Gyr ($T_{\rm dJs}$, $T_{\rm cluster}^{\rm Jstack}$, $T_{\rm dir}^{e3}$, $T_{\rm shape}^{e3}$, $T_{\rm align}^{ei}$ and $T_{\rm dei}$); ii), timescales related to the end of the  violent halo mass assembly phase and KPP satellites rotational stabilisation, with values distributing around $\sim6$ Gyr ($T_{\rm no-fast}$ and $T_{\rm Krot}^{\rm eq}$); and iii) a timescale for KPP satellite turn-around, taking place at earlier Universe ages and showing no relation with KPP formation processes ($T_{\rm apo1}$). In turn, the distribution of $T_{\rm Krot}^{0.5}$ values lies between blocks i) and ii) marking the transition between KPP formation and its stabilisation.

Regarding the the timescales having median values around $\sim$ 4 Gyr, some of them qualify processes of early KPP  formation ($T_{\rm cluster}^{\rm Jstack}$ and $T_{\rm dJs}$), while other timescales refer to the dark matter structuration in the surroundings of the HS system formation site ($T_{\rm dei}$, $T_{\rm dir}^{e3}$).

This coevality (among timescales qualifying dark matter and satellite mass flows) strongly suggests that  we witness the infall of (proto-)satellites as part of the mass flows that, later on, will form a sheet -- a sheet that, later on, will on its turn (almost) disappear as it feeds filaments and clumps (including satellites).

It should be recalled that current models (e.g., the Adhesion Model) predict the existence of a mathematical plane that functions as a primary (i.e., the earliest one) attractor for large-scale mass migrant flows.
However, there is not a physical such plane as a pre-existing entity. Hence, rather than satellites accreted over a pre-existing sheet, we witness the co-eval organization of mass density into a flatened structure, and of (proto-)satellites into an early KPP.

As said above, in Fig.~\ref{ch5:fig_VI:CDF-timescales} and \ref{ch5:fig_VI:Timescales_visual_distrib}
KPP-HS systems have been split into purple and green populations. In most cases the CDF medians of both populations are coeval, as seen in Fig.~\ref{ch5:fig_VI:Timescales_visual_distrib}. Among those that are not coeval, three cases stand out: $T_{\rm align}^{ei}$, $T_{\rm dir}^{e3}$ and $T_{\rm dei}$. 
Below we carry out more detailed analyses  from an individual perspective for the timescales:

\begin{enumerate}[label = {\roman*)}, wide, left=0pt, labelsep=0em,labelindent=0pt]

    \item $T_{\rm align}^{ei}$ shows a broader global range compared to the other timescales, displaying a bimodal distribution. On one hand, a peak occurs at a Universe age around $3-4$ Gyr, showing a broad distribution around $4.3$ Gyr (see purple vertical bands in Fig.~\ref{ch5:fig_VI:CDF-timescales}). This corresponds to systems where KPP satellites reorient their poles and align them with $\vec{e}_3(t)$. On the other hand, a second peak (green CDF), which involves $\sim25\%$ of KPP-HS systems, peaks at T$_{\rm uni}\sim 1$ Gyr, showing a tighter distribution of $1.0^{+0.7}_{-0.2}$ Gyr. These low $T_{\rm align}^{e_i}$ values correspond to systems whose KPP satellite poles align since high redshift with the LV principal directions. An example of such a system was previously identified in \citetalias{Gamez-Marin2024} for KPP1 satellites in the PDEVA-5004 zoom-in simulation.

    \item 
    The median time for KPP formation, $T_{\rm cluster}^{\rm Jstack}$, has a P24 distribution peaking at $3.9$ Gyr. 
    Considering their range of values, we see that the building up of satellite clustering generally occurs before the fast, violent phase of host halo mass assembly in KPP-HS systems is over (marked by $T_{\rm no-fast}$).

    \item $T_{\rm dir}^{e3}$ marks the Universe age when the LV's direction of maximum compression $\vec{e}_3(t)$ becomes fixed within a threshold of 10\%. We find a prominent increase at T$_{\rm uni}<1$ Gyr in its CDF, corresponding to LVs that have this direction fixed from very early cosmic times, whereas the rest of the systems show variable $T_{\rm dir}^{e3}$ values widely spread across time. Overall values (represented by a black point in Fig.~\ref{ch5:fig_VI:Timescales_visual_distrib}) have a median of $3.8$ Gyr, accompanied by a large dispersion. 
    We see that the green population is delayed with respect to the purple one by $\sim1.5$ Gyr, as indicated by their medians.

    \item $T_{\rm cluster}^{\rm Jstack}$  and $T_{\rm align}^{ei}$ purple populations show roughly coeval values (i.e. once values below $\sim$ 2 Gyr, represented by green points, are excluded).
    These timescales are similar to $T_{\rm shape}^{e3}$ and $T_{\rm dei}$ (although slightly higher) for the purple population. 
    These similarities between timescales suggest that mass flows along the $\vec{e}_3(t)$ principal direction, as traced by (proto-)satellites, significantly influence the shaping of KPPs, particularly those whose satellite poles cluster around $\vec{e}_3(t)$, extending results of \citetalias{Gamez-Marin2024}.

    \item As mentioned, the timescales related to satellite incorporation onto the (mathematical) planes defined by $\vec{J}_{\rm stack}$ and $\vec{e}_i$, $T_{\rm dJs}$ and $T_{\rm dei}$, display similar values for the whole population (black points in Fig.~\ref{ch5:fig_VI:Timescales_visual_distrib}).  This is indicative of the coincidence that exists between the $\vec{J}_{\rm stack}$ axis and the $\vec{e}_i(t)$ principal direction with which it aligns.    
\end{enumerate}

These findings highlight the interplay between different timescales and their role in shaping KPPs, offering valuable insights into the processes driving satellite orbital pole clustering and alignment with the CW.


\section{Discussion}
\label{sec:Dicussion}


\subsection{Comparison with other works}
\label{ch5:sec:Comparison_zoomin_timescales}

\paragraph*{Comparison with Aq-C$^\alpha$ and PDEVA-5004 zoom-in simulation results}\mbox{\\}

An important point to dilucidate is how well the results on the distributions of  timescales relevant for KPP formation presented in Tab.~\ref{ch5:tab:DataTable2} and Fig.~\ref{ch5:fig_VI:CDF-timescales} compare with the values obtained from the zoom-in simulations analyses presented in \citetalias{Gamez-Marin2024}, summarized in its tab.~2. To answer this question, zoom-in simulation results have been drawn in their corresponding panels in Fig.~\ref{ch5:fig_VI:CDF-timescales} as coloured diamonds with error bars\footnote{Note that some panels do not display the values for the zoom-in-simulations, as these timescales were not computed in \citetalias{Gamez-Marin2024}.}. We see that the consistency is satisfactory, as most diamonds from zoom-in simulations lie within the interquatile ranges of their analogous values coming from the TNG50 analysis.
Only KPP1  in PDEVA-5004 (red diamond) shows a slightly earlier $T_{\rm cluster}^{\rm Jstack}$ than the interquartil range, but it still remains within the overall distribution of timescales displayed by KPPs from the P24 sample.

In addition, other parameter values found in the analyzed zoom-in simulations  
are also consistent with the ones obtained here.
For example, the KPP in Aq-C$^{\alpha}$ and KPP2 in PDEVA-5004 have $\kappa_{\rm rot}^{\rm eq}$ values $~\sim0.65$, and in the case of the KPP1 in PDEVA-5004 is $\kappa_{\rm rot}^{\rm eq}\sim0.7$, see fig.~3 in \citetalias{Gamez-Marin2024}. These values lie within the distribution of $\kappa_{\rm rot}^{\rm eq}$ of the KPP-HS systems under study, which show a distribution with median and percentiles of $\kappa_{\rm rot}^{\rm eq}=0.68^{+0.08}_{-0.05}$ (see Tab.~\ref{ch5:tab:DataTable2}). 

As for the role of the local environment in the formation of KPPs, we have seen that most KPP-HS systems listed in Tab.~\ref{ch5:tab:DataTable2} have their satellite orbital poles aligned to the $\vec{e}_3$ principal direction (67.4\%), while some have them aligned to the $\vec{e}_2$ or $\vec{e}_1$ principal directions (19.5\% and 4.3\%, respectively). We discuss these two possibilities in turn.
 
KPP formation in P24 HS systems is studied in  Sec.~\ref{ch5:sec:vel_ei_LVs}  above, where it has been found that satellite orbital pole alignment with the $\vec{e}_3$ principal direction comes about through two possible processes: 
a), the pole alignment with $\vec{e}_3$ is in place from very early times, and the satellite velocity vector lies close to the $\vec{e}_3$-plane and is preferentially aligned to the $\vec{e}_2$ or $\vec{e}_1$ principal directions (as explained in point v) in Sec.~\ref{ch5:sec:vel_ei_LVs}), or b), the satellite trajectory is bent as a consequence of the $\vec{e}_3$-plane collapse  (driven by the local CW evolution at high redshift) (see points ii) to iv) of Sec.~\ref{ch5:sec:vel_ei_LVs}). 
In the bending case, the satellite velocity is first more or less aligned with the $\vec{e}_3$ principal direction at very early times, i.e., the satellite velocity has an important component along $\vec{e}_3$, but it is not fully parallel to this principal direction.  This alignment pattern is the same as that described for KPP satellites with pole alignment around the $\vec{e}_3$ principal direction  in the Aq-C$^{\alpha}$ and PDEVA-5004 zoom-in simulations in \citetalias{Gamez-Marin2024}. 

The process of satellite orbital pole alignment with the $\vec{e}_2(t)$ (or $\vec{e}_1(t)$) principal direction is described in point i) of Sec.~\ref{ch5:sec:vel_ei_LVs}. (Proto-)satellite  velocity is initially  preferentially aligned with the $\vec{e}_3$ axis, falling  rather perpendicularly  onto  the $\vec{e}_3$-plane.  No bending occurs and the satellite orbital pole maintains its initial alignment to the $\vec{e}_2$ (or $\vec{e}_1$) axis.  
This alignment pattern has been found in  the case of a sizable subset of satellites in the PDEVA-5004 simulation (the KPP1 system) in \citetalias{Gamez-Marin2024}, whose orbital poles are aligned with the $\vec{e}_1$ direction, as well as in  some satellites in the Aq-C$^{\alpha}$ simulation, with orbital poles aligned with the $\vec{e}_2$ eigenvector. 
Therefore, we confirm in this analysis of the P24 sample the role that the CW plays as driver of early KPP formation, put forward in \citetalias{Gamez-Marin2024} relative to two zoom-in simulations \citep{Gamez-Marin2024}.

\paragraph*{Are GSE-like mergers necessary to explain the early formation of populated KKP systems?}\mbox{\\}

We discuss recent findings by \citet{RodriguezCardoso2026}, who suggest that a massive Gaia--Sausage--Enceladus (GSE)-like merger---modeled as an early (first infall by $z \approx 2$), radial binary event that results in approximately 40\% of the host's final DM mass, followed by a quiet mass assembly thereafter \citep[see][ their fig.~3]{Rey:2022} -- may be a necessary driver for forming populated and thin \textit{early} KPPs, in addition to the orbital pole  alignments driven by the CW dynamics  described in the present study. 
Specifically, the authors suggest that halo effects induced by the merger may be an important ingredient driving KPP formation.

To test the universality of this claim, we examine counter-examples within our sample of KPP-HS systems with  a large satellite population, say $N_{\rm sat} \ge 25$, without such merger histories.
By definition, early KPP-HS systems have been selected to present a high $f_{\rm KPP}>25\%$ (with some of them showing $f_{\rm KPP}>30\%$, as in the case of KPP-HS systems listed in the first block of Table 1 in \citetalias{GamezMarin2025_PaperV}).

With the exception of the KPP-HS system with ID\# 447914, all KPP-HS systems listed in Table \ref{ch5:tab:DataTable2}
with  $N_{\rm sat} \ge 25$ satellites (15 KPP-HS systems, see their ID number and corresponding parameter values in Table 1 of \citetalias{GamezMarin2025_PaperV}) 
assemble  less (and many of them significantly less) than 40\% of their mass by T$_{\rm uni} = 3$~Gyr and are not quiescent thereafter.
As shown in Table 1 of \citetalias{GamezMarin2025_PaperV}, all KPP-HS systems  with $N_{\rm sat} \ge 25$ satellites 
present an elevated degree of satellite co-orbitation, as quantified by 
the satellite co-orbiting   fraction curve $f_{\rm sat}^{\alpha_{\rm co-orbit}}(t)$\footnote{For a given early KPP system, this is the fraction of its $N_{\rm sat}$ satellites that at Universe age $t$ are within an angle $\alpha_{\rm co-orbit}=36.^{\circ}87$ from the system's co-orbitation axis.} -- see  examples in Appendix \ref{app:appendix} here, and Fig. 4b in \citetalias{GamezMarin2025_PaperV} as well
--,  or by  the co-orbitation excess parameter (see $f_{\rm sat}^{\rm exc}(z=0)$ column in Table 1  and Fig. 4c of the same paper). This co-orbitation is consistent with the MW's one $f_{\rm MW}^{\alpha_{\rm co-orbit}} \simeq$  0.44$\pm 0.12$, from data in \citet{Taibi24}.
Moreover, detailed inspection of the temporal evolution of these KPPs shows that they are 
positional planes with low $c/a$ and high $b/a$ axis ratios over extended periods of cosmic time (see e.g. ID\# 433829 in Fig.~7 of \citetalias{GamezMarin2025_PaperV}). 
Notably, systems with  ID\# 432106, 419618 and 429555  reproduce MW-like co-rotation ($f_{\rm co\text{-}rot} \sim 0.8$) as well.

Importantly, early KPP satellite orbital pole clustering, quantified by the $T_{\rm cluster}^{\rm Jstack}$ timescale (see Fig.~12), happens well before  satellite infall onto the host halo, quantified by the $T_{\rm inf}$ timescale (see Fig.~11 in \citetalias{GamezMarin2025_PaperV}). 
This does not fit into a scenario  with halo effects as the main drivers of early KPP formation in the P24 HS sample.

We therefore conclude that the GSE-like mass-assembly pathway is not unique; diverse mass assembly histories leading to the early formation of populated, thin KPPs arise naturally 
 when exploring generic CW initial conditions.


\subsection{Robustness under LV size change}
\label{ch5:sec:K_LV_change}

Results on LV evolution and its effects on satellite orbital pole clustering, presented in the previous sections, have been obtained using a LV radius at $z_{\rm high}$ of $R_{\rm LV}=K \times R_{\rm 200c}$, with $K=15$.

The size of LVs must be high enough that each HS system is conveniently sampled, including all the satellite-to-be structures and the surrounding medium, and that, at the same time, the particular properties of each HS system can be individually  followed along its corresponding LV evolution. Thus, this size cannot be too large either, as the inclussion of other massive structures different from the analyzed KPP-HS system can affect these measurements. 

Changing the LV size implies that the corresponding TOI matrices change, potentially resulting in different principal directions (driving orbital pole clustering of KPP satellites at high redshift) and different principal axes (responsible for LV shape evolution).
In this section, we briefly analyze the effects on KPP satellite alignments when decreasing the LV scale to $K=10$ (involving an LV mass $\sim3$ times lower) instead of the reference value $K=15$. 

\begin{figure}
\centering
\includegraphics[width=0.85\linewidth]{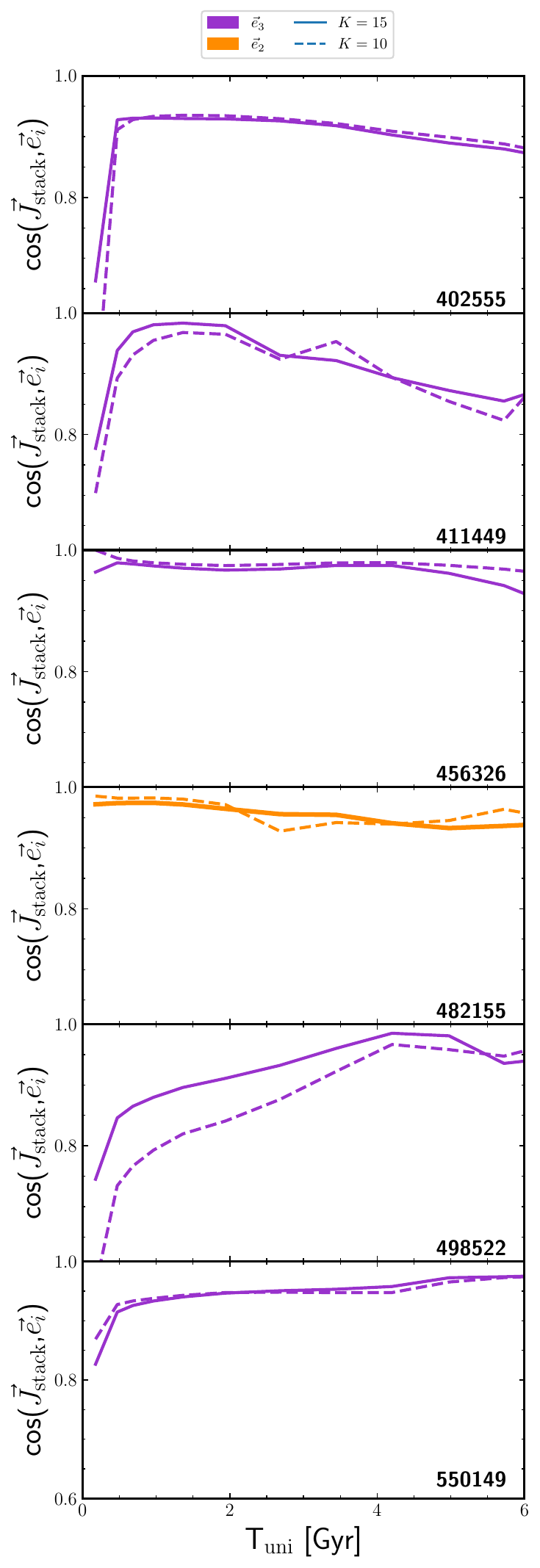}
\caption{Examples of the alignment signal between the $\vec{J}_{\rm stack}$ axial vector and the principal direction of the LV with which it aligns for 6 KPP-HS systems. Continuous (dashed) lines represent the alignment with the principal directions of the LVs defined assuming a scale of $K=15$ ($K=10$). Colour codes stand for the LV principal direction around which the KPP satellite poles cluster. The KPP-HS ID\# is shown on the right, bottom corner in each panel.}
\label{ch5:fig_VI:Kscale_comparison}
\end{figure}

In Fig.~\ref{ch5:fig_VI:Kscale_comparison} we plot the values of cos($\vec{J}_{\rm stack},e_i^{K=15}$) (continuous lines) and cos($\vec{J}_{\rm stack},e_i^{K=10}$) (dashed lines), where ``$i$'' stands for the LV principal direction the KPP set is aligned to,  
for six KPP-HS systems. The T$_{\rm uni}$ range drawn is limited to early Universe ages, when satellite pole 
clustering is built up.  It can be appreciated that the alignments 
are very similar for these principal axes driving clustering. This implies that the scenario for orbital pole clustering  based on the $\vec{J}_{\rm stack}$-plane and the alignment of this vector with the clustering-driving principal direction, still holds when  the LV size is decreased by choosing $K=10$. Therefore, the principal-direction-driven scenario for orbital pole clustering  is robust under changes of LV scale within a factor $\sim3$ in volume.


\section{Summary and Conclusions}
\label{sec:SummConclu}

In this section we summarise the main results and conclusions of this work on the formation of kinematically persistent planes of satellites (KPPs). Definitions of acronyms and frequently used terms are provided in the Glossary (see Table~\ref{tab:Glossary}).


\subsection{A brief summary on the physical origins of early KPPs in the P24 sample}
\label{sec:ResultSummary}
 
After the identification in \citet{GamezMarin2025_PaperV} (Paper V) of 46 early KPPs
of satellites,  out of 190 host-satellite (HS) systems in the  so-called P24 sample (MW/M31-like galaxies, selected and described in \cite{Pillepich24_MWM31}   from the TNG50 simulation), in this paper we look for the physics behind the formation of these 46 KPPs. Specifically, we analyse the formation of these structures in the context of the early formation and evolution of the local environment around these systems.

Early KPP identification, in a given HS system, has been carried out through the so-called `Scanning of Stacked Orbital Poles Method' method, where $\vec{J}_{\rm stack}$ is the fixed in time axial vector around which a maximum of the satellites in
the HS system under consideration orbit, or, put in other words, the axial vector around which a maximum number of satellite orbital poles cluster within an aperture angle of $\alpha_{\rm crit} = 36.^{\circ}87$ along a cosmic period of at least half the Universe age since a Universe age T$_{\rm uni}\simeq6$ Gyr.

A TOI analysis of the deformations suffered by the so-called Lagrangian Volumes (LVs) -- spherical volumes built up at $\zhigh \sim 20$ around the position  
of the c.o.m., at $\zhigh$,  of the dark matter particles bound to the halo at $z=0$, with radii $R_{\rm LV}=15 \times R_{200c}(z=0)$ -- yields the LV principal directions $\vec{e}_i(t)$ ($i=1,2,3$) and the ellipsoid-of-inertia axes $a(t) > b(t) >c(t) $ evolution.

Our analyses indicate  that 67.4\% of  KPPs have their satellite orbital poles aligned with $\vec{e}_3$,  19.6\% to the $\vec{e}_2$, and 2.2\% with the $\vec{e}_1$ principal directions. Moreover, in 8.7\% cases no such alignments appear. These alignments are statistically distinguishable from random alignments, and from those of satellites outside the KPP structure.

We further show that orbital pole clustering essentially results from satellite motion becoming roughly confined to the $\vec{e}_3$-plane (the plane defined by the $\vec{e}_3(t)$ vector and going through the LV's c.o.m.). The $\vec{e}_3$-plane is, for most of the KPP-HS systems  under our study, close to the plane  normal to the $\vec{J}_{\rm stack}$  vector, the so-called $\vec{J}_{\rm stack}$-plane.
Thereby, in this paper we verify and extend the main results obtained in \citetalias{Gamez-Marin2024}, where results for only two zoom-in disc-like systems are presented.

The high-redshift evolution of the local CW (characterised by the LVs) results in the formation of flattened structures. 
In this context, the $\vec{e}_3$-plane represents the mathematical idealization of this structure.
KPP satellite orbital pole clustering essentially results from satellites roughly orbiting within a plane normal to their corresponding $\vec{J}_{\rm stack}$ vector, which coincides with the $\vec{e}_3$-plane in $67.4\%$ of the cases.

In \citetalias{GamezMarin2025_PaperV}, we used the $\kappa_{\rm rot}$ morphological parameter as a descriptor of KPP  rotation (in-plane and circularized motion). We proved that $\kappa_{\rm rot}$ increases along evolution, until it reaches a maximum stable value $\kappa_{\rm rot}^{\rm eq}$, which remains constant except for fluctuations. This confirms that KPP structures generally evolve from dispersion-dominated at early times ($\kappa_{\rm rot} < 0.5$), until a value of $\kappa_{\rm rot}=0.5$ is reached, i.e., they become rotation-dominated, or disky-like, at a Universe age  $T_{\rm Krot}^{0.5}$.

In this paper, we have extended the analysis of the velocity space of KPP satellites by introducing the $\kappa_{\rm z}(t)$ and $\kappa_{\rm rad}(t)$ kinematic  functions. They represent the fraction of kinetic energy involved in parallel and  radial motion, respectively, relative to the $\vec{J}_{\rm stack}$ axial vector, defined in a cylindrical coordinate system centred at the host c.o.m. and velocity, and with vertical axis $\vec{J}_{\rm stack}$. 
We show that the $\kappa_z(t)$ curve decreases until it becomes very low, keeping roughly constant from that moment onwards until $z=0$. 
This implies that, 
for KPP satellites, the fraction of  kinetic energy in vertical motions is very low, i. e., their motion is essentially confined in the $\vec{J}_{\rm stack}$-plane. This establishment of KPP satellites on this plane eventually leads to their orbital poles to cluster as their trajectories are constrained within this plane. 

We also show that, once $\kappa_z(t)$ becomes irrelevant, the $\kappa_{\rm rad}(t)$ curve generally attains low values as well. At this point, most kinetic energy comes from in-plane, rotational motion, which translates into $\kappa_{\rm rot}(t)$  reaching its maximum value $\kappa_{\rm rot}^{\rm eq}$ 
at the characteristic timescale \(T_{\rm Krot}^{\rm eq}\), at which point the KPP satellite system enters into a stationary Keplerian dynamical regime with maximum circularity.

To complement this  analysis, we have also studied the incorporation of KPP satellites onto the $\vec{J}_{\rm stack}$-plane from a positional perspective by means of the median KPP satellite  comoving distance $d_{\rm Js}(t)$ to this plane. We observed that, in most cases, the $d_{\rm Js}(t)$ curve has a clear two-phase shape with a marked 'knee', where the distance reaches a minimum that is kept onward, except for fluctuations. 
Other systems show a slower change of their $d_{\rm Js}(t)$ curves. Nonetheless, they still reach a minimum value that is also kept across time. 
The fact that processes related to the local environment development and the establishment of KPP satellites in their $\vec{e}_i$-plane also show a well-defined two phase scenario is indicative of the possible role that the local CW formation and evolution has on the clustering of KPP satellite poles, i.e., KPP formation.

The shape of the $d_{\rm Js}(t)$ and $d_{ei}(t)$ curves points to the timescales for KPP satellite incorporation onto the $\vec{J}_{\rm stack}$ and $\vec{e}_i$-planes, $T_{\rm dJs}$ and $T_{\rm dei}$ respectively.
We found that $T_{\rm cluster}^{\rm Jstack}$ (the timescale for orbital pole clustering) 
and these timescales are roughly coeval, with both distributions peaking at $\sim4$ Gyr (see median and percentile values in their corresponding CDF plots in Fig.~\ref{ch5:fig_VI:CDF-timescales}), during the fast phase of mass assembly of host halos (i.e. before the halo mass assembly stabilisation time $T_{\rm no-fast}$). This quasi-coevality between timescales suggests that the clustering of orbital poles is related to the settling of KPP satellites into the
$\vec{e}_i$-plane, a plane whose formation is ultimately a consequence of early mass flows taking place in the local CW,
and that, on their turn, drive the establishment of satellites onto a preferential plane.

An interesting aspect of understanding the physical origin of KPP configurations is finding the factors responsible for the clustering and stabilisation of satellite orbital poles, and why either clustering or stabilisation fails in nonKPP-HS systems. One possible differentiating factor is the effect of sustained dynamical activity within the HS system. Our analysis highlights a key difference between KPP-HS and nonKPP-HS systems: while KPP-HS systems exhibit low dynamical activity after T$_{\rm uni}\sim6$ Gyr, nonKPP-HS systems undergo intense late activity, exceeding that of their KPP-HS counterparts (central panel of fig.~4 in \citetalias{GamezMarin2025_PaperV}).


\subsection{Conclusions}

These are the main conclusions reached from the P24 sample analyses on the physical processes behind early KPP  formation:

\begin{enumerate}[label = {\arabic*)}, wide, left=0pt,labelindent=0pt]
    \item A common two-phase scenario is observed across different processes related to both the evolution of the environment -- characterised by dark matter particles forming LVs, and their subsequent deformation, shaping, and orientation -- and the evolution of (proto-)satellite trajectories as they settle into their final planar configuration.
    This two-phase evolution is marked by the early, fast evolution of the LV shapes, driven by the dynamics of the local density. This is indicated by the coevality between the most relevant timescales related to the evolution of the local environment, and those concerning KPP formation, presenting median values around T$_{\rm uni}\sim4$ Gyr.
    
    \item Clustering of satellite orbital poles in KPPs results from their coherent motion within a plane close to normal to the $\vec{e}_3(t)$,  the $\vec{e}_2(t)$ and (less frequently) to the $\vec{e}_1(t)$ principal directions. In other words, KPP formation results from satellite pole alignments with these respective directions.

    \item Two different formation patterns can be distinguished  as the origin  of these high redshift alignments. i) The first pattern involves  persistent satellite orbital pole alignments with $\vec{e}_3(t)$, and, consequently, their velocity vectors lie on the $\vec{e}_3$-plane after bending. KPPs in this case consist of satellites with very early preferential infall along the $\vec{e}_3(t)$ direction prior to bending and that are  successfully bent into the $\vec{e}_3$-plane around T$_{\rm uni}\sim4$ Gyr -- this is, by far, the most frequent pattern of early KPP formation --, along with satellites with almost in-plane trajectories (i.e., velocities preferentially aligned with $\vec{e}_2(t)$ from high redshift). 
ii) The second pattern  refers to satellites whose orbital poles, after KPP formation, are preferentially aligned with the $\vec{e}_2(t)$ principal direction -- and thereby, with motion in the $\vec{e}_2(t)$-plane -- and whose velocity, prior to KPP formation, has a dominant component  in the $\vec{e}_3(t)$ direction. The median velocity of KPP satellite members remains roughly in the same plane after KPP formation. Only two examples exist where the KPP satellites move roughly in the  $\vec{e}_1(t)$-plane
after their formation.

No examples have been found where KPP satellite members move initially preferentially along the $\vec{e}_1(t)$ direction (corresponding to the $a(t)$ axis, the stretching one). 
Nor that move initially along the $\vec{e}_2(t)$ axis and after KPP formation orbit onto a plane close to the $\vec{e}_2$-plane.

    \item We identify three main timescale blocks for KPP-HS systems, which are related to: i) KPP formation and plane collapse at $\sim4$ Gyr ($T_{\rm dJs}$, $T_{\rm cluster}^{\rm Jstack}$, $T_{\rm dir}^{e3}$, $T_{\rm shape}^{e3}$, $T_{\rm align}^{ei}$ and $T_{\rm dei}$), ii) Fast halo mass assembly completion and KPP satellite stabilisation at $\sim6$ Gyr ($T_{\rm no-fast}$ and $T_{\rm Krot}^{\rm eq}$), and iii) KPP satellite turn-around, occurring at early cosmic times, $T_{\rm apo1}$. 

    \item Coevality among timescales corresponding to LV evolution and to KPP formation gives support to a scenario where the same physical processes responsible for high redshift wall-like structure formation (as formulated in the Adhesion Model) are also responsible for high redshift satellite trajectory bending onto these planes, thereby establishing the physical basis of the most common pattern of early KPP formation.
    
    \item The analysis of the origin of the 46 KPPs in the P24 sample  adds diversity to the parameters describing their origin, quantified as statistical  dispersion. While timescale dispersions are generally low among different HS systems, interesting exceptions are: i), the $T_{\rm align}^{ei}$ distribution is bimodal, reflecting the distinction between those {HS systems where satellite pole alignments demand} a bending event to emerge, from those that do not need such an event; ii), $T_{\rm dir}^{e3}$ catches two different populations too: those HS systems where the $\vec{e}_3$ principal axis freezes out very early, and those for which this timescale is spread over the cosmic ages. 
    This diversity underlies the various KPP formation patterns at high redshift summarized in point (3) above, distinct from the most frequent one.   
\end{enumerate}

It can be seen that \textit{early} KPPs are fossil remnants of spatial planar configurations showing up at an early epoch, shaped by the anisotropic collapse of mass into flattened structures. 
In this work we have highlighted how the early evolution of the density field around galaxy formation sites drives KPP formation. Other complementary perspectives for a more advanced understanding of KPP appearance are needed as well, for example why in 62.6\% of eligible HS systems no \textit{early} KPPs are identified. 
This work lays the groundwork for further exploration. 
Understanding low-$z$ formation pathways, such as the 
effects of late dynamical activity, including the late accretion of an LMC-like galaxy with its own sub-satellites system,
may provide deeper insight into their origins and stability.

 \section*{Acknowledgements}
The authors thank A. Pillepich and the IllustrisTNG team for creating and publicly releasing the MW/M31 selection in TNG50. 
All authors thank the Ministerio de Ciencia e Innovación (Spain)  for financial support under Project grant PID2021-122603NB-C21,
as well as Project  PID2024-156100NB-C21  financed by MICIU/AEI
/10.13039/501100011033 / FEDER, EU.
MGM acknowledges support from the MINECO/FEDER funding (Spain) through a FPI fellowship associated to PGC2018-094975-C21 grant, and thanks  Pedro Cataldi and Susana Pedrosa for their help and comments with the merger trees,  as well as Patricia Tissera for her hospitality when visiting PUCC, Chile. He also thanks the Institute for Computational Cosmology, Durham University, for  useful discussions during his research stay there.
ISS acknowledges support by the European Research
Council (ERC) through Advanced Investigator grant to C.S. Frenk, DMIDAS (GA 786910), and from the Science and Technology Facilities Council STFC ST/P000541/1 and ST/X001075/1.
DSR acknowledges financial support from the Ministerio de Ciencia e Innovación (Spain) through grant CNS2024‑154550, funded by MI-CIU/AEI/10.13039/501100011033.
This work has received funding from the European Union's HORIZON-MSCA-2021-SE-01 Research and Innovation programme under the Marie Sklodowska-Curie grant agreement number 101086388 - Project acronym: LACEGAL.

\vspace{-0.5cm}
\section*{Data Availability}
The simulation data used in this study are publicly available at \url{https://www.tng-project.org/data/downloads/TNG50-1/}. The analysis codes  underlying this work may be shared upon reasonable request, within the framework of a scientific collaboration  with the co-authors.

\bibliographystyle{mnras}
\bibliography{bib_THESIS_MGM} 


\clearpage
\onecolumn

\appendix

\clearpage
\section{Glossary}
\label{app:appendixA}

\begin{center}
\captionsetup{
  type=table,
  justification=raggedright,
  singlelinecheck=false,
  labelfont=bf
}
\caption{Summary of terms used in this work other than Timescales (defined in Tab.~\ref{ch5:tab:DataTable2}).}
\label{tab:glossary}

\begin{tabularx}{\textwidth}{lX @{}}
\hline
  \textbf{Term} & \textbf{Definition} \\
\hline
HS system & Host-satellite system, formed by a central galaxy and its satellites, identified through the selection criteria specified in Sec. \ref{ch4:sec:Simus}\\
Attraction basin & A region of space where the mass (dark matter, gas, stars and their structures) within it is gravitationally dragged towards a common, high mass density volume, such as a galaxy group or cluster. \\
LV & Lagrangian Volume, see  Section \ref{ch5:sec:LVs_intro}.\\
TOI         & Reduced Tensor of Inertia, $I_{i,j}^{\rm r}$, \citep[][ see Eq.~\ref{reducedI}]{Cramer}. 
\\
 $\vec{e}_i(t)$, $i=1,2,3$  &  The $i$-th principal direction of the TOI  at Universe age $t$. \\
$a(t), b(t), c(t)$ & The major, intermediate and minor principal axes of the TOI  at Universe age $t$.\\
$\vec{J}_{\rm stack}$ for a given HS system & Axial, fixed-in-time vector around which a maximum of the satellite orbital poles in an HS system cluster within an aperture angle of $\alpha_{\rm crit} = 36.^{\circ}87$, along a period of at least 4 Gyr since a Universe age T$_{\rm uni}\simeq 6$ Gyr.  \\
The $\vec{J}_{\rm stack}$-plane for an HS system & Plane normal to the fixed-in-time axial vector $\vec{J}_{\rm stack}$, that passes through the central galaxy's centre of mass. .\\
$\vec{e}_i$-planes, $i$=1,2,3, for an HS system & Same as the $\vec{J}_{\rm stack}$-plane, but considering the $\vec{e}_i(t)$ ($i=1,2,3$) principal directions for the HS system at Universe age $t$ instead of the $\vec{J}_{\rm stack}$ fixed-in-time direction.  \\
KPP of satellites in a given HS system &  Set of satellites, with fixed identities along time, whose orbital poles stay within an  aperture angle of  $\alpha_{\rm crit}$ from $\vec{J}_{\rm stack}$, along  a time interval of, at least, 4 Gyr between
 T$_{\rm uni}\simeq 6$ Gyr (a typical value for the halo mass stabilisation timescale) up to $z=0$.\\
CS$(t)$ set for an HS  system at time $t$ & Set of co-orbiting satellites, i.e., satellites that, at time $t$, are within an angle of $\alpha_{\rm crit}$ from the corresponding $\vec{J}_{\rm stack}$. The CS$(t)$ set consists of the satellites belonging to the kinematically persistent plane (KPP)  in this HS system, plus those non-KPP satellites that at time $t$ happen to temporarily co-orbit around $\vec{J}_{\rm stack}$.\\
KPP-HS system &  An HS system where a kinematically persistent plane (KPP) of satellites has been identified according with the selection criteria specified in Sec. \ref{sec:KPP-identification}.\\
nonKPP-HS system &  An HS system where no KPP has been identified, following the selection criteria explained in Sec. \ref{sec:KPP-identification}.\\
$N_{\rm sat}$, $N_{\rm KPP}$  &   Number of satellites in a given HS system; number of satellites in the KPP of this HS system. \\
$f_{\rm KPP}$  &   Fraction of KPP satellites, $N_{\rm KPP}$/$N_{\rm sat}$, in a given HS system.\\
$\kappa_{\rm rot, i}(t)$  ($i$-th  satellite in an HS system) & The fraction of its mechanical energy, at Universe age $t$, in rotational motion (i.e., in plane and along the azimutal angle in a cylindrical coordinate system with  $\vec{J}_{\rm stack}$ as the $Z$ axis).\\
$\kappa_{\rm z, i}(t)$, $\kappa_{\rm rad, i}(t)$ for the $i$-th  satellite & Same as $\kappa_{\rm rot, i}(t)$ for the satellite velocity component along 
$\vec{J}_{\rm stack}$ and normal to it (radial component), respectively, at Universe age $t$.\\
$\kappa_{\rm rot}(t)$, $\kappa_{\rm z}(t)$, $\kappa_{\rm rad}(t)$  for a satellite set &  The median values
of  $\kappa_{\rm rot, i}(t)$, $\kappa_{\rm z, i}(t)$ and $\kappa_{\rm rad, i}(t)$, respectively, 
for the satellite set members, at Universe age $t$. \\
$\kappa_{\rm rot}^{\rm eq}$ for a satellite set & Equilibrium value its $\kappa_{\rm rad}(t)$ curve reaches (except for fluctuations) at Universe age $T_{\rm Krot}^{\rm eq}$.\\
$d_{k, \rm Js}(t)$; $d_{\rm Js}(t)$ for a satellite set & Comoving distance from the $k$-th satellite set member centre of mass, to the $\vec{J}_{\rm stack}$-plane, at Universe age $t$. $d_{\rm Js}(t)$ is the median value for the satellite subset members at Universe age $t$.\\
$T(t)$ &  Triaxiality parameter at Universe age $t$, see definition in Eq. \ref{TshapeDef}.\\
$d_{k, ei}(t)$, $d_{ei}(t)$, $i$ = 1,2,3 &  Same as $d_{k, \rm Js}(t)$ and $d_{\rm Js}(t)$, respectively, but considering  the corresponding $ \vec{e}_i(t)$-planes ($i=1,2,3$) instead of the $\vec{J}_{\rm stack}$-plane. \\
$A_{j,i}$. $A_i$, $i$ = 1,2,3 for a satellite set &  Alignment signals of the velocity vector, $\vec{v}_{j}$, $j$=1,..., $N_{\rm set}$,  with the 3 principal directions $\vec{e}_i(t)$, $i=1,2,3$: cos$(\vec{v}_{j}$,$\vec{e}_i)\equiv A_{j,i}$. $A_i$, $i$ = 1,2,3 are the medians for the satellite set. $N_{\rm set}$ is the number of satellites in the set. \\
\hline  
\end{tabularx}
\label{tab:Glossary}
\end{center}

\section{Complementary Figures}
\label{app:appendix}

\begin{center}
\captionsetup{
  type=figure,
  justification=raggedright,
  singlelinecheck=false,
  labelfont=bf
}
\includegraphics[width=0.24\linewidth]{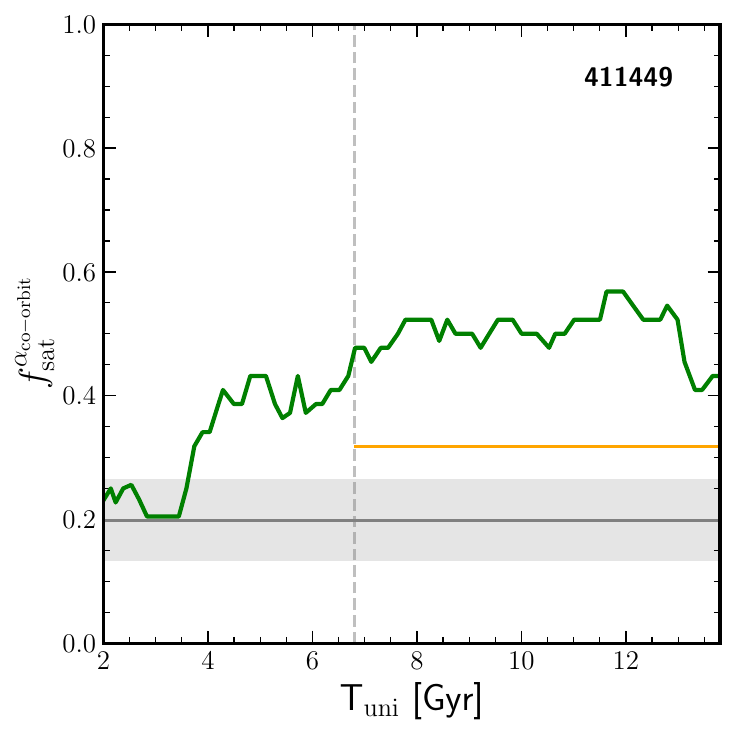}
\includegraphics[width=0.24\linewidth]{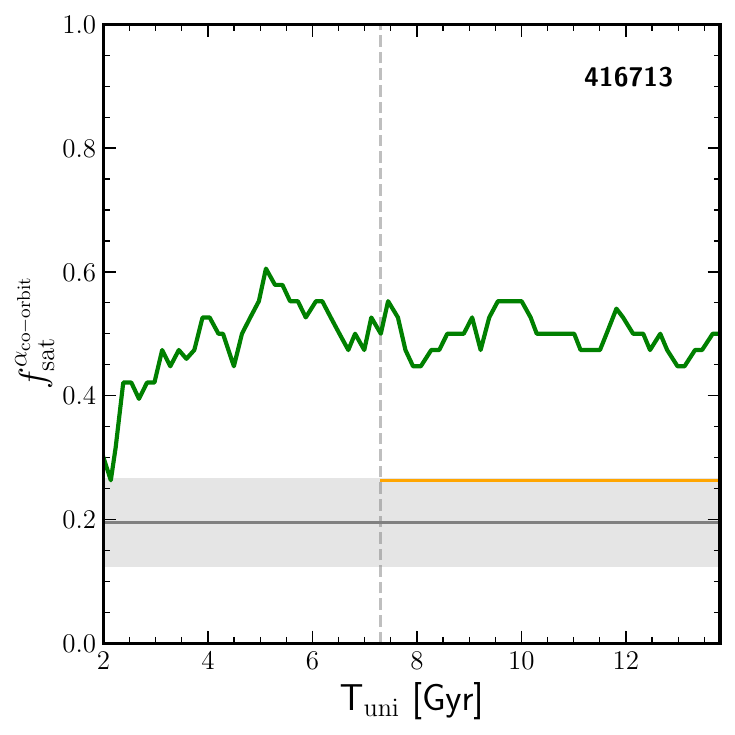}
\includegraphics[width=0.24\linewidth]{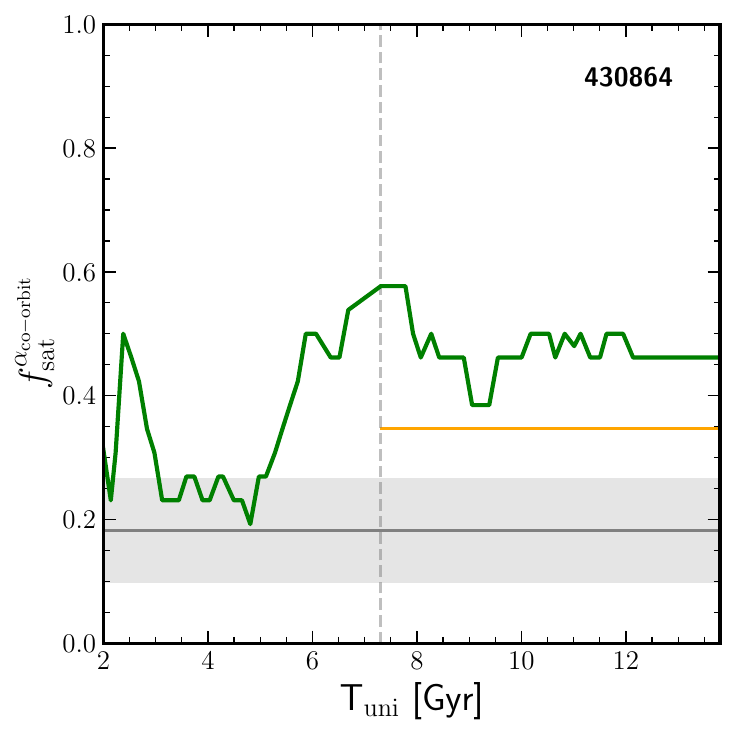}
\includegraphics[width=0.24\linewidth]{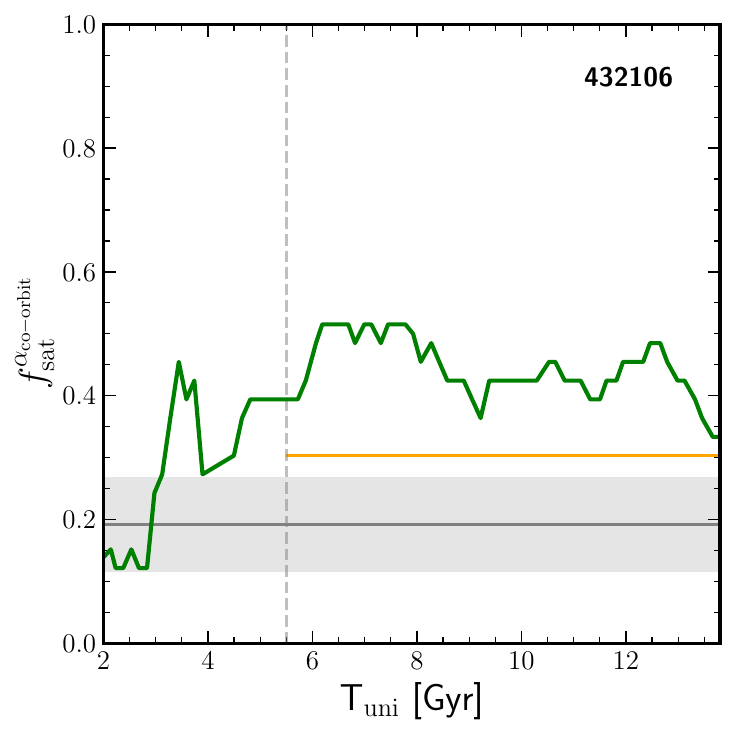}
\caption{The the satellite co-orbiting   fraction curve $f_{\rm sat}^{\alpha_{\rm co-orbit}}(t)$ for four early KPP systems (green line). Yellow horizontal lines stand for the $f_{\rm KPP}$ fraction, while vertical dashed lines mark the $T_{\rm no-fast}$ timescale. ID\# are 411449, 416713, 430864, and 432106. Black lines and gray shadows are results for isotropized versions of the satellite orbital poles distribution. These examples show a higher degree of co-orbitation  than the estimated value for the Milky Way satellite system; see text.} 
\label{fig:coorbi-examples}
\end{center}


\bsp	
\label{lastpage}
\end{document}